\documentclass[11pt]{article}
\usepackage{axodraw}
\usepackage{epsfig}
\usepackage{amsfonts}
\usepackage{amsmath}
\usepackage{bbm}
\allowdisplaybreaks[1]

\input pix.sty

\newcommand{\ko}{k_0}
\newcommand{\km}{k_-}
\newcommand{\kp}{k_+}
\newcommand{\aL}{a^{ }_\rmii{L}}
\newcommand{\aR}{a^{ }_\rmii{R}}
\renewcommand{\eq}{eq.~}
\renewcommand{\eqs}{eqs.~}
\renewcommand{\se}{sec.~}

\renewcommand{\fig}{fig.~}
\renewcommand{\figs}{figs.~}

\newcommand{\Nc}{N_{\rm c}}

\newcommand{\gammaE}{\gamma_\rmii{E}}

\newcommand{\rmO}{{\mathcal{O}}}
\newcommand{\bmu}{\bar\mu}

\def\lsi{\raise0.3ex\hbox{$<$\kern-0.75em\raise-1.1ex\hbox{$\sim$}}}
\def\gsi{\raise0.3ex\hbox{$>$\kern-0.75em\raise-1.1ex\hbox{$\sim$}}}
\newcommand{\lsim}{\mathop{\lsi}}
\newcommand{\gsim}{\mathop{\gsi}}

\newcommand{\nF}{n_\rmii{F}}
\newcommand{\nB}{n_\rmii{B}}
 \renewcommand{\nF}[1]{n_\rmii{F{#1}}}
 \renewcommand{\nB}[1]{n_\rmii{B{#1}}}

\newcommand{\rmii}[1]{{\mbox{\tiny\rm{#1}}}}

\newcommand{\re}{\mathop{\mbox{Re}}}
\newcommand{\im}{\mathop{\mbox{Im}}}

\newcommand{\Tint}[1]{{\hbox{$\sum$}\!\!\!\!\!\!\!\int\,}_{\!\!\!\!\raise-0.9ex\hbox{$\scriptstyle{#1}$}}}
\newcommand{\Tinti}[1]{{{\Sigma}\!\!\!\!\raise0.3ex\hbox{$\int$}_\rmii{${#1}$}}}

\newcommand{\bi}{\begin{itemize}}
\newcommand{\ei}{\end{itemize}}

\newcommand{\hide}[1]{ }
\newcommand{\bsl}[1]{\,\slash\!\!\!\!{#1}\,}

 \newcommand{\mufac}{ } 
\def\TAsc(#1,#2)(#3,#4,#5)%
{\SetWidth{2.0}\CArc(#1,#2)(#3,#4,#5)\SetWidth{1.0}}
\def\Lwidth{3}

\def\TAgl(#1,#2)(#3,#4,#5){\SetWidth{2.0}\PhotonArc(#1,#2)(#3,#4,#5){\Lwidth}%
{6.283 #3 mul 360 div #4 #5 sub #4 #5 sub mul sqrt mul Tdensity mul}%
\SetWidth{1.0}}
\def\TLgl(#1,#2)(#3,#4){\SetWidth{2.0}\Photon(#1,#2)(#3,#4){\Lwidth}
{#1 #3 sub #1 #3 sub mul #2 #4 sub #2 #4 sub mul add sqrt Tdensity mul}%
\SetWidth{1.0}}

\renewcommand{\picc}[1]{\;\parbox[c]{60pt}{\begin{picture}(60,30)(0,0)
\SetWidth{1.0}\SetScale{1.3} #1 \end{picture}}\; }
\def\Lwidth{1.3}

\newcommand{\picv}[1]{\;\parbox[c]{80pt}{\begin{picture}(80,70)(0,-5)
\SetWidth{1.0}\SetScale{1.0} #1 \end{picture}}\; }
\def\Generic{\picv{%
 \CArc(40,30)(30,0,360)%
 \Lqu(0,30)(10,30)%
 \Lqu(70,30)(80,30)%
 \Line(40,0)(40,60)%
 \Text(-5,30)[r]{$\scriptstyle \sigma_0, K$}%
 \Text(85,30)[l]{$\scriptstyle \sigma_0, K$}%
 \Text(60,60)[l]{$\scriptstyle \sigma_2, Q$}%
 \Text(27,30)[l]{$\scriptstyle \sigma_5, \;\, Q-P$}
 \Text(20,0)[r]{$\scriptstyle \sigma_4, P-K$}%
 \Text(20,60)[r]{$\scriptstyle \sigma_1, P$}%
 \Text(60,0)[l]{$\scriptstyle \sigma_3, Q-K$}%
}}
\makeatletter \@addtoreset{equation}{section} \makeatother
\renewcommand{\theequation}{\arabic{section}.\arabic{equation}}
\makeatletter
\renewcommand\section{\@startsection {section}{1}{\z@}%
                                   {-5.5ex \@plus -1ex \@minus -.2ex}
                                   {2.3ex \@plus.2ex}%
                                   {\normalfont\large\bfseries}}
\renewcommand\subsection{\@startsection{subsection}{2}{\z@}%
                                     {-3.25ex\@plus -1ex \@minus -.2ex}%
                                     {1.5ex \@plus .2ex}%
                                     {\normalfont\normalsize\bfseries}}
\renewcommand\thesection {\@arabic\c@section}
\renewcommand\thesubsection   {\thesection.\@arabic\c@subsection}
\renewcommand{\@seccntformat}[1]{%
\csname the#1\endcsname.\hspace{1.0em}}
\makeatother

\begin{document}

\flushbottom

\begin{titlepage}

\begin{flushright}
\vspace*{1cm}
\end{flushright}
\begin{centering}
\vfill

{\Large{\bf
 Thermal right-handed neutrino production rate \\[2mm] 
 in the relativistic regime
}} 

\vspace{0.8cm}

M.~Laine 

\vspace{0.8cm}

{\em
Institute for Theoretical Physics, 
Albert Einstein Center, University of Bern, \\ 
Sidlerstrasse 5, CH-3012 Bern, Switzerland\\}


\vspace*{0.8cm}

\mbox{\bf Abstract}
 
\end{centering}

\vspace*{0.3cm}
 
\noindent
The production rate of right-handed neutrinos from a Standard Model
plasma at a temperature above a hundred GeV is evaluated up to NLO 
in Standard Model couplings. The results apply in the so-called
relativistic regime, referring parametrically to a mass $M \sim \pi T$, 
generalizing thereby previous NLO results which only apply in the
non-relativistic regime $M \gg \pi T$.  The non-relativistic expansion
is observed to converge for $M \gsim 15 T$, but the smallness of 
{\em any} loop corrections allows it to be used in practice already
for $M \gsim 4 T$. In the latter regime any non-covariant dependence
of the differential rate on the spatial momentum is shown to be 
mild. The loop expansion breaks down in the ultrarelativistic
regime $M \ll \pi T$, but after a simple mass resummation it
nevertheless extrapolates reasonably well towards a result obtained
previously through complete LPM resummation, apparently confirming a
strong enhancement of the rate at high temperatures (which facilitates
chemical equilibration). When combined with other ingredients
the results may help to improve upon the accuracy of leptogenesis
computations operating above the electroweak scale.

\vfill

 
\vspace*{1cm}
  
\noindent
August 2013

\vfill

\end{titlepage}

%
\section{Introduction}

The neutrino sector is arguably the least precisely charted 
among the different parts of experimentally accessible 
particle physics. Whereas it is well established that at least two
dominantly left-handed neutrinos have masses, and that the 
mass eigenstates are misaligned with the weak interaction 
eigenstates, already the absolute value of 
the mass scale remains poorly constrained. There is also feeble 
empirical handle on the dynamics of neutrino mass generation, 
although a see-saw mechanism involving Majorana masses of
gauge-singlet right-handed neutrinos is a natural candidate. 
Of course right-handed neutrinos can be introduced in any case,
but then a large parameter space of Yukawa couplings and 
Majorana masses, the latter unbounded from above, remains available to 
phenomenological consideration. 

The big volume of the parameter space suggests seeking 
for cosmological constraints on neutrino properties. 
Apart from the well-studied significance of left-handed
neutrinos to the overall expansion rate of the Universe
through the pressure and energy density that they exert, 
it is also possible that right-handed neutrinos
have cosmological significance. For instance, they could play a role
in explaining two outstanding cosmological mysteries, the existence
of a matter-antimatter asymmetry~\cite{yanagida} 
(for reviews, see e.g.\ refs.~\cite{lepto1,lepto2})
as well as the existence of particle dark matter~\cite{shifuller}
(for a review see e.g.\ ref.~\cite{numsm}). 

The present paper is related to developing theoretical tools for studying 
right-handed neutrinos within a cosmological environment. It has been 
understood recently~\cite{bb0,bb1,bb2} that in the so-called
{\em ultrarelativistic} regime $\pi T \gg M$, where $T$ denotes
the temperature and $M$ a right-handed neutrino Majorana mass parameter, 
the thermal loop expansion breaks down and needs to be resummed with
techniques analogous to those that were previously developed in the context 
of photon production from a hot QCD plasma~\cite{amy1,amy2,amy3}.
In contrast,  in the so-called {\em non-relativistic} regime $\pi T \ll M$,
next-to-leading order (NLO) corrections can be computed and are
in general small~\cite{salvio,nonrel}, in accordance with 
expectations based on the Operator Product Expansion (OPE)~\cite{sch}.  
This leaves open the question of how these two qualitatively 
very different regimes interpolate to each other. 

The purpose of the present paper is to present a theoretically 
consistent computation of the right-handed neutrino production 
rate in the so-called {\rm relativistic} regime, $\pi T \sim M$.
The principal tools needed for this 
have been developed in ref.~\cite{master}, 
and here we assemble the full results. 
We also inspect under which conditions the results 
go over to those of the limiting ultrarelativistic and 
non-relativistic cases. In addition the structure of the 
differential production rate is analyzed with the goal of 
suggesting a numerically affordable and yet
relatively accurate approximation scheme that may be 
used in practical applications (however the study of practical applications
is postponed to future work). Another project with partly similar
goals has recently been  outlined in ref.~\cite{gh}.

After summarizing the setup in \se\ref{se:setup}, the basic 
theoretical results, together with comparisons with the non-relativistic
and ultrarelativistic regimes, are presented in \se\ref{se:results}. 
Examples of numerical results for differential production spectra are 
shown in \se\ref{se:spectra}, whereas in \se\ref{se:rate} the total 
production rate is considered. Some conclusions and an outlook 
are offered in \se\ref{se:concl}.

%
\section{Setup}
\la{se:setup}

Solving the Liouville - von Neumann equation
for the time evolution of the density matrix of a coupled system 
of Standard Model particles and right-handed neutrinos to leading 
non-trivial order in neutrino Yukawa couplings (but to all 
orders in Standard Model couplings), for times sufficiently
small that the right-handed neutrinos do not
chemically equilibrate~\cite{hadronic}, 
it is found that their ``differential production rate'' can be written as
\be
 \frac{{\rm d}N_+ (\mathcal{K})}{{\rm d}^4\mathcal{X} {\rm d}^3\vec{k}}
 \; \equiv \;
 \left.
 \frac{{\rm d}N (\mathcal{K})}{{\rm d}^4\mathcal{X} {\rm d}^3\vec{k}}
 \right|^{ }_{
 \frac{{\rm d}N (\mathcal{K})}{{\rm d}^3\vec{x}\, {\rm d}^3\vec{k}}
 \; \approx \; 
 0 
 }
 \; = \;
 \frac{2\nF{}(\ko)}{(2\pi)^3 } \; \Gamma(\mathcal{K})
 \;. \la{Gamma_plus}
\ee
Here a ``width'' has been defined as
\be
 \Gamma(\mathcal{K}) \equiv 
 \frac{1}{\ko}
 \im \Pi^{ }_\rmii{R}(\mathcal{K})
 =
 \frac{1}{\ko}
 \im\bigl\{ \Pi_\rmii{E}(K) \bigr\}_{k_n \to -i [\ko + i 0^+]}
 \;, \la{master2}
\ee
where $\Pi_\rmii{E}$ is a gauge-invariant and Lorentz-singlet 2-point 
correlation function 
of the ``currents'' that right-handed neutrinos couple to,
\be
 \Pi_\rmii{E}(K) \equiv |h_{\nu\rmii{B}}|^2 \, 
 \tr\Bigl\{ 
   i \bsl{K}  
   \Bigl[
     \int_0^{1/T} \! {\rm d}\tau \int_\vec{x} e^{i K\cdot X}
   \Bigl\langle
     (\tilde{\phi}^\dagger \aL \ell)(X) \; 
     (\bar{\ell}  \aR \tilde\phi)(0)
   \Bigr\rangle^{ }_T 
   \Bigr]
 \Bigr\}
 \;, \la{PiE_def}
\ee
and $\Pi_\rmii{R}$ is the corresponding retarded real-time correlator
(its imaginary part equals the spectral function). 
Moreover, 
$h_{\nu\rmii{B}}$ denotes a bare neutrino Yukawa coupling 
(or, more generally, elements of a Yukawa matrix); 
Euclidean variables are denoted by 
$X \equiv (\tau,\vec{x})$, 
$K \equiv (k_n,\vec{k})$; 
the corresponding Minkowskian ones by
$\mathcal{X} \equiv (t,\vec{x})$, 
$\mathcal{K} \equiv (\ko,\vec{k})$; 
the metric conventions are
$K^2 = k_n^2 + k^2$,  
$\mathcal{K}^2 = \ko^2 - k^2$, with $k \equiv |\vec{k}|$;
$k_n$ stands for a fermionic Matsubara frequency, 
reflecting the fact that spin-$\fr12$ fields are antiperiodic
across the Euclidean time direction of extent $1/T$. (Definitions
of other variables appearing in \eq\nr{PiE_def} can be 
found in ref.~\cite{nonrel}.)
A corresponding integrated ``total production rate'' is 
\be
 \gamma_+(T)
 \equiv  \frac{{\rm d}N_+}{{\rm d}^4\mathcal{X}}
 = \int \! \frac{ {\rm d}^3\vec{k} }{(2\pi)^3 }
 \; {2\nF{}(\ko)} \; \Gamma(\mathcal{K})
 \;. \la{gamma}
\ee

We note in passing that an opposite case of a ``differential decay rate'', 
defined by taking the thermal average of a spin-summed 
rate for the disappearance 
of a right-handed neutrino of momentum $\vec{k}$, 
can be expressed in terms of the same 
function $\Gamma(\mathcal{K})$ that appears in \eq\nr{master2}:
\be
 \frac{{\rm d}N_- (\mathcal{K})}{{\rm d}^4\mathcal{X} {\rm d}^3\vec{k}}
 \; = \;
 - \frac{2 [ 1 - \nF{}(\ko)] }{(2\pi)^3 } \; \Gamma(\mathcal{K})
 \;. \la{Gamma_minus}
\ee

The observables above are particularly simple because they involve a sum
over the spin states of the right-handed neutrinos. A more general problem
concerns the determination of the self-energy matrix of the right-handed
neutrinos, given to leading order in neutrino Yukawa couplings by 
\eq\nr{PiE_def} without a Dirac contraction with $ i \bsl{K} $. 
An NLO discussion of this observable 
in the non-relativistic regime can be found in ref.~\cite{selfE}.

Returning to \eq\nr{PiE_def}, one of the strengths of the imaginary-time
formulation of thermal field theory 
is that the expression obtained for $\Pi_\rmii{E}$ can be 
significantly simplified through substitutions of loop momenta before 
taking the cut leading to $\Gamma$. In fact, as shown in 
ref.~\cite{nonrel}, $\Pi_\rmii{E}$ can be represented in terms 
of a small number of ``master'' sum-integrals. For the specific
case of naive dimensional regularization of 
the $\gamma_5$-matrix, the expression reads
\ba
  \frac{\Pi_\rmii{E}}{|h_{\nu\rmii{B}}|^2 } & = & 2  
  \, \Bigl(
   \widetilde{\mathcal{J}}_\rmi{a} 
   -\mathcal{J}^{ }_\rmi{a} 
   -\mathcal{J}^{ }_\rmi{b}
  \Bigr)
  \nn
  & + & 12 \lambda_\rmii{B}
  \Bigl(
   -\mathcal{I}^{ }_\rmi{b}
   +\mathcal{I}^{ }_\rmi{c}
   +\mathcal{I}^{ }_\rmi{d}
  \Bigr) 
  \nn
  & + & 2  h_{t\rmii{B}}^2 \Nc 
  \Bigl[
   2\Bigl( \widetilde{\mathcal{I}}_\rmi{b}  
  -  \widetilde{\mathcal{I}}_\rmi{c}  
  -  \widetilde{\mathcal{I}}_\rmi{d} \Bigr)
  +  \widetilde{\mathcal{I}}_\rmi{e}  
  -  \widetilde{\mathcal{I}}_\rmi{f}  
  +  \widetilde{\mathcal{I}}_\rmi{h}    
  \Bigr] 
  \nn
  & + &  \frac{ g_{1\rmii{B}}^2 + 3 g_{2\rmii{B}}^2 }{2}
  \Bigl[
  -        \mathcal{I}^{ }_\rmi{b}
    +  2 \Bigl( \widetilde{\mathcal{I}}^{ }_\rmi{e}
     -  \mathcal{I}^{ }_\rmi{e}
     +  \mathcal{I}^{ }_\rmi{g}
     +  \mathcal{I}^{ }_\rmi{j} \Bigr)
     - 4 \Bigl( \mathcal{I}^{ }_\rmi{h}
     +  \widehat{\mathcal{I}}^{ }_\rmi{h}  \Bigr)   
  \nn 
  & + & 
    ({D-1})  \Bigl(
       \mathcal{I}^{ }_\rmi{c}
      +\mathcal{I}^{ }_\rmi{d}     
    \Bigr) 
   + 
    ({D-2}) \Bigl(
      \overline{\mathcal{I}}^{ }_\rmi{c}
      -\overline{\mathcal{I}}^{ }_\rmi{d}     
      -\widetilde{\mathcal{I}}^{ }_\rmi{b}
      -\widehat{\mathcal{I}}^{ }_\rmi{c}
      +\widehat{\mathcal{I}}^{ }_\rmi{d}     
      +\widehat{\mathcal{I}}^{ }_\rmi{h'}     
    \Bigr)
  \Bigr]
  \;, \la{PiE}
\ea
where $\lambda_{\rmii{B}}$, 
$h_{t\rmii{B}}$, $g_{1\rmii{B}}$, $g_{2\rmii{B}}$ denote
the bare Higgs, top Yukawa, U(1) gauge, and SU(2) 
gauge couplings, respectively; $\Nc = 3$ is the number of colours;   
and $D \equiv 4-2\epsilon$ is the space-time dimensionality. The definitions
of the independent master sum-integrals 
$\mathcal{J}^{ }_\rmi{a}, ... $ are listed in appendix~A. 
Renormalization of this bare expression is achieved through 
\ba
 |h_{\nu\rmii{B}}|^2 & = &  |h_\nu(\bmu)|^2 \mu^{2\epsilon} 
 \,\mathcal{Z}_\nu \;, \quad 
  \mathcal{Z}_\nu \; \equiv \; 
  1 +  \frac{1}{(4\pi)^2\epsilon}
   \Bigl[
     h_t^2 \Nc - \fr34 (g_1^2 + 3 g_2^2) 
   \Bigr] 
   \;,  \la{Znu}
\ea
where $\mu$ is a scale parameter related to dimensional regularization
(in the following inconsequential factors 
$\mu^{\pm2\epsilon}$ are omitted); 
the $\msbar$ scale is defined as 
$
 \bmu^2 \equiv 4\pi\mu^2 e^{-\gammaE} 
$; 
and $h_t$, $g_1$, $g_2$ denote the renormalized
top Yukawa, U(1) gauge, and SU(2) 
gauge couplings, respectively.

Taking a cut like in \eq\nr{master2} leads to what we term 
master spectral functions:
\be
 \rho^{ }_{\mathcal{I}^{ }_\rmi{x}} \equiv
 \im [ \mathcal{I}^{ }_\rmi{x} ]^{ }_{k^{ }_n \to -i [\ko + i 0^+]} 
 \;. \la{cut}
\ee
Numerical results for all of these are listed
in appendix~B, apart from $\rho^{ }_{\mathcal{I}_\rmii{j}}$; 
the case $\rho^{ }_{\mathcal{I}_\rmii{j}}$,
together with the general methodology used, 
were discussed in ref.~\cite{master}.
In the next section we collect the results obtained
after inserting the master 
spectral functions into the imaginary part of \eq\nr{PiE}.

%
\section{Main results}
\la{se:results}

%
\subsection{Strict NLO expression}

Each of the master spectral functions can be split into two parts: 
\be
 \rho^{ }_{\mathcal{I}^{ }_\rmi{x}} 
 = 
  \rho^{\rmi{vac}}_{\mathcal{I}^{ }_\rmi{x}} 
 + 
  \rho^{\rmii{$T$}}_{\mathcal{I}^{ }_\rmi{x}} 
 \;. \la{splitup}
\ee
The first term must include all divergences, and may be 
chosen to include
finite parts as well. We note that although denoted by 
$
  \rho^{\rmi{vac}}_{\mathcal{I}^{ }_\rmi{x}} 
$, 
this structure does have an overall temperature
dependence, of the same functional form as the leading-order (LO) 
result which it renormalizes. The purely thermal part 
$
  \rho^{\rmii{$T$}}_{\mathcal{I}^{ }_\rmi{x}} 
$
is, in contrast, finite and of a more complicated functional form. 
The divergences 
of the vacuum parts cancel against those in 
$\mathcal{Z}^{ }_\nu$, \eq\nr{Znu}.
Subsequently, with the choices of 
$
  \rho^{\rmi{vac}}_{\mathcal{I}^{ }_\rmi{x}} 
$
explained in appendix~B, 
we obtain a finite renormalized expression
for the imaginary part of the retarded correlator: 
\ba
 \frac{\im \Pi^{ }_\rmii{R}}{|h_\nu(\bmu)|^2} & = & 
 \frac{{M}^2 T}{8\pi k} 
 \ln \biggl[ \frac{\sinh(\kp / T)}{\sinh(\km / T)}\biggr]
 \nn & + &   
 12 \lambda \Bigl\{ -\rho^\rmii{$T$}_{\mathcal{I}^{ }_\rmii{b}}
 + \rho^\rmii{$T$}_{\mathcal{I}^{ }_\rmii{d}} \Bigr\}
 \nn & + &   
 2 h_t^2 \Nc 
 \biggl\{
   2 \Bigl[ \rho^\rmii{$T$}_{\widetilde{\mathcal{I}}^{ }_\rmii{b}}
       -  \rho^\rmii{$T$}_{\widetilde{\mathcal{I}}^{ }_\rmii{d}} \Bigr]
   - \rho^\rmii{$T$}_{\widetilde{\mathcal{I}}^{ }_\rmii{f}}
   + \rho^\rmii{$T$}_{\widetilde{\mathcal{I}}^{ }_\rmii{h}}
 \nn & & \; - \, 
   \frac{\pi {M}^2}{(4\pi)^4 k}
   \int_{\km}^{\kp} \! {\rm d}p \, 
   \frac{ \nF{}(\ko -p) \nB{}(p) }{\nF{}(\ko)}
   \biggl[
     \ln \frac{(\kp - p)(p - \km)\bmu^2 }{k^2 M^2} + \fr{11}{2} 
   \biggr]
 \biggr\}
 \nn & + &   
 \frac{g_1^2 + 3 g_2^2}{2}
 \biggl\{
    - \rho^\rmii{$T$}_{{\mathcal{I}}^{ }_\rmii{b}} 
   + 2 \Bigl[
   - \rho^\rmii{$T$}_{\widetilde{\mathcal{I}}^{ }_\rmii{b}} 
   + \rho^\rmii{$T$}_{\widehat{\mathcal{I}}^{ }_\rmii{d}} 
   - \rho^\rmii{$T$}_{\overline{\mathcal{I}}^{ }_\rmii{d}} 
     \Bigr]
   + 3 \rho^\rmii{$T$}_{{\mathcal{I}}^{ }_\rmii{d}}
  \nn & & \; + \, 
     2 \Bigl[ 
     \rho^\rmii{$T$}_{{\mathcal{I}}^{ }_\rmii{g}}
    + \rho^\rmii{$T$}_{\widehat{\mathcal{I}}^{ }_\rmii{h'}}
    + \rho^\rmii{$T$}_{{\mathcal{I}}^{ }_\rmii{j}}
     \Bigr] 
    - 4 \Bigl[ 
       \rho^\rmii{$T$}_{{\mathcal{I}}^{ }_\rmii{h}}
    + \rho^\rmii{$T$}_{\widehat{\mathcal{I}}^{ }_\rmii{h}}
     \Bigr] 
 \nn & & \; + \, 
   \frac{3 \pi {M}^2}{(4\pi)^4 k}
   \int_{\km}^{\kp} \! {\rm d}p \, 
   \frac{ \nF{}(\ko -p) \nB{}(p) }{\nF{}(\ko)}
   \biggl[
     \ln \frac{(\kp - p)(p - \km)\bmu^2 }{k^2 {M}^2} + \fr{41}{6} 
   \biggr]
 \biggr\}
 \;. \hspace*{10mm} \la{imNLO}
\ea
Here we have defined 
\be
 k^{ }_\pm \equiv \frac{\ko \pm k}{2}
 \;, \quad 
 M^2 \equiv \mathcal{K}^2 = 4 \kp \km > 0 
 \;, \la{kpm}
\ee 
and $\nB{}$, $\nF{}$ are the Bose and Fermi distributions.  

\begin{figure}[t]


\centerline{%
 \epsfysize=7.5cm\epsfbox{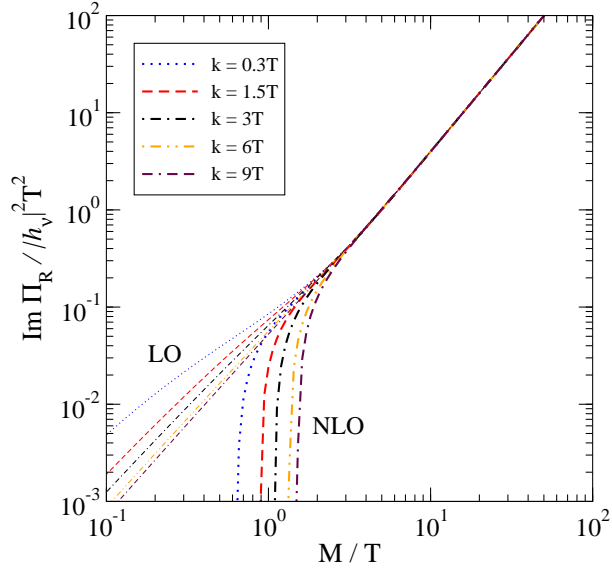}%
}

\caption[a]{\small
The expression from \eq\nr{imNLO}, in units of $T^2$, 
for $\ko^2 \equiv k^2 + M^2$. The couplings and the renormalization
scale are fixed as specified in appendix~C. The loop expansion breaks down 
at $M \sim T$.  
}

\la{fig:imNLO}
\end{figure}

A numerical evaluation of this expression is shown in 
\fig\ref{fig:imNLO} (parameters and the renormalization scale are 
chosen as explained in appendix~C). Results are displayed at several momenta 
on both sides of $k = 3T$. 
It is clear that the naive loop expansion breaks down for $M\sim T$. 
Based on this plot one might conclude that loop corrections decrease
the production rate but, as will become apparent in 
\se\ref{ss:ur}, such a conclusion
is premature.   

%
\subsection{Non-relativistic limit}

In the non-relativistic limit ${M}^2 \gg (\pi T)^2$
\eq\nr{imNLO} can be represented
in more explicit form: up to and including $\rmO(T^4/{M}^2)$, 
this ``OPE'' expression reads~\cite{nonrel}
\ba
 \frac{\im \Pi^{ }_\rmii{R}}{|h_\nu(\bmu)|^2} & = & 
 \frac{{M}^2}{8\pi}
 \biggl\{
  1 - \frac{12\lambda}{{M}^2} \int_\vec{p} \frac{\nB{}}{p}
 \la{imASY} \\
 & & \; - \, 
 h_t^2\Nc 
 \biggl[
   \frac{1}{(4\pi)^2} 
     \biggl( 
      \ln\frac{\bmu^2}{{M}^2} + \fr72
     \biggr)
  + \frac{k_0^2 + k^2/3}{{M}^6} \int_\vec{p} \frac{4 p\, \nF{}}{3}
 \biggr]
 \nn & & \; + \, 
 (g_1^2 + 3 g_2^2)
 \biggl[
   \frac{3}{4(4\pi)^2} 
     \biggl( 
      \ln\frac{\bmu^2}{{M}^2} + \fr{29}6
     \biggr)
  + \frac{k_0^2 + k^2/3}{{M}^6} \int_\vec{p} 
  \frac{p\, (17 \nF{} - 16 \nB{}) }{3}
 \biggr]
 \biggr\}  
 \;, \nonumber
\ea
where the integrals over the phase space distributions have 
elementary forms: 
\be
 \int_\vec{p} \frac{\nB{}}{p} = \frac{T^2}{12}
 \;, \quad 
 \int_\vec{p} \frac{\nF{}}{p} = \frac{T^2}{24}
 \;, \quad 
 \int_\vec{p} p\, \nB{} = \frac{\pi^2T^4}{30}
 \;, \quad
 \int_\vec{p} p\, \nF{} = \frac{7\pi^2T^4}{240}
 \;. \la{nBnF}
\ee

An interesting question is how low the temperature should be 
in order for \eq\nr{nBnF} to yield an accurate representation 
of the full result. It has been pointed out in ref.~\cite{master} 
that for a particular 2-loop master spectral function, the 
non-relativistic approximation is only accurate for $M \gsim 25 T$.
However, in \eq\nr{imNLO} the 2-loop contributions are suppressed
by couplings and loop factors, whereas for the 1-loop term the 
thermal corrections are exponentially small for $\kp,\km \gg \pi T$.
Therefore, a somewhat better convergence may be expected. 

\begin{figure}[t]


\centerline{%
 \epsfysize=7.5cm\epsfbox{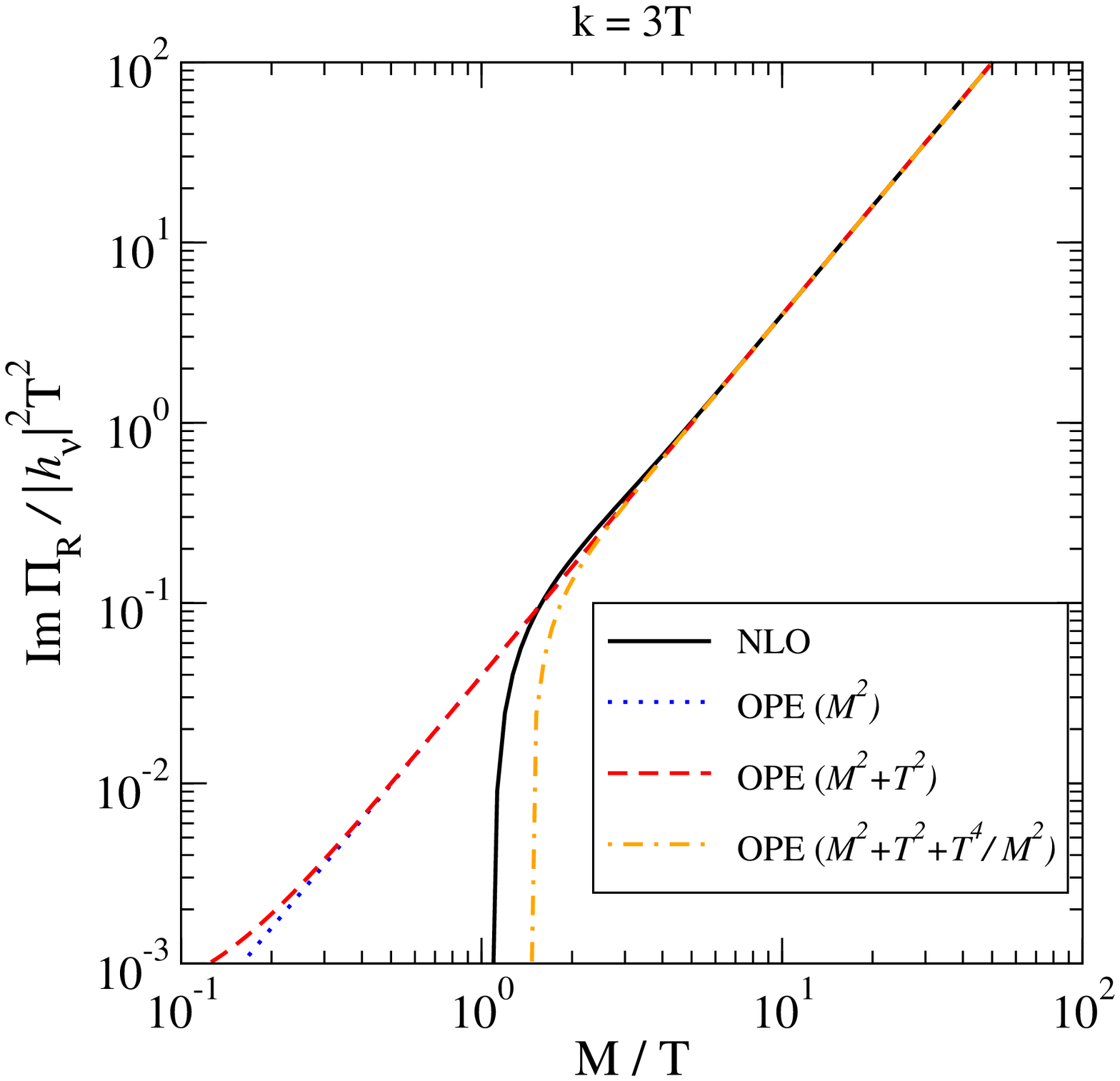}%
~~~\epsfysize=7.5cm\epsfbox{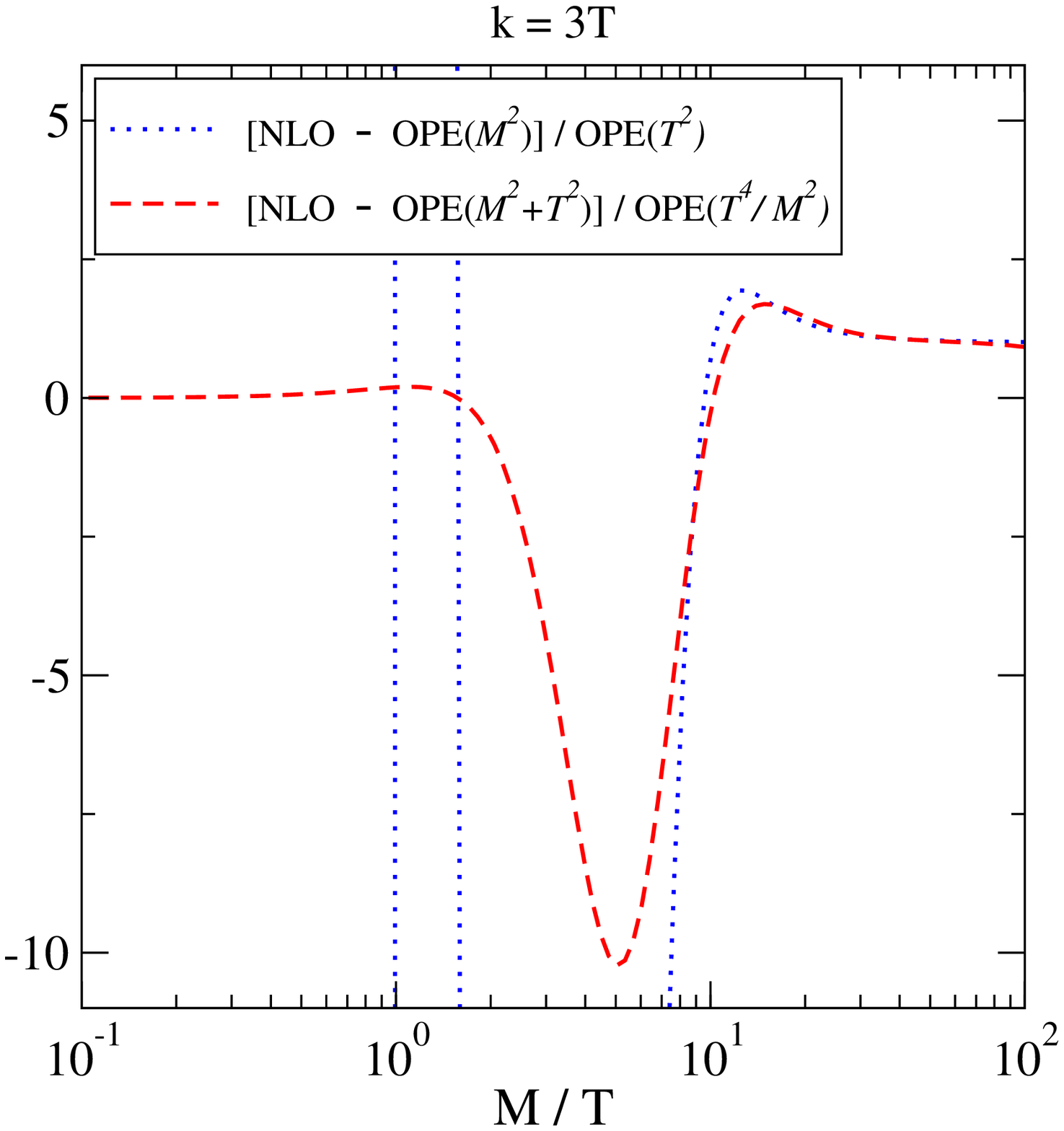}
}

\caption[a]{\small
Left: The ``OPE'' expression from \eq\nr{imASY}, up to 
three consecutive orders
as indicated in the parentheses, 
compared with the ``NLO'' result from \eq\nr{imNLO}, 
for $\ko^2 \equiv k^2 + M^2$. 
Right: The relative difference between \eqs\nr{imNLO}, \nr{imASY}. 
The couplings and the renormalization scale are fixed as
specified in appendix~C. On the resolution of the logarithmic
plot it seems that the OPE expression is accurate for all 
$M \gsim 4 T$, however as shown by the right panel  
discrepancies with respect to the unexpanded expression are correctly 
represented only for $M \gsim 15 T$.   
}

\la{fig:imASY}
\end{figure}

The full and non-relativistic 
results are compared in \fig\ref{fig:imASY} for $k = 3 T$.
We observe that 
the OPE-asymptotics 
appears to join
the full expression 
at $M \gsim 4 T$. This is somewhat of 
an optical illusion, however; as shown by the right panel, 
even the sign of the thermal correction is correctly 
reproduced only for $M \gsim 10 T$, and the relative error
decreases only for $M \gsim 15 T$. In any case the 
convergence is better in the full result than in  
the particular individual 2-loop master spectral function 
studied in ref.~\cite{master}.  

%
\subsection{Towards ultra-relativistic limit}
\la{ss:ur}

Let us return to the breakdown of the loop expansion, 
as illustrated in \fig\ref{fig:imNLO}. What happens is
that for $k \sim \pi T$ but $M \ll \pi T$, 
i.e.\ $2\km = \sqrt{k^2 + M^2} - k \approx M^2/(2 k) \ll \pi T$, 
the LO term is of magnitude
$
 \im \Pi_\rmii{R}^\rmii{LO} / |h_\nu|^2 \sim M^2 \ln (T/M)
$
whereas the NLO term is of magnitude
$
 \im \Pi_\rmii{R}^\rmii{NLO} / |h_\nu|^2 \sim  g^2 M^2 T^2 / \km^2 
 \sim  g^2 T^4 / M^2
$,
where $g^2$ denotes a generic coupling. 
The relative magnitude of the correction is 
$\sim g^2 T^4 / M^4$, 
and consequently the loop expansion requires resummation for  
$ M \lsim {g}^{\frac{1}{2}} T$. 
The most divergent NLO correction
comes with a negative sign; it is related to 
Higgs mass thermal resummation, as will be discussed presently. 

For an even smaller $M \lsim gT$, further resummations are needed. 
A Hard Thermal Loop (HTL)
resummation was presented in ref.~\cite{kmt}, however HTL resummation alone 
does not lead to a consistent 
weak-coupling expansion for the present observable 
in the ultrarelativistic regime~\cite{bb0}. 
Indeed a systematic computation requires 
a Landau-Pomeranchuk-Migdal (LPM) resummation~\cite{bb1,bb2}. 
This amounts to a solution of a Schr\"odinger-type equation
with a light-cone potential, which implements a 
resummation of ladder diagrams, representing multiple soft
scatterings taking place within the average ``formation time''
of the ultrarelativistic right-handed neutrino being produced.  

In the present study, we will not implement LPM resummation
(comments on this are however made in \se\ref{se:concl}). 
Rather, we follow the convention of capturing a sub-series
of higher order corrections through the assignment of thermal 
masses to otherwise massless particles. The concept of a thermal mass
is ambiguous for particles of non-zero spin, 
depending e.g.\ on whether soft ($k \ll \pi T$)
or hard ($k \gsim \pi T$) excitations are considered. For the 
present problem the latter kinematics is the relevant one, 
and then the thermal masses are those sometimes called the 
``asymptotic'' ones (for a concise summary see ref.~\cite{sch2}). 
It turns out that the Higgs mass thermal resummation (where 
no ambiguities appear) indeed consistently removes 
the dominant divergences from $\im \Pi_\rmii{R}^{ }$
in the regime $M \sim {g}^{\frac{1}{2}} T$, 
as we now show. 

Let us start by simply 
inserting the masses $m_\phi, m_\ell$ for the Higgs and for 
the left-handed leptons, respectively, and subsequently
define a Euclidean correlator through 
\be 
 \Pi_\rmii{E}^\rmi{tree}(K) \equiv 
 4 |h_\nu|^2 \Tint{P} \frac{K\cdot(P-K)}{[P^2 + m_\phi^2][(P-K)^2 + m_\ell^2]}
 \;.
\ee
This 1-loop sum-integral is labelled a ``tree-level'' contribution because
of frequent conventions in literature:
its cut corresponds to $1\leftrightarrow 2$ scatterings. 
Carrying out the Matsubara sum and taking the cut, one finds three channels, 
for $M > m_\phi + m_\ell$, $m_\phi > M + m_\ell$, and $m_\ell > M + m_\phi$, 
respectively. For the actual values relevant for the Standard Model, 
\ba
  m_\phi^2 & = & \frac{T^2}{16}
 \Bigl(
   g_1^2 + 3g_2^2 + \fr43 h_t^2 \Nc + 8 \lambda 
 \Bigr)
 \;, \la{mphi} \\ 
  m_\ell^2 & = & \frac{T^2}{16}
 \bigl(
   g_1^2 + 3g_2^2 
 \bigr)
 \;, \la{mell}
\ea
only the first two channels can get realized. 
In each channel, the angular
integral between the directions of $\vec{p}$ and $\vec{k}$ can be carried
out by taking $E^{ }_{pk}\equiv \sqrt{(\vec{p-k})^2 + m_\ell^2}$ as an 
integration variable, and subsequently the integral over the radial 
direction can also be performed, by taking $E_p \equiv \sqrt{p^2 + m_\phi^2}$
as a variable. For $\ko = \sqrt{k^2 + M^2}$ the result reads\footnote{%
 Equivalent expressions can be found in literature. 
 } 
\ba
 \frac{\im \Pi_\rmii{R}^\rmi{tree} }{|h_\nu|^2}  & = & 
 \frac{(M^2 - m_\phi^2 + m_\ell^2) T}{8\pi k}
 \ln\left\{
   \frac{\sinh\Bigl(\frac{E_\rmii{max}}{2 T} \Bigr)
         \cosh\Bigl(\frac{\ko - E_\rmii{min}}{2 T} \Bigr) }
        {\sinh\Bigl(\frac{E_\rmii{min}}{2 T} \Bigr)
         \cosh\Bigl(\frac{\ko - E_\rmii{max}}{2 T} \Bigr) }  
 \right\}
\nn & & \; \times
 \Bigl[ 
  \theta  \bigl( M - m_\phi - m_\ell \bigr) 
 - \theta \bigl( m_\phi - m_\ell - M \bigr) 
 - \theta \bigl( m_\ell - m_\phi - M \bigr) 
 \Bigr] 
 \;, \la{imTREE}
\ea
where 
\ba
 E_\rmi{max(min)} & \equiv & 
 \frac{\ko (M^2 + m_\phi^2 - m_\ell^2) 
 \pm 
 k \Delta(M,m_\phi,m_\ell)  }{2 M^2}
 \;, \\[2mm]
 \Delta(M,m_\phi,m_\ell) & \equiv & 
 \sqrt{ M^4 - 2 M^2 (m_\phi^2 + m_\ell^2) + (m_\phi^2 - m_\ell^2)_{ }^2 }
 \;.  
\ea

Suppose now that
we are in the regime $m_\phi^2, m_\ell^2 \ll M^2$, 
and expand to first order in the small masses. Then 
\be
 E_\rmi{min} \approx \km + \frac{m_\phi^2}{4\km} - \frac{m_\ell^2}{4 \kp}
 \;, \quad
 E_\rmi{max} \approx \kp + \frac{m_\phi^2}{4\kp} - \frac{m_\ell^2}{4 \km}
 \;,
\ee
and \eq\nr{imTREE} becomes
\ba
 \frac{\im \Pi_\rmii{R}^\rmii{tree} }{|h_\nu|^2}  & \approx &  
 \frac{( M^2 - m_\phi^2 + m_\ell^2 ) T}{8\pi k} \ln 
 \biggl\{ \frac{\sinh(\kp/T)}{\sinh(\km/T)} \biggr\}
 \nn & & \; + \,   
 \frac{{M}^2}{8\pi k}
 \biggl\{
   m_\phi^2 \,
    \biggl[
      \frac{1 + \nB{}(\kp) - \nF{}(\km)}{4\kp} -  
      \frac{1 + \nB{}(\km) - \nF{}(\kp)}{4\km}
    \biggr] 
 \nn & & \qquad + \,  
   m_\ell^2 \,
    \biggl[
      \frac{1 + \nB{}(\km) - \nF{}(\kp)}{4\kp} -  
      \frac{1 + \nB{}(\kp) - \nF{}(\km)}{4\km}
    \biggr] 
 \biggr\} + \rmO(m_{\phi,\ell}^4)
 \;. \nn \la{ur_exp}
\ea
Inserting the expressions from \eqs\nr{mphi}, \nr{mell} this  
is seen to agree {\em exactly} with the sum of all  
$\rho^{ }_{\mathcal{I}^{ }_\rmii{b}}$'s and
$\rho^{ }_{\mathcal{I}^{ }_\rmii{d}}$'s in \eq\nr{imNLO}, cf.\ 
\eqs\nr{rho_Ib}, \nr{rho_Id}. These master spectral functions 
include the  
only quadratically divergent structures in the limit  
$\km \ll \pi T$ as can be deduced from the right panels
of \figs\ref{fig:Ib}--\ref{fig:Ihp}. The most divergent 
terms are $\sim M^2 \nB{}(\km) / \km$ and are related to 
the Higgs mass resummation as 
is clearly visible from the second row of \eq\nr{ur_exp}. 

We now define a ``resummed'' result by 
accounting for all $\rho^{ }_{\mathcal{I}^{ }_\rmii{b}}$'s and
$\rho^{ }_{\mathcal{I}^{ }_\rmii{d}}$'s
of \eq\nr{imNLO} through the thermal masses: 
\ba
 \frac{\im \Pi^\rmi{resum}_\rmii{R}}{|h_\nu(\bmu)|^2} & \equiv & 
 \frac{\im \Pi^\rmi{tree}_\rmii{R}}{|h_\nu(\bmu)|^2}
 \nn & + &   
 2 h_t^2 \Nc 
 \biggl\{
   - \rho^\rmii{$T$}_{\widetilde{\mathcal{I}}^{ }_\rmii{f}}
   + \rho^\rmii{$T$}_{\widetilde{\mathcal{I}}^{ }_\rmii{h}}
 \nn & & \; - \, 
   \frac{\pi {M}^2}{(4\pi)^4 k}
   \int_{\km}^{\kp} \! {\rm d}p \, 
   \frac{ \nF{}(\ko -p) \nB{}(p) }{\nF{}(\ko)}
   \biggl[
     \ln \frac{(\kp - p)(p - \km)\bmu^2 }{k^2 {M}^2} + \fr{11}{2} 
   \biggr]
 \biggr\}
 \nn & + &   
 \frac{g_1^2 + 3 g_2^2}{2}
 \biggl\{
     2 \Bigl[ 
     \rho^\rmii{$T$}_{{\mathcal{I}}^{ }_\rmii{g}}
    + \rho^\rmii{$T$}_{\widehat{\mathcal{I}}^{ }_\rmii{h'}}
    + \rho^\rmii{$T$}_{{\mathcal{I}}^{ }_\rmii{j}}
     \Bigr] 
    - 4 \Bigl[ 
       \rho^\rmii{$T$}_{{\mathcal{I}}^{ }_\rmii{h}}
    + \rho^\rmii{$T$}_{\widehat{\mathcal{I}}^{ }_\rmii{h}}
     \Bigr] 
 \nn & & \; + \, 
   \frac{3 \pi {M}^2}{(4\pi)^4 k}
   \int_{\km}^{\kp} \! {\rm d}p \, 
   \frac{ \nF{}(\ko -p) \nB{}(p) }{\nF{}(\ko)}
   \biggl[
     \ln \frac{(\kp - p)(p - \km)\bmu^2 }{k^2 {M}^2} + \fr{41}{6} 
   \biggr]
 \biggr\}
 \;, \hspace*{10mm} \la{imRESUM}
\ea
where $\im \Pi_\rmii{R}^\rmi{tree}$ is the tree-level
result from \eq\nr{imTREE}. Note that the remaining master
spectral functions continue to be evaluated without masses
(in these spectral functions masses amount to higher-order corrections). 

\begin{figure}[t]


\centerline{%
 \epsfysize=7.5cm\epsfbox{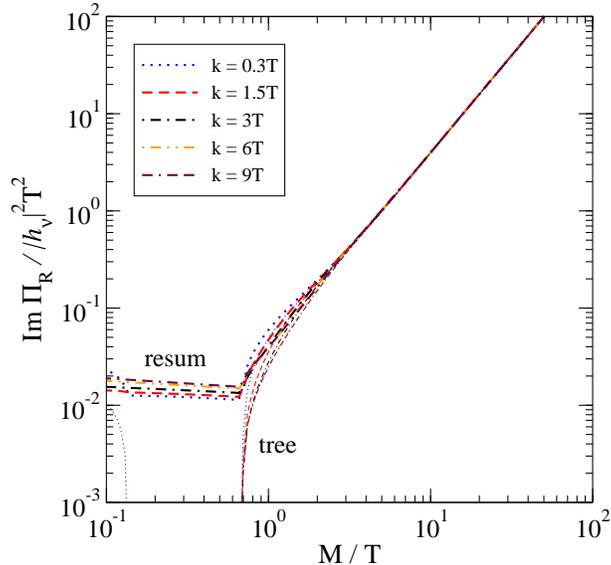}%
}

\caption[a]{\small
The mass-resummed expression from \eq\nr{imRESUM} (thick lines), 
in units of $T^2$, for $\ko^2 \equiv k^2 + M^2$, versus the 
tree-level result from \eq\nr{imTREE} (thin lines). 
The couplings and the renormalization
scale are fixed as specified in appendix~C. 
The cusp is expected to be 
removed through higher-order corrections, but the overall
magnitude of the mass-resummed result is already in qualitative 
agreement with the ultrarelativistic results of refs.~\cite{bb1,bb2}
(cf.\ \fig\ref{fig:rateRESUM}). 
}

\la{fig:imRESUM}
\end{figure}

The tree and mass-resummed spectral functions are
shown in \fig\ref{fig:imRESUM}. It is apparent that the
downwards divergence seen in \fig\ref{fig:imNLO} is a reflection
of thermal mass generation; after this effect has been taken
into account, the other NLO terms show an enhancement. However 
the mass resummation implemented does not capture all the terms 
that need to be resummed for $M \lsim gT$; 
indeed, as has been demonstrated 
with the case of hot QCD~\cite{agz} and more
recently with the problem at hand~\cite{bb1,bb2}, cusps 
such as those seen in \fig\ref{fig:imRESUM}
are also removed through a systematic resummation of all 
corrections pertinent to the ultrarelativistic regime.\footnote{%
  The mass-resummed result of the current study
  still diverges logarithmically for $M/T \to 0$; 
  this originates from the master spectral function 
  $\rho^\rmii{$T$}_{\widehat{\mathcal{I}}^{ }_\rmii{h'}}$,  
  cf.\ \fig\ref{fig:Ihp}(right).  
  The divergence is also removed by resummations.
 } 
It is nevertheless interesting that the order of magnitude of the 
mass-resummed result is not unlike that found 
in refs.~\cite{bb1,bb2} (cf.\ \se\ref{se:rate}).

%
\section{Spectra and spectral functions}
\la{se:spectra}

\begin{figure}[t]


\centerline{%
\epsfysize=7.5cm\epsfbox{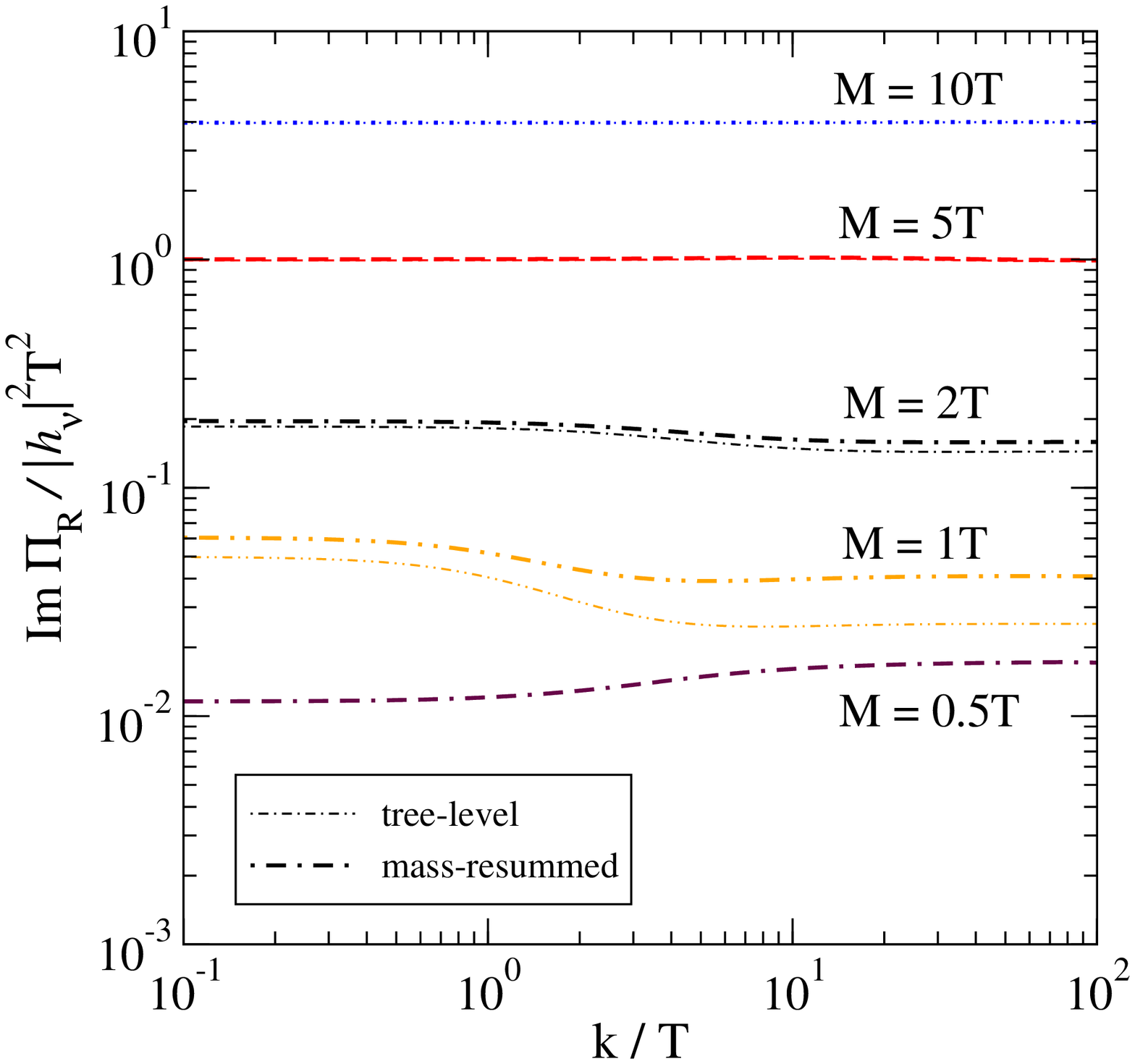}%
~~~\epsfysize=7.5cm\epsfbox{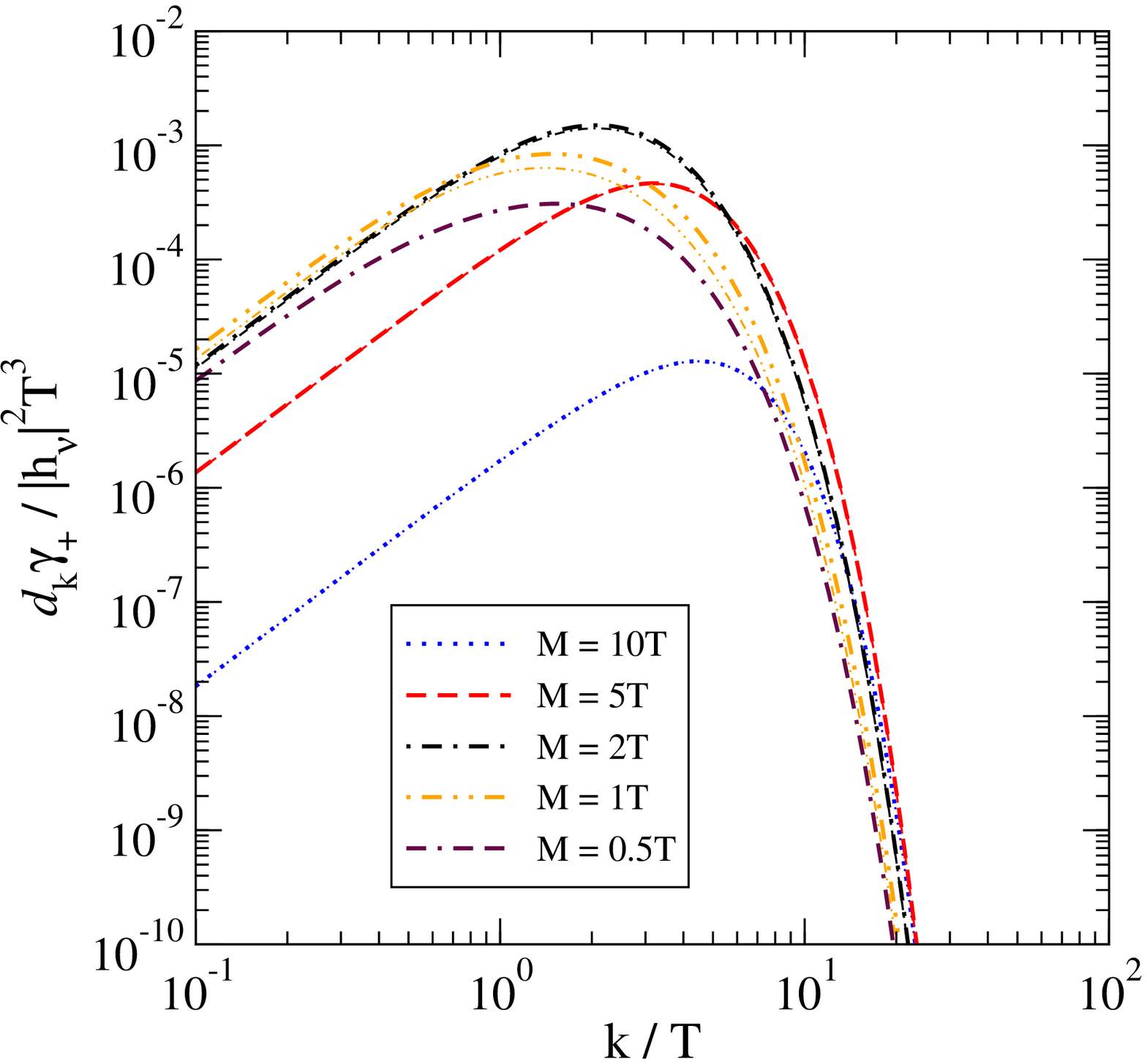}
}

\caption[a]{\small
Left: The $k$-dependence of \eq\nr{imRESUM} (thick lines), for selected
$M$ and $\ko^2 \equiv k^2 + M^2$. Thin lines indicate the ``tree-level''
result from \eq\nr{imTREE}. In the regime of validity of the 
computation, i.e.\ $M \gsim \pi T$,
$k$-dependence is quite modest. 
Right: The corresponding production spectra,
$
  \partial_k\, \gamma_+
$
from \eq\nr{dGamma}. 
The couplings and the renormalization scale are fixed as
specified in appendix~C.
}

\la{fig:spectraRESUM}
\end{figure}

We have already observed that, for a fixed $\mathcal{K}^2 = M^2$, 
the non-covariant dependence of $\im \Pi_\rmii{R}$
on the spatial momentum $k$ is small, cf.\ \fig\ref{fig:imRESUM}.
This is illustrated again in \fig\ref{fig:spectraRESUM}(left), 
for a number of different $M/T$. 
Subsequently we plot the whole spectra according
to \eq\nr{gamma}, i.e.\ 
\be
 \partial_k\, \gamma_+ \equiv 
 \frac{k^2 \nF{}(\sqrt{k^2 +M^2})}{\pi^2 \sqrt{k^2 + M^2}} 
 \, \im \Pi_\rmii{R}
 \;, \la{dGamma}
\ee
for a few selected $M/T$, 
in \fig\ref{fig:spectraRESUM}(right). Obviously the latter results display
a much stronger $k$-dependence than $\im \Pi_\rmii{R}$, 
however this emerges through 
the trivial ``kinematic'' structures shown in \eq\nr{dGamma}, 
rather than complicated plasma physics determining $\im \Pi_\rmii{R}$.

%
\section{Total production rate}
\la{se:rate}

\begin{figure}[t]


\centerline{%
\epsfysize=7.5cm\epsfbox{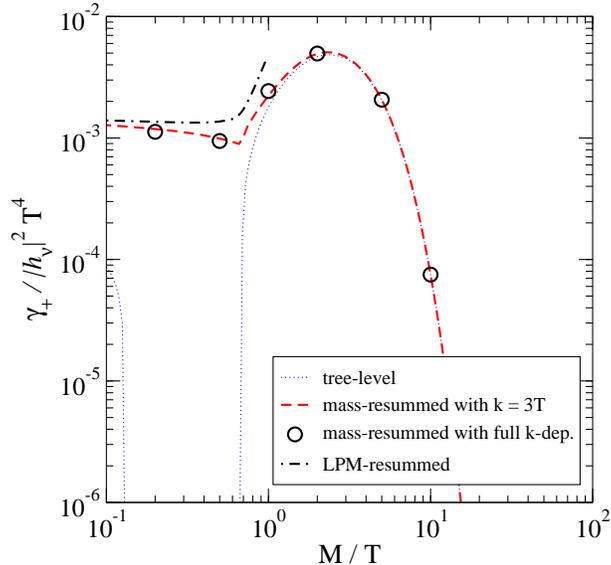}%
}

\caption[a]{\small
The total production rate based on evaluating 
$\im \Pi_\rmii{R}$ at $k = 3T$ and taking it 
as a constant otherwise (red dashed line), compared with results
obtained with the full $k$-dependence included (black circles).  
The couplings and the renormalization scale 
are fixed as specified in appendix~C.
For $M \lsim T$ we also compare with the complete LPM-resummed
result from ref.~\cite{bb2} (dash-dotted line). 
}

\la{fig:rateRESUM}
\end{figure}

According to \eq\nr{gamma}, 
the total right-handed neutrino production rate reads 
\be
 \gamma_+
 = 
 \int_0^\infty \! {\rm d}k 
 \;  \partial_k\, \gamma_+ 
 \;, \la{total_rate}
\ee
where $\partial_k\, \gamma_+$ is the differential 
production rate from \eq\nr{dGamma}. 
Given the small $k$-dependence as seen in 
\fig\ref{fig:spectraRESUM}(left), a good approximation for the total 
production rate can be obtained by evaluating the ``expensive''
$\im \Pi_\rmii{R}$ only at some typical momentum, 
for instance $k\sim 3 T$. The accuracy of this approximation
is illustrated in \fig\ref{fig:rateRESUM}, and found to be 
in general excellent. It should be noted
that in the non-relativistic regime the actual average momentum
is $k \sim \sqrt{2 M T}$ rather than $k \sim 3 T$, but the 
approximation does not lose its accuracy, because the 
$k$-dependence of $\im\Pi_\rmii{R}$ becomes even less
significant for $M \gg \pi T$ 
(cf.\ \fig\ref{fig:spectraRESUM}(left)).

In \fig\ref{fig:rateRESUM} the total production rate is also 
compared with the LPM-resummed result from ref.~\cite{bb2}. 
Although our expression is not reliable for $M \ll \pi T$ and 
the result of ref.~\cite{bb2} is not reliable for $M \gsim \pi T$, 
it is remarkable how well the two appear to extrapolate towards
each other. (Our results are closer to the systematic analysis of 
ref.~\cite{bb2} than the phenomenological approach of ref.~\cite{bg2}.)
In principle it should also be possible to combine the two results
into an expression applicable for a general $M/\pi T$ 
(some comments are made
in \se\ref{se:concl}), however implementing this in practice 
necessitates a dedicated separate study. 

%
\section{Conclusions and outlook}
\la{se:concl}
 
The purpose of this paper has been to extend previous NLO results
for the right-handed neutrino production rate up to higher temperatures, 
into the so-called relativistic regime in which the temperature is of 
a similar magnitude as the mass of the right-handed neutrinos. 
In the so-called strong washout scenario of leptogenesis, 
the right-handed neutrinos equilibrate initially, whereby no lepton
asymmetry exists in a certain temperature range; it is generated at
low temperatures when the right-handed neutrinos chemically decouple
and can subsequently decay. It is conceivable that most of the 
decays take place in a non-relativistic regime, however it is not 
clear a priori how accurate computations based on the non-relativistic
approximation are, because thermal effects are only 
power-suppressed~\cite{sch}. The results obtained here suggest
that corrections are substantial for $T \gsim M / 4$. If 
a significant contribution arises from this range, then the results
of the current study may be used for a more precise analysis. 

One finding of the current investigation is that in the relativistic
and non-relativistic regimes, the non-covariant dependence of the retarded
correlator denoted by $\im\Pi^{ }_\rmii{R}$ on the spatial momentum 
with respect to the heat bath is quite modest 
(cf.\ \fig\ref{fig:spectraRESUM}(left)). Therefore the ``expensive'' part
of the computation needs to be carried out only for a specific 
chosen $k$, for instance $k = 3 T$, in order to determine the overall
magnitude of $\im\Pi^{ }_\rmii{R}$. When inserted into the proper
overall relations, this information is sufficient for determining
the total production rate with good accuracy 
(cf.\ \fig\ref{fig:rateRESUM}).

Apart from the strong washout scenario of leptogenesis, 
another possibility is that the right-handed neutrinos never 
equilibrate chemically; one then speaks of a weak washout scenario
(cf.\ e.g.\ ref.~\cite{jsg}). 
In this case even ultrarelativistic temperatures play a role, 
and resummations are needed for consistent results~\cite{bb1,bb2}. 
Although the results of the current paper lose their validity when 
approaching the ultrarelativistic regime, they do allow us to 
anticipate some features of the corresponding expressions, such 
as that there is no gap between the two possible $1\leftrightarrow 2$ 
channels permitted by tree-level kinematics with thermal masses 
(cf.\ \fig\ref{fig:imRESUM}). In addition the total 
rates of the two approaches extrapolate towards each other
surprisingly well, even though both computations eventually 
leave their ranges of validity
(cf.\ \fig\ref{fig:rateRESUM}).
The physics conclusion from the enhanced 
rate at high temperatures is that chemical equilibration 
is more likely than naively expected, which
may reduce the parameter space for weak-washout leptogenesis. 

In principle, it should be possible to combine the current NLO
results, valid in the non-relativistic and relativistic regimes, 
with the resummed LO result of the ultrarelativistic regime~\cite{bb1,bb2}. 
Some care is however needed for avoiding double-counting in doing so.
Matching computations of a necessary
type have been carried out with HTL resummation previously, but 
the case of LPM resummation seems to 
represent a bigger challenge.\footnote{%
 Schematically, omitting Lorentz-violating structures, the NLO computation
 we have carried out is of the form 
 $\im \Pi_\rmii{R,UV}^{ }
  \sim {M}^2 [\phi^{ }_1 (T^2 / {M}^2) + 
 g^2 \phi^{ }_2 (T^2 / {M}^2) + ... ]$
 whereas any resummed result contains all orders in $g^2$: 
 $\im \Pi_\rmii{R,IR}^{ } \sim 
 {M}^2 [ \chi^{ }_1 (T^2 / {M}^2, g^2) + 
 ... ]$.
 For combining the two, the resummed result has to be re-expanded 
 in $g^2$, 
 in order to cancel the terms from $\im \Pi^{ }_\rmii{R,UV}$ that it 
 resums: 
 $\im \Pi_\rmii{R,full}^{ } \sim {M}^2 \{
 \chi^{ }_1  (T^2 / {M}^2, g^2) +
 \phi^{ }_1 (T^2 / {M}^2) - \chi^{ }_1 (T^2 / {M}^2, 0) + 
 g^2 [ \phi^{ }_2 (T^2 / {M}^2) 
 - \partial^{  }_{g^2} \chi^{ }_1 (T^2 / {M}^2, 0) ] + ... \}
 $.
 The determination of the latter subtraction term
 is not quite trivial, because $\chi^{ }_1  (T^2 / {M}^2, g^2)$
 contains parts only available through a numerical solution of 
 an inhomogeneous Schr\"odinger-type equation and is otherwise 
 complicated as well. 
 }  
Nevertheless the problem may be worth giving a go; 
for practical applications a result valid  
for all temperatures would clearly be quite convenient. 
(Ultimately NLO corrections
should also be worked out for the LPM regime; 
they are likely to be suppressed only by
$\sqrt{g^2}$ there~\cite{km}.)

Another possible direction for future research is the application 
of the methods and master spectral functions discussed here to 
other problems of cosmological relevance. For instance, the determination
of the gravitino production rate from a hot Standard Model plasma is 
a problem perhaps meriting a further look. 

%
\section*{Acknowledgements}

I am grateful to Dietrich B\"odeker for helpful discussions. 
This work was partly supported by the Swiss National Science Foundation
(SNF) under grant 200021-140234.

%
\appendix
\renewcommand{\thesection}{Appendix~\Alph{section}}
\renewcommand{\thesubsection}{\Alph{section}.\arabic{subsection}}
\renewcommand{\theequation}{\Alph{section}.\arabic{equation}}

%
\section{Definitions of master sum-integrals}
\la{app:A}

Denoting by $\Tinti{P}$ and $\Tinti{\{P\}}$
sum-integrals over bosonic and  fermionic
Matsubara four-momenta, the master sum-integrals entering
the computation are defined as follows~\cite{nonrel}:
\ba
 \mathcal{J}^{ }_\rmi{a} & \!\!\equiv\!\! & 
 \Tint{P} \frac{1}{P^2}
 \;,
 \\
 \widetilde{\mathcal{J}}^{ }_\rmi{a} & \!\!\equiv\!\! & 
 \Tint{\{P\}} \frac{1}{P^2}
 \;,
 \\ 
 \mathcal{J}^{ }_\rmi{b} & \!\!\equiv\!\! & 
 \Tint{P} \frac{K^2}{P^2(P-K)^2}
 \;, \la{def_Jb}
 \\
 \mathcal{I}_\rmi{b} & \!\!\equiv\!\! & 
 \Tint{PQ} \frac{1}{Q^2P^2(P-K)^2}
 \;, \la{def_Ib}
 \\
 \widetilde{\mathcal{I}}_\rmi{b} & \!\!\equiv\!\! & 
 \Tint{P\{Q\}} \frac{1}{Q^2P^2(P-K)^2}
 \;,
 \\
 \mathcal{I}_\rmi{c} & \!\!\equiv\!\! & 
 \Tint{PQ} \frac{1}{Q^2P^4}
 \;, \la{def_Ic}
 \\
 \widetilde{\mathcal{I}}_\rmi{c} & \!\!\equiv\!\! & 
 \Tint{P\{Q\}} \frac{1}{Q^2P^4}
 \;,
 \\
 \widehat{\mathcal{I}}_\rmi{c} & \!\!\equiv\!\! & 
 \Tint{\{P\}Q} \frac{1}{Q^2P^4}
 \;, 
 \\
 \overline{\mathcal{I}}_\rmi{c} & \!\!\equiv\!\! & 
 \Tint{\{PQ\}} \frac{1}{Q^2P^4}
 \;, 
 \\
 \mathcal{I}_\rmi{d} & \!\!\equiv\!\! & 
 \Tint{PQ} \frac{K^2}{Q^2P^4(P-K)^2}
 \;, \la{def_Id}
 \\
 \widetilde{\mathcal{I}}_\rmi{d} & \!\!\equiv\!\! & 
 \Tint{P\{Q\}} \frac{K^2}{Q^2P^4(P-K)^2}
 \;,
 \\
 \widehat{\mathcal{I}}_\rmi{d} & \!\!\equiv\!\! & 
 \Tint{\{P\}Q} \frac{K^2}{Q^2P^4(P-K)^2}
 \;,
 \\
 \overline{\mathcal{I}}_\rmi{d} & \!\!\equiv\!\! & 
 \Tint{\{PQ\}} \frac{K^2}{Q^2P^4(P-K)^2}
 \;,
 \\
 \mathcal{I}_\rmi{e} & \!\!\equiv\!\! & 
 \Tint{PQ} \frac{1}{Q^2P^2(P-Q)^2}
 \;, \la{def_Ie}
 \\
 \widetilde{\mathcal{I}}_\rmi{e} & \!\!\equiv\!\! & 
 \Tint{P\{Q\}} \frac{1}{Q^2P^2(P-Q)^2}
 \;, 
 \\
 \mathcal{I}_\rmi{f} & \!\!\equiv\!\! & 
 \lim^{ }_{\lambda \to 0}
 \Tint{PQ} \frac{1}{Q^2[(Q-P)^2 + \lambda^2](P-K)^2}
 \;, \la{def_If}
 \\
 \widetilde{\mathcal{I}}_\rmi{f}\, & \!\!\equiv\!\! & 
 \lim^{ }_{\lambda \to 0}
 \Tint{P\{Q\}} \frac{1}{Q^2[(Q-P)^2 + \lambda^2](P-K)^2}
 \;, 
 \\ 
 \mathcal{I}_\rmi{g} & \!\!\equiv\!\! & 
 \Tint{PQ} \frac{K^2}{P^2(P-K)^2Q^2(Q-K)^2}
 \;, \la{def_Ig}
 \\
 \mathcal{I}_\rmi{h} & \!\!\equiv\!\! & 
 \lim^{ }_{\lambda \to 0}
 \Tint{PQ} \frac{K^2}{Q^2P^2[(Q-P)^2+\lambda^2](P-K)^2}
 \;, \la{def_Ih}
 \\
 \widetilde{\mathcal{I}}_\rmi{h} & \!\!\equiv\!\! & 
 \lim^{ }_{\lambda \to 0}
 \Tint{P\{Q\}} \frac{K^2}{Q^2P^2[(Q-P)^2+\lambda^2](P-K)^2}
 \;, 
 \\
 \widehat{\mathcal{I}}_\rmi{h} & \!\!\equiv\!\! & 
 \lim^{ }_{\lambda \to 0}
 \Tint{\{P\}Q} \frac{K^2}{Q^2P^2[(Q-P)^2+\lambda^2](P-K)^2}
 \;, 
 \\ 
 \widehat{\mathcal{I}}_\rmi{h'} \! & \!\!\equiv\!\! & 
 \lim^{ }_{\lambda \to 0}
 \Tint{\{P\}Q} \frac{2K\cdot Q}{Q^2P^2[(Q-P)^2+\lambda^2](P-K)^2}
 \;, \la{def_Ihp}
 \\
 \mathcal{I}_\rmi{j} & \!\!\equiv\!\! & 
 \lim^{ }_{\lambda \to 0}
 \Tint{PQ} \frac{K^4}{Q^2P^2[(Q-P)^2+\lambda^2](P-K)^2(Q-K)^2} 
 \;.  \la{def_Ij} \hspace*{1cm}
\ea
In order to handle the different statistics simultaneously, 
we introduce a generic labelling of lines 
(with individual propagators omitted or doubled in some cases):
\be
 \Generic
 \hspace*{12mm} \;.  \la{labelling}
\ee
The labels $\sigma_0,...,\sigma_5$ equal $+1$ for bosons and 
$-1$ for fermions. Given fermion number conservation, 
only two of the indices are independent, and 
the triple $(\sigma_1\sigma_4\sigma_5)$ has been chosen for this task; 
subsequently
\be
    \sigma_0 = \sigma_1 \sigma_4 
    \;, \quad 
    \sigma_2 = \sigma_1 \sigma_5 \;, \quad
    \sigma_3 = \sigma_4 \sigma_5 \;. \la{indices}
\ee  

The spectral functions are obtained from \eq\nr{cut}
which is commensurate with the spectral representation
\be
 \int_{-\infty}^{\infty} \! \frac{{\rm d} \ko }{\pi} 
 \frac{ \rho^{ }_{\mathcal{I}^{ }_\rmi{x}} }{\ko - i k_n} 
 = \mathcal{I}^{ }_\rmi{x}
 \;. \la{spectral}
\ee
In practice, the spectral function can be read from 
an imaginary-time correlator by partial fractioning 
its dependence on $k_n$ and by then replacing
\be
  \frac{1}{-i k_n + C} \rightarrow \pi \delta( - \ko + C)
  \;.
 \la{cut2}
\ee
Note that structures containing no or polynomial $k_n^2$-dependence yield 
a vanishing spectral function according to \eq\nr{cut}; in these cases 
\eq\nr{spectral} needs to be modified by ``contact terms''
(a discussion can be found e.g.\ in ref.~\cite{sch}).

%
\section{Results for master spectral functions in time-like domain}
\la{app:all}

In the following spectral functions are listed for the range $\ko > k > 0$.
Results for $\ko \to -\ko$ follow from antisymmetry, whereas the space-like 
domain $0 < \ko < k$ has not been worked out here, although  
thermal spectral functions can be non-zero there as well.

%
\subsection{$\rho^{ }_{\mathcal{J}^{ }_\rmii{a}}$}

Since $\mathcal{J}_\rmi{a}$ and $\widetilde{\mathcal{J}}_\rmi{a}$ 
are independent of the external momentum, there is no cut:  
\ba
 \rho^{ }_{\mathcal{J}_\rmi{a}} & \!\! = \!\! & 
 \rho^{ }_{\widetilde{\mathcal{J}}_\rmi{a}}
 \; = \;
 0
 \;. \la{asy_Ja}
\ea

%
\subsection{$\rho^{ }_{\mathcal{J}^{ }_\rmii{b}}$}
\la{ss:Jb}

For the case in \eq\nr{def_Jb} we get, 
after carrying out the Matsubara sum,
\ba
 \mathcal{J}_\rmi{b}
 & = &  
  \int_{\vec{p}} \frac{K^2} 
  { 4 \epsilon^{ }_p \epsilon^{ }_{pk} }
  \biggl\{ 
  \biggl[ 
   \frac{1}{i k_n + \epsilon^{ }_p + \epsilon^{ }_{pk}} + 
   \frac{1}{- i k_n + \epsilon^{ }_p + \epsilon^{ }_{pk}}
  \biggr]
  \Bigl[
   1  + n^{ }_{\sigma_1}(\epsilon^{ }_p) + n^{ }_{\sigma_4}(\epsilon^{ }_{pk})
  \Bigr]
  \nn & & \quad + \, 
  \biggl[ 
   \frac{1}{i k_n - \epsilon^{ }_p + \epsilon^{ }_{pk}} + 
   \frac{1}{- i k_n - \epsilon^{ }_p + \epsilon^{ }_{pk}}
  \biggr]
  \Bigl[
    n^{ }_{\sigma_1}(\epsilon^{ }_p)  - n^{ }_{\sigma_4}(\epsilon^{ }_{pk}) 
  \Bigr]
  \biggr\}
  \;,
\ea
where we used the labelling of \eq\nr{labelling}, denoted
\be
  \epsilon^{ }_p \equiv p \equiv |\vec{p}| \;, \quad
  \epsilon^{ }_{pk} \equiv |\vec{p-k}| \;, \la{energies1}
\ee
and  defined
\be
 n^{ }_\sigma (\epsilon) \equiv \frac{\sigma }{ e^{\epsilon/T} - \sigma } 
 \;, \quad
 n^{-1}_\sigma(\epsilon) = \sigma e^{\epsilon/T} - 1
 \;, \quad
 \int^{\epsilon} \! {\rm d}\epsilon' \, n_\sigma (\epsilon')
 = T \ln \bigl(1 - \sigma e^{- \epsilon/T} \bigr)
 \;, \la{ni}
\ee
which satisfies $n^{ }_+ = \nB{}$ and $n^{ }_- = - \nF{}$.

Taking the cut, only one of the four channels contributes for 
$\ko > k$, and the spectral function reads 
\be
 \rho^{ }_{\mathcal{J}^{ }_\rmii{b}} 
 = 
 - \int_\vec{p} 
 \frac{\pi\mathcal{K}^2}{4 \epsilon^{ }_p \epsilon^{ }_{pk}}
 \delta(\ko - \epsilon^{ }_p - \epsilon^{ }_{pk})
 \Bigl[ 
   1 
    + n^{ }_{\sigma_1}(\epsilon^{ }_p) 
   + n^{ }_{\sigma_4}(\epsilon^{ }_{pk})
 \Bigr]
 \;. \la{rho_Jb_1}
\ee
It is often
convenient to employ the alternative representation
\be
   1
  + n^{ }_{\sigma_1}(\epsilon^{ }_p) 
 + n^{ }_{\sigma_4}(\ko - \epsilon^{ }_{p})
 = 
 n_{\sigma_0}^{-1}(\ko) \, 
 n^{ }_{\sigma_4}(\ko - p) n^{ }_{\sigma_1}(p)
 \;, \la{rewrite}
\ee
where we made use of $\sigma_1 \sigma_4 = \sigma_0$.

Because the leading-order contribution gets 
multiplied by a counterterm, we need to determine it up to $\rmO(\epsilon)$.
The integration measure reads,  
in $d=3-2\epsilon$ spatial dimensions, 
\be
 \int_\vec{p} = 
 \frac{(4\pi)^{\epsilon}}{4\pi^2 \Gamma(1-\epsilon)}
 \int_0^\infty \! {\rm d}p \, p^{2-2\epsilon}
 \int_{-1}^{+1} \! {\rm d}z \, (1-z^2)^{-\epsilon}  
 \;,
\ee
where $z \equiv \vec{p}\cdot\vec{k} / pk$. The integral over $z$ 
can be converted into one over $\epsilon^{ }_{pk}$ through
\be
 {\rm d}z = - \frac{\epsilon^{ }_{pk} {\rm d} \epsilon^{ }_{pk}}{p k} 
 \;, 
\ee
and the Dirac-$\delta$ gets realized for $\km < p < \kp$,  with
$k_\pm$ defined according to \eq\nr{kpm}.
Recalling the constraint
$
  \delta(\ko - p - \epsilon^{ }_{pk})
$, 
the function
appearing in the angular integration is conveniently expressed as
\be
 1 - z^2 = \frac{\mathcal{K}^2 (\kp - p)(p - \km)}{k^2 p^2}
 \;.
\ee
Introducing the $\msbar$ scheme scale parameter, $\bmu$, by 
inserting 
\be
 1 =  \mu^{-2\epsilon}\bmu^{2\epsilon} 
 \frac{e^{\epsilon\gammaE}}{(4\pi)^{\epsilon}}
 \;, 
\ee
and suppressing the inconsequential $\mu^{-2\epsilon}$, 
we thereby obtain 
\ba
 \rho^{ }_{\mathcal{J}^{ }_\rmii{b}} 
 & = & 
 - \frac{\pi\mathcal{K}^2}{(4\pi)^2 k} 
 \frac{e^{\epsilon\gammaE}}{\Gamma(1-\epsilon)}
 \int_{\km}^{\kp}
 \! {\rm d}p \,   
 \frac{ n^{ }_{\sigma_4}(\ko - p) n^{ }_{\sigma_1}(p) }{n^{ }_{\sigma_0}(\ko)}
 \biggl[ \frac{\bmu^2 k^2} 
 {\mathcal{K}^2 (\kp - p)(p - \km)} \biggr]^{\epsilon}
 \nn & = & 
 - \frac{\pi\mathcal{K}^2}{(4\pi)^2 k} 
 \int_{\km}^{\kp}
 \! {\rm d}p \,   
 \frac{ n^{ }_{\sigma_4}(\ko - p) n^{ }_{\sigma_1}(p) }{n^{ }_{\sigma_0}(\ko)}
 \nn & & \hspace*{2cm} \times \, 
 \biggl[ 
   1 + 
   \epsilon\ln \frac{\bmu^2}{\mathcal{K}^2} 
   + \epsilon \ln \frac{k^2}{ (\kp - p)(p - \km)} 
   + \rmO(\epsilon^2)
 \biggr]
 \;. \la{rho_Jb_final}
\ea

The remaining integral is easily carried out in the 
term of $\rmO(\epsilon^0)$: 
\ba
 \rho^{ }_{\mathcal{J}^{ }_\rmii{b}} 
 & = & 
 -\frac{\pi\mathcal{K}^2}{(4\pi)^2 k}
 \biggl\{ 
   T \ln \biggl( 
     \frac{e^{{\kp} / {T}}+\sigma_0 e^{-\kp/T}-\sigma_1 - \sigma_4}
     {e^{{\km} / {T}}+\sigma_0 e^{-\km/T}-\sigma_1 - \sigma_4}
   \biggr)
   \biggl[ 
   1 + \epsilon\biggl( \ln\frac{\bmu^2}{\mathcal{K}^2} + 2 \biggr)
   \biggr]
 \nn & & \; + \, \epsilon\,  
 \int_{\km}^{\kp}
 \! {\rm d}p \,   
 \frac{ n^{ }_{\sigma_4}(\ko - p) n^{ }_{\sigma_1}(p) }
  { n^{ }_{\sigma_0}(\ko) }
 \biggl[ 
   \ln \frac{k^2}{ (\kp - p)(p - \km)}  - 2 
 \biggr]
 \biggr\} + \rmO(\epsilon^2)
 \;. \hspace*{10mm} \la{Jb_eps0}
\ea
For $\kp,\km \gg \pi T$, the second row vanishes up to 
exponentially small corrections so that, 
in accordance with ref.~\cite{nonrel},  
\ba
 \rho^{ }_{\mathcal{J}_\rmi{b}} & \stackrel{\kp,\km \gg \pi T}{\approx} & 
 -\frac{\mufac\mathcal{K}^2}{16\pi} \biggl[ 
   1 + \epsilon\biggl( \ln\frac{\bmu^2}{\mathcal{K}^2} + 2 \biggr)
     + \rmO\bigl( \epsilon^2  \bigr)
 \biggr] 
 \;. \la{asy_Jb}
\ea
The specific statistics relevant for the current paper are
\ba
 {\mathcal{J}}^{ }_\rmi{b} & \Leftrightarrow & 
 (\sigma_1\sigma_4\sigma_5 | \sigma_0) = (+-+|-)
 \;.
\ea
Here and in the following, the values of selected non-independent indices 
as obtained from \eq\nr{indices} have also been indicated to the right
of the vertical line. 

%
\subsection{$\rho^{ }_{\mathcal{I}^{ }_\rmii{b}}$}

The spectral function corresponding to \eq\nr{def_Ib} can be written
generically as
\be
 \rho^{ }_{\mathcal{I}^{ }_\rmii{b}} 
  = 
 - \frac{\rho^{ }_{\mathcal{J}^{ }_\rmii{b}} 
 }{\mathcal{K}^2}
 \int_\vec{q} \frac{n^{ }_{\sigma_2}(q)}{q}
 = 
  \frac{T}{16\pi k}
   \ln \biggl( 
     \frac{e^{{\kp} / {T}}+\sigma_0 e^{-\kp/T}-\sigma_1 - \sigma_4}
     {e^{{\km} / {T}}+\sigma_0 e^{-\km/T}-\sigma_1 - \sigma_4}
   \biggr)
 \int_\vec{q} \frac{n^{ }_{\sigma_2}(q)}{q}
 \;. \la{rho_Ib} 
\ee
For $\kp,\km \gg \pi T$ this goes over into~\cite{nonrel}
\ba
 \rho^{ }_{\mathcal{I}_\rmii{b}} & \stackrel{\kp,\km \gg \pi T}{\approx} &
 \frac{1}{16\pi} 
 \int_\vec{q} \frac{n^{ }_{\sigma_2}(q)}{q}
 \;, \la{asy_Ib}
\ea
up to exponentially small corrections. The value of the remaining
integral is given in \eq\nr{nBnF}. 
The specific statistics for the current problem are  
\ba
 \mathcal{I}^{ }_\rmi{b} & \Leftrightarrow &  
 (\sigma_1\sigma_4\sigma_5 | \sigma_2 \sigma_0) = (+-+|+-)
 \;, \la{Ib_mpp} \\
 \widetilde{\mathcal{I}}^{ }_\rmi{b} & \Leftrightarrow &  
 (\sigma_1\sigma_4\sigma_5 | \sigma_2 \sigma_0) = (+--|--)
 \;. \la{Ib_mpm}
\ea
Defining a typical thermal momentum through 
\be
 k^2_\rmi{av}(M) \equiv \frac{\int_0^\infty \! {\rm d}k \, k^4 
 \exp(-\frac{\sqrt{k^2 + M^2}}{T}) }{\int_0^\infty \! {\rm d}k \, k^2 
 \exp(-\frac{\sqrt{k^2 + M^2}}{T})} = 
 \frac{3 M T K^{ }_3(\fr{M}{T})}{K^{ }_2(\fr{M}{T})}
 \;, \la{kav}
\ee
numerical results for $\rho^{ }_{\mathcal{I}^{ }_\rmii{b}}$ are
shown in \fig\ref{fig:Ib}.

\begin{figure}[t]


\centerline{%
 \epsfysize=7.0cm\epsfbox{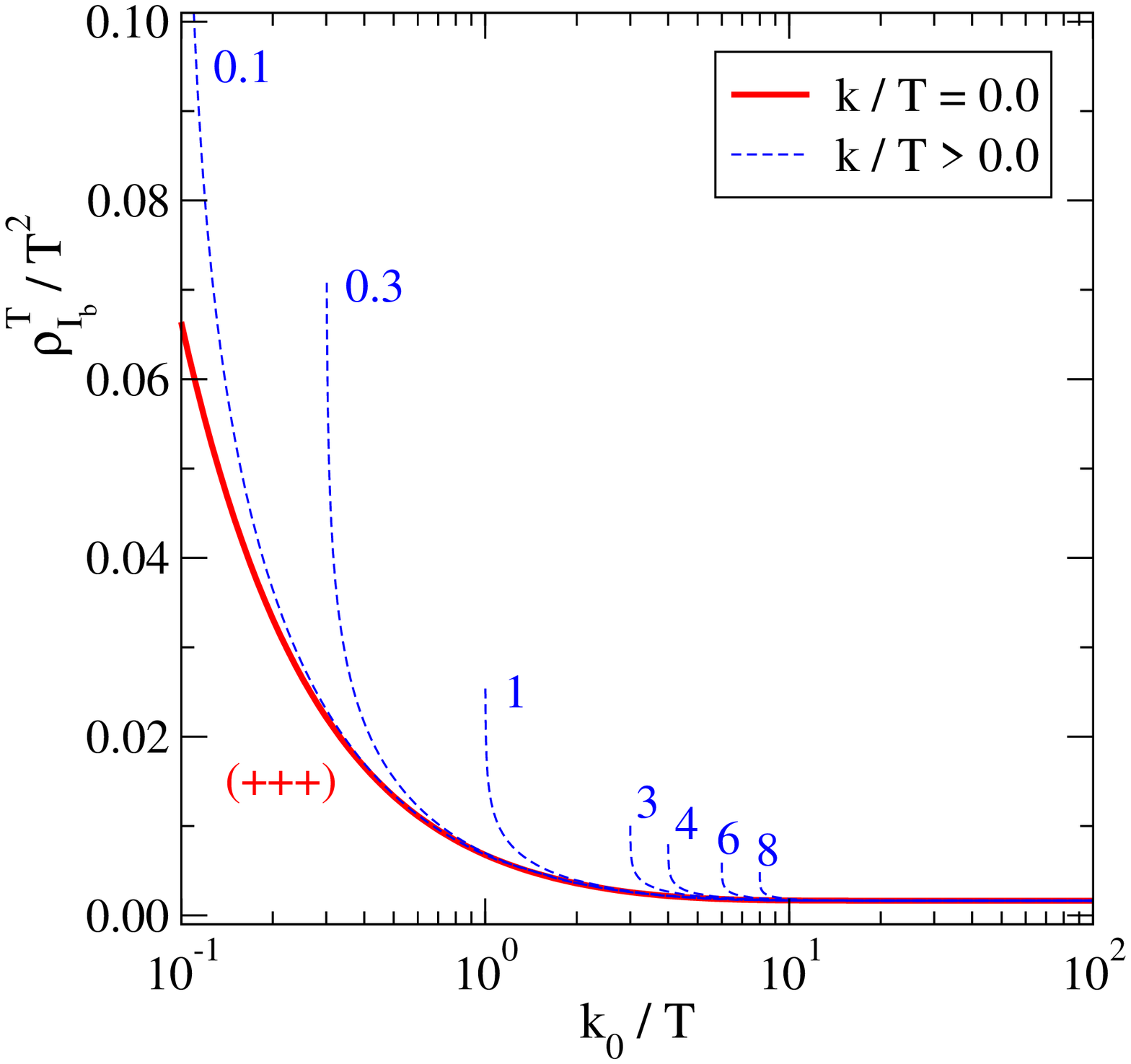}%
~~~\epsfysize=7.6cm\epsfbox{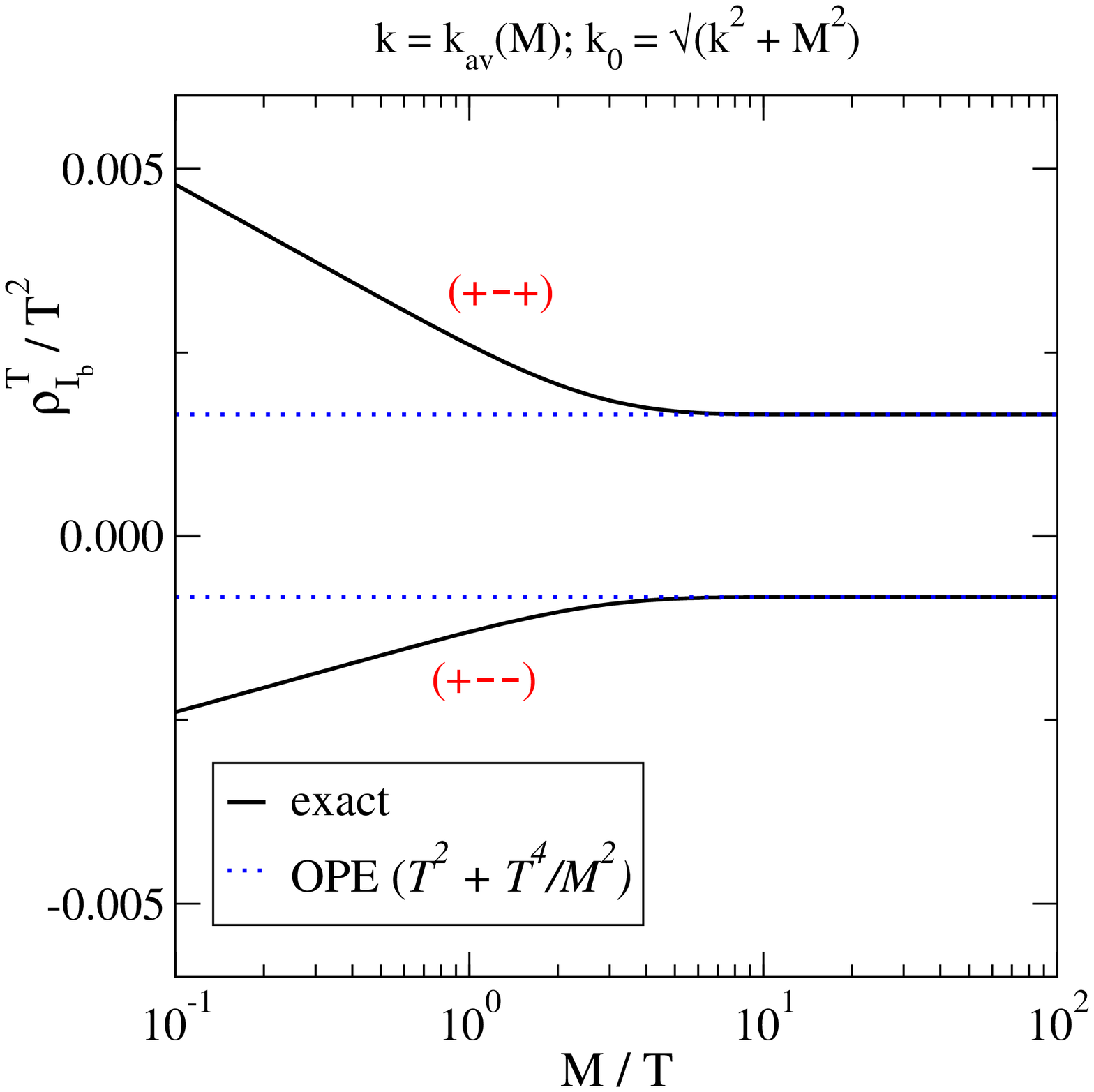}
}

\caption[a]{\small
Left: The spectral function 
$\rho^\rmii{$T$}_{\mathcal{I}^{ }_\rmii{b}} \equiv 
 \rho^{ }_{\mathcal{I}^{ }_\rmii{b}}$ 
with the purely bosonic statistics $(\sigma_1\sigma_4\sigma_5) = (+++)$,
for $\ko \ge k + 0.001 T$, 
compared with the zero-momentum limit determined in ref.~\cite{bulk_wdep}.
Right: The spectral function 
$\rho^{ }_{\mathcal{I}^{ }_\rmii{b}}$ with the momentum of 
\eq\nr{kav} and statistics of 
\eqs\nr{Ib_mpp}, \nr{Ib_mpm} as a function of $M/T$, 
compared with the OPE-asymptotics from \eq\nr{asy_Ib}. 
}

\la{fig:Ib}
\end{figure}

%
\subsection{$\rho^{ }_{\mathcal{I}^{ }_\rmii{c}}$}

Since all versions of $\mathcal{I}_\rmi{c}$ are
independent of the external momentum, there is no cut:  
\ba
 \rho^{ }_{\mathcal{I}_\rmi{c}} & \!\! = \!\! & 
 \rho^{ }_{\widetilde{\mathcal{I}}_\rmi{c}}
 \; = \; 
 \rho^{ }_{\widehat{\mathcal{I}}_\rmi{c}}
 \; = \; 
 \rho^{ }_{\overline{\mathcal{I}}_\rmi{c}}
 \; = \; 
 0
 \;. \la{asy_Ic}
\ea

%
\subsection{$\rho^{ }_{\mathcal{I}^{ }_\rmii{d}}$}

The derivation of the spectral function corresponding to \eq\nr{def_Id} 
follows from that for $\rho^{ }_{\mathcal{J}^{ }_\rmii{d}}$ in 
\se\ref{ss:Jb}; we simply give
the line with momentum $P$ a mass, $\lambda$, and take a derivative
with respect to the mass. If we change variables from $p$ to 
\be
  E_p \equiv \sqrt{p^2 + \lambda^2}
  \;,
\ee 
then $\lambda$ only appears  
in the boundaries of the $E_p$-integration. 
No terms of $\rmO(\epsilon)$ are needed, so   
the general result can be expressed as
\ba
 \rho^{ }_{\mathcal{I}^{ }_\rmii{d}} & = &  
 \frac{\pi\mathcal{K}^2}{(4\pi)^2 k}
 \int_\vec{q} \frac{n^{ }_{\sigma_2}(q)}{q}
 \frac{{\rm d}}{{\rm d}\lambda^2}
 \biggl\{ 
    \int_{\km + \frac{\lambda^2}{4\km}}^{\kp + \frac{\lambda^2}{4\kp}}
    \! {\rm d} E_p \, 
    \Bigl[ 1 + n^{ }_{\sigma_1}(E_p) + n^{ }_{\sigma_4}(\ko - E_p)  \Bigr]
 \biggr\}_{\lambda = 0}
 \nn
 & = &
 \frac{\pi\mathcal{K}^2 n^{-1}_{\sigma_0} (k_0) }{(4\pi)^2 k}
 \biggl[
   \frac{ n^{ }_{\sigma_1}(\kp) n^{ }_{\sigma_4}(\km)}{4\kp} 
 -
   \frac{ n^{ }_{\sigma_1}(\km) n^{ }_{\sigma_4}(\kp)}{4\km}
 \biggr] 
 \int_\vec{q} \frac{n^{ }_{\sigma_2}(q)}{q}
 \;. \la{rho_Id}
\ea
The value of the remaining
integral is given in \eq\nr{nBnF}. 
For $\kp,\km\gg \pi T$ the asymptotics reads~\cite{nonrel}
\ba
 \rho^{ }_{\mathcal{I}_\rmii{d}} & \stackrel{\kp,\km \gg \pi T}{\approx} &
 - \frac{1}{16\pi} 
 \int_\vec{q} \frac{n^{ }_{\sigma_2}(q)}{q}
 \;, \la{asy_Id}
\ea
with exponentially small corrections. 
The specific statistics for the current problem are 
\ba
 \mathcal{I}^{ }_\rmi{d} & \Leftrightarrow &  
 (\sigma_1\sigma_4\sigma_5 | \sigma_2\sigma_0) = (+-+ | +-)
 \;, \la{Id_mpp} \\
 \widetilde{\mathcal{I}}^{ }_\rmi{d} & \Leftrightarrow &  
 (\sigma_1\sigma_4\sigma_5 | \sigma_2\sigma_0) = (+-- | --)
 \;, \\ 
 \widehat{\mathcal{I}}^{ }_\rmi{d} & \Leftrightarrow & 
 (\sigma_1\sigma_4\sigma_5 | \sigma_2\sigma_0) = (-+- | +-)
 \;, \\ 
 \overline{\mathcal{I}}^{ }_\rmi{d} & \Leftrightarrow & 
 (\sigma_1\sigma_4\sigma_5 | \sigma_2\sigma_0) = (-++ | --)
 \;. \la{Id_pmm}
\ea
A numerical evaluation in shown in \fig\ref{fig:Id}.

\begin{figure}[t]


\centerline{%
 \epsfysize=7.0cm\epsfbox{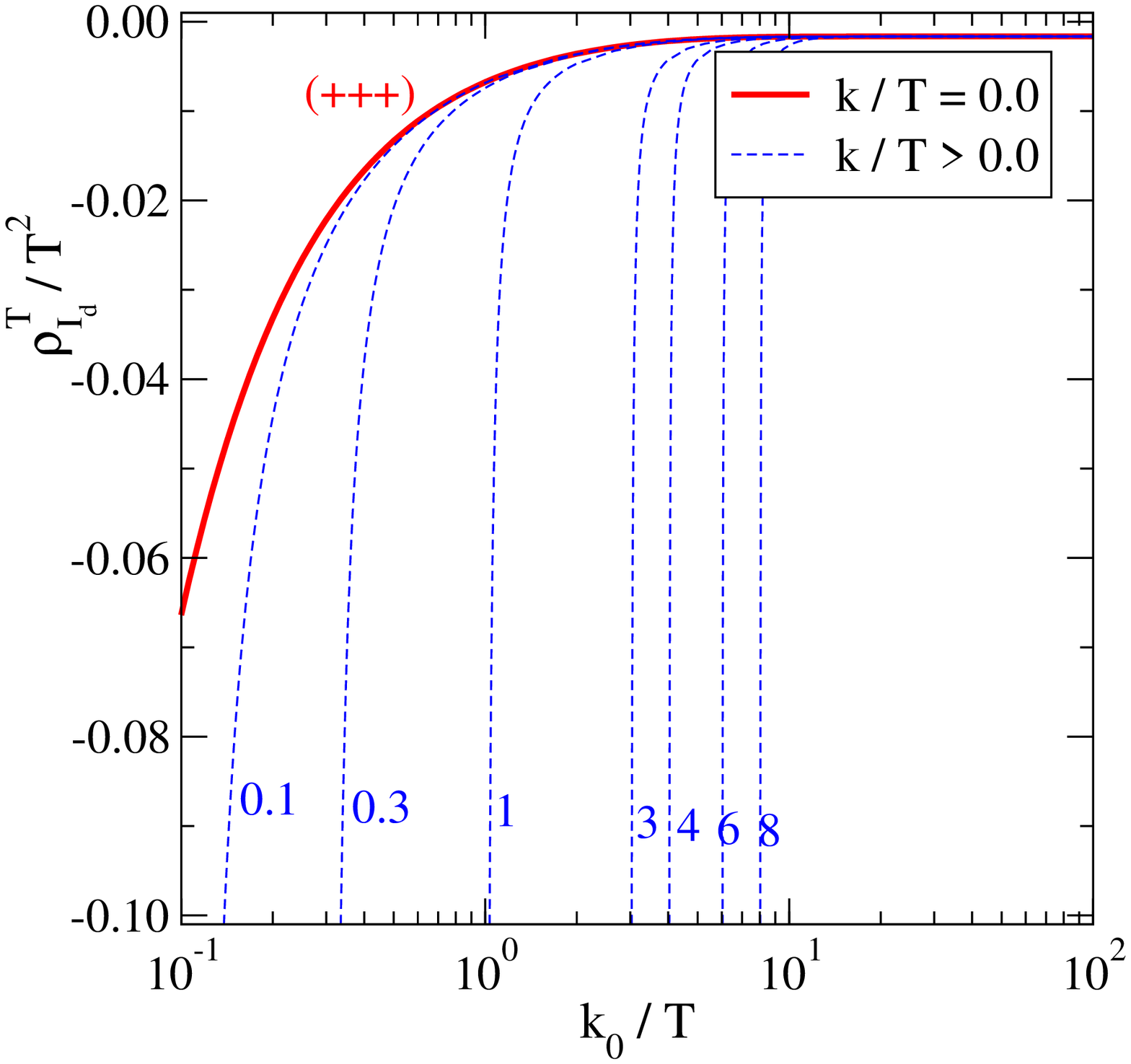}%
~~~\epsfysize=7.6cm\epsfbox{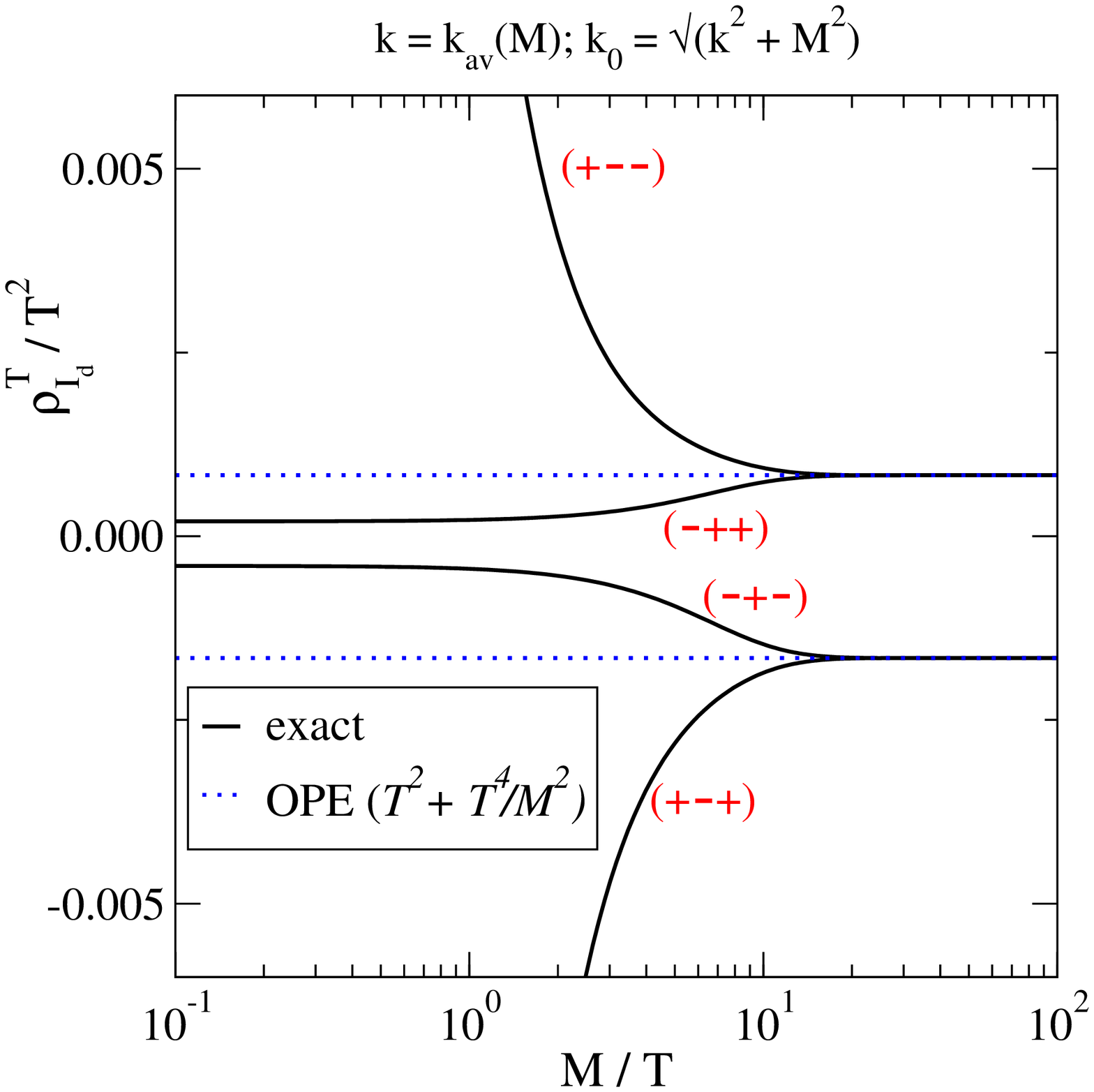}
}

\caption[a]{\small
Left: The spectral function 
$\rho^{ }_{\mathcal{I}^{ }_\rmii{d}} \equiv 
 \rho^\rmii{$T$}_{\mathcal{I}^{ }_\rmii{d}} $ 
with the purely bosonic statistics $(\sigma_1\sigma_4\sigma_5) = (+++)$, 
for $\ko \ge k + 0.001 T$, 
compared with the zero-momentum limit determined in ref.~\cite{bulk_wdep}.
Right: The spectral function 
$\rho^{ }_{\mathcal{I}^{ }_\rmii{d}}$ with the momentum of 
\eq\nr{kav} and statistics of 
\eqs\nr{Id_mpp}--\nr{Id_pmm} as a function of $M/T$, 
compared with the OPE-asymptotics from \eq\nr{asy_Id}. 
}

\la{fig:Id}
\end{figure}

%
\subsection{$\rho^{ }_{\mathcal{I}^{ }_\rmii{e}}$}

Since both versions of $\mathcal{I}_\rmi{e}$ are
independent of the external momentum, there is no cut:  
\ba
 \rho^{ }_{\mathcal{I}_\rmi{e}}  & \!\! = \!\! & 
 \rho^{ }_{\widetilde{\mathcal{I}}_\rmi{e}}
 \; = \;
 0 
 \;. \la{asy_Ie}
\ea

%
\subsection{$\rho^{ }_{\mathcal{I}^{ }_\rmii{f}}$}

After carrying out the Matsubara sums, the expression for 
$\mathcal{I}_\rmi{f}$ reads
\ba
 \mathcal{I}_\rmi{f}
 & = &  \lim_{\lambda\to 0}
  \int_{\vec{p},\vec{q}} \frac{1} 
  { 8 \epsilon^{ }_q \epsilon^{ }_{pk} E^{ }_{qp} }
  \biggl\{ 
  \nn &&
  \frac{1}{-i k_n  + \epsilon^{ }_{pk} + \epsilon^{ }_q + E^{ }_{qp}}
  \; \Bigl( 
   \bigl[1 
   + n^{ }_{\sigma_4}(\epsilon^{ }_{pk})
   + n^{ }_{\sigma_2}(\epsilon^{ }_q) \bigr] 
   \bigl[1 + n^{ }_{\sigma_5}(E^{ }_{qp}) \bigr] 
   + n^{ }_{\sigma_4}(\epsilon^{ }_{pk})
      n^{ }_{\sigma_2}(\epsilon^{ }_q)
  \Bigr)
  \nn & + &
  \frac{1}{-i k_n  - \epsilon^{ }_{pk} + \epsilon^{ }_q + E^{ }_{qp}}
  \; \Bigl( 
  n^{ }_{\sigma_4}(\epsilon^{ }_{pk})
  \bigl[1 + n^{ }_{\sigma_2}(\epsilon^{ }_q)  +
     n^{ }_{\sigma_5}(E^{ }_{qp}) \bigr] 
  -  n^{ }_{\sigma_2}(\epsilon^{ }_q)  n^{ }_{\sigma_5}(E^{ }_{qp}) \Bigr)
  \nn & + &
  \frac{1}{-i k_n  + \epsilon^{ }_{pk} - \epsilon^{ }_q + E^{ }_{qp}}
  \; \Bigl( 
   n^{ }_{\sigma_2}(\epsilon^{ }_q) 
  \bigl[1 
    + n^{ }_{\sigma_4}(\epsilon^{ }_{pk})
   + n^{ }_{\sigma_5}(E^{ }_{qp})  \bigr] 
  -   n^{ }_{\sigma_4}(\epsilon^{ }_{pk}) 
      n^{ }_{\sigma_5}(E^{ }_{qp})
  \Bigr)
  \nn & + &
  \frac{1}{-i k_n  + \epsilon^{ }_{pk} + \epsilon^{ }_q - E^{ }_{qp}}
  \; \Bigl( 
   n^{ }_{\sigma_5}(E^{ }_{qp})
  \bigl[1 
    + n^{ }_{\sigma_4}(\epsilon^{ }_{pk})
   + n^{ }_{\sigma_2}(\epsilon^{ }_q) \bigr] 
  - n^{ }_{\sigma_4}(\epsilon^{ }_{pk} )
  n^{ }_{\sigma_2}(\epsilon^{ }_q)
    \Bigr) \; \biggr\}
   \nn 
 & + & (ik_n \to -i k_n)  
 \;. \la{If_full}
\ea
The corresponding spectral function is obtained from \eq\nr{cut2}.
In addition to the energies of \eq\nr{energies1}, a further
variable appears here which contains the infrared regulator $\lambda$: 
\be
 E^{ }_{qp} \equiv \sqrt{(\vec{q-p})^2 + \lambda^2}
 \;. \la{energies2}
\ee
The four channels of \eq\nr{If_full} represent real corrections
and were labelled (r1)--(r4) in ref.~\cite{master}; 
the $2\leftrightarrow 2$ and 
$1\leftrightarrow 3$ scatterings that they represent 
are of the type illustrated in \fig{3} of ref.~\cite{master} but 
with internal propagators carrying the indices
$\sigma_1, \sigma_3$ shrunk to points. 

In the notation of ref.~\cite{master}, 
\eq(4.28) now reads
\ba
 \Bigl\langle 
   \Phi^{ }_\rmi{r1}(\ko - p_0 | q | p_0 - q | \cdot)
 \Bigr\rangle
 & = & - 
 \Bigl\langle 
   \Phi^{ }_\rmi{r2}(p_0 - \ko  | q | p_0 - q | \cdot)
 \Bigr\rangle
 \nn 
 & = &  - 
 \Bigl\langle 
   \Phi^{ }_\rmi{r3}(\ko - p_0 | -q | p_0 - q | \cdot)
 \Bigr\rangle
 \nn 
 & = &  - 
 \Bigl\langle 
   \Phi^{ }_\rmi{r4}(\ko - p_0 | q | q - p_0 | \cdot)
 \Bigr\rangle
 \nn[2mm]
 & = & 
 \, \frac{ n^{ }_{\sigma_4} (\ko - p_0) 
 \, n^{ }_{\sigma_2} (q) 
 \, n^{ }_{\sigma_5} (p_0 - q) }{2  n^{ }_{\sigma_0}(\ko)  }
 \;. \hspace*{1cm} 
\ea
The integration over
$p$, with ranges as specified in ref.~\cite{master},
is trivial. Factoring out
\be
 \frac{\pi 
 \, n^{ }_{\sigma_4} (\ko - p) 
 \, n^{ }_{\sigma_2} (q) 
 \, n^{ }_{\sigma_5} (p - q)
 }{(4\pi)^4 k\,   n^{ }_{\sigma_0}(\ko) }
 \;,
\ee 
the $\lambda\to 0$ limits (in the different regimes as specified
in ref.~\cite{master}) read: 
\ba
 (\mbox{a}) = ({\mbox{l}}): & & 4(p-q)
  \;, \la{If_range_a} \\
 (\mbox{b}): & & 4(\kp - q)
  \;, \\
 (\underline{\mbox{b}}): & & 4(p-\km)
  \;, \\
 (\mbox{c}) = - ({\mbox{h}}) = - (\underline{\mbox{h}})
 = -   ({\mbox{j}}): & & 4(\kp - \km)
  \;, \\
 (\tilde{\mbox{c}}): & & 4 \km
  \;, \\
 (\mbox{d}): & & 4(\kp  + q - p)
  \;, \\
 (\mbox{e}) =  ({\mbox{f}}): & & 4(\kp + \km - p )
  \;, \\
 (\underline{\mbox{e}}) =  (\underline{\mbox{f}}): & & 4 q
  \;,  \la{If_range_e} \\
 ({\mbox{g}}) =  (\underline{\mbox{g}}): & & 4 (\km + q - p)
  \;, \\
 ({\mbox{i}}) = (\underline{\mbox{k}}): & & 4 (\km - q )
  \;,  \la{If_range_i} \\
 (\underline{\mbox{i}}) = ({\mbox{k}}): & & 4(p - \kp)
  \;.  \la{If_range_uk}
\ea

For $\kp, \km \gg \pi T$, the ultraviolet asymptotics of 
the spectral function reads~\cite{nonrel}
\ba
 \rho^{ }_{\mathcal{I}_\rmi{f}} &  \stackrel{\kp,\km \gg \pi T}{\approx} & 
   \frac{\pi \mufac\mathcal{K}^2}{(4\pi)^4} \times \fr12  
 +  
 \int_\vec{p} \biggl\{
 \frac{n^{ }_{\sigma_4} + n^{ }_{\sigma_5} + n^{ }_{\sigma_2}}{16\pi p}
 \biggr\}
 \;. \hspace*{5mm} \la{asy_If}
\ea
The specific statistics for the current problem are 
\ba
 \mathcal{I}^{ }_\rmi{f} & \Leftrightarrow &  
 (\sigma_1\sigma_4\sigma_5 | \sigma_2\sigma_0) = (+-+|+-)
 \;, \la{If_mpp} \\
 \widetilde{\mathcal{I}}^{ }_\rmi{f} & \Leftrightarrow &  
 (\sigma_1\sigma_4\sigma_5 | \sigma_2\sigma_0) = (+--|--)
 \;. \la{If_mpm}
\ea

\begin{figure}[t]


\centerline{%
 \epsfysize=7.0cm\epsfbox{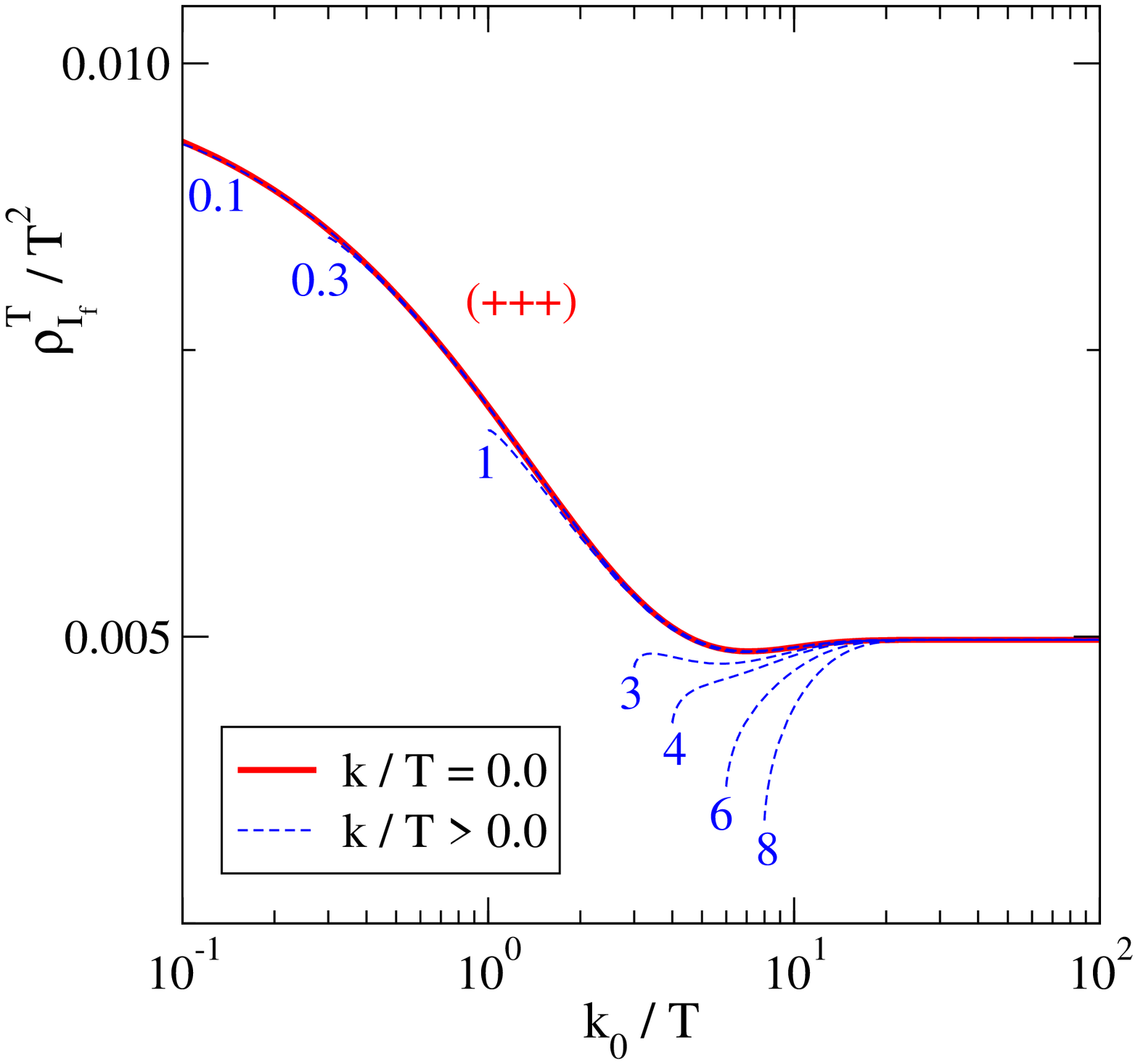}%
~~~\epsfysize=7.6cm\epsfbox{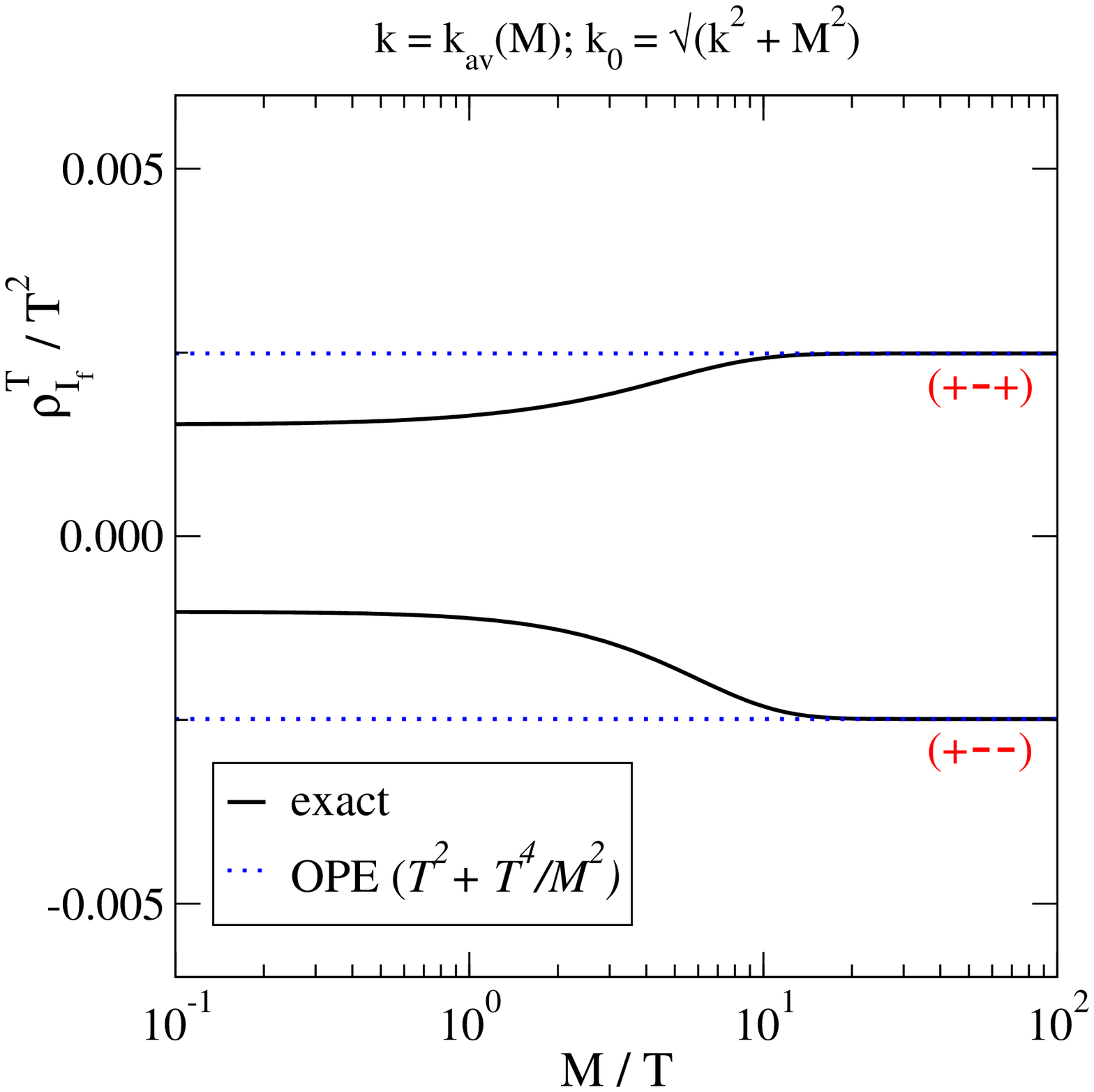}
}

\caption[a]{\small
Left: The thermal part of 
$\rho^{ }_{\mathcal{I}^{ }_\rmii{f}}$ 
with the purely bosonic statistics $(\sigma_1\sigma_4\sigma_5) = (+++)$, 
for $\ko \ge k + 0.001 T$, 
compared with the zero-momentum limit determined in ref.~\cite{bulk_wdep}.
Right: The thermal part of 
$\rho^{ }_{\mathcal{I}^{ }_\rmii{f}}$ with the momentum of 
\eq\nr{kav} and statistics of 
\eqs\nr{If_mpp}, \nr{If_mpm} as a function of $M/T$, 
compared with the OPE-asymptotics from \eq\nr{asy_If}. 
}

\la{fig:If}
\end{figure}

\noindent
For numerical evaluation, we have reflected the final 
integral to the domain defined in \fig{6} of ref.~\cite{master}. 
The corresponding integrand is not shown explicitly, since no 
substantial cancellations take place in the reflections. 
In addition, we always separate a ``vacuum-like'' part
as in \eq\nr{splitup}, 
with a coefficient given by the leading term in \eq\nr{asy_If}:
\ba
 \rho^{\rmi{vac}}_{\mathcal{I}^{ }_\rmi{f}} & \equiv & 
 \frac{\pi \mathcal{K}^2 T }{(4\pi)^4 k}
    \ln \biggl( 
     \frac{e^{{\kp} / {T}}+\sigma_0 e^{-\kp/T}-\sigma_1 - \sigma_4}
     {e^{{\km} / {T}}+\sigma_0 e^{-\km/T}-\sigma_1 - \sigma_4}
   \biggr)
 \times \fr12 
 \;. \la{If_thermal_splitup}
\ea
It is the thermal part which is plotted numerically in \fig\ref{fig:If},
and compared with the OPE asymptotics from \eq\nr{asy_If} as well as 
with the $k\to 0$ limit from ref.~\cite{bulk_wdep}.

%
\subsection{$\rho^{ }_{\mathcal{I}^{ }_\rmii{g}}$}

For $\mathcal{I}^{ }_\rmii{g}$ of \eq\nr{def_Ig}, Matsubara sums lead to 
\ba
 \mathcal{I}_\rmi{g}
 & = &  
  \int_{\vec{p}} \frac{K^2} 
  { 4 \epsilon^{ }_p \epsilon^{ }_{pk}  }
  \; \biggl\{ 
  \biggl[ 
   \frac{1}{i k_n + \epsilon^{ }_p + \epsilon^{ }_{pk}} + 
   \frac{1}{- i k_n + \epsilon^{ }_p + \epsilon^{ }_{pk}}
  \biggr]
  \Bigl[
   1  + n^{ }_{\sigma_1}(\epsilon^{ }_p) + n^{ }_{\sigma_4}(\epsilon^{ }_{pk})
  \Bigr]
  \nn & & \qquad\quad + \, 
  \biggl[ 
   \frac{1}{i k_n - \epsilon^{ }_p + \epsilon^{ }_{pk}} + 
   \frac{1}{- i k_n - \epsilon^{ }_p + \epsilon^{ }_{pk}}
  \biggr]
  \Bigl[
    n^{ }_{\sigma_1}(\epsilon^{ }_p)  - n^{ }_{\sigma_4}(\epsilon^{ }_{pk}) 
  \Bigr]
  \biggr\} 
  \nn && \times \,  
   \int_{\vec{q}} \frac{1} 
  { 4 \epsilon^{ }_q \epsilon^{ }_{qk}  }
  \biggl\{ 
  \biggl[ 
   \frac{1}{i k_n + \epsilon^{ }_q + \epsilon^{ }_{qk}} + 
   \frac{1}{- i k_n + \epsilon^{ }_q + \epsilon^{ }_{qk}}
  \biggr]
  \Bigl[
   1 + n^{ }_{\sigma_2}(\epsilon^{ }_{q})
     + n^{ }_{\sigma_3}(\epsilon^{ }_{qk}) 
  \Bigr]
  \nn & & \qquad\quad + \, 
  \biggl[ 
   \frac{1}{i k_n - \epsilon^{ }_q + \epsilon^{ }_{qk}} + 
   \frac{1}{- i k_n - \epsilon^{ }_q + \epsilon^{ }_{qk}}
  \biggr]
  \Bigl[
    n^{ }_{\sigma_2}(\epsilon^{ }_q)  - n^{ }_{\sigma_3}(\epsilon^{ }_{qk}) 
  \Bigr]
  \biggr\}
  \;.  \la{Ig_full}
\ea
Taking the cut like in \eq\nr{cut2}, 
the result factorizes into a product of a structure
like in \eq\nr{rho_Jb_1}, and a principal value integral over the 
other part; the latter corresponds to a virtual loop 
correction, of the type illustrated in \fig{3} of ref.~\cite{master}
but with the line carrying the index $\sigma_5$ shrunk to a point.  
It has a divergent vacuum contribution, 
\be
  \int_{\vec{q}} \frac{1} 
  { 4  \epsilon^{ }_q \epsilon^{ }_{qk}  }
  \mathbbm{P} 
  \biggl[ 
   \frac{1}{\ko + \epsilon^{ }_q + \epsilon^{ }_{qk}} + 
   \frac{1}{- \ko + \epsilon^{ }_q + \epsilon^{ }_{qk}}
  \biggr]
 =   
 \frac{1 \mufac}{(4\pi)^2}
 \biggl(
  \frac{1}{\epsilon} + \ln\frac{\bmu^2}{\mathcal{K}^2} + 2 
 \biggr)
 \;,
\ee
as well as a finite thermal part. In the latter the angular
integral is doable if we substitute integration variables so 
as to always have $\epsilon^{ }_p$ or $\epsilon^{ }_q$ as the
argument of the phase space distribution. Recalling the 
$\rmO(\epsilon)$-part from \eq\nr{Jb_eps0}, we get 
\ba
 \rho^{\rmi{vac}}_{\mathcal{I}^{ }_\rmi{g}} & \equiv & 
 -\frac{\pi \mathcal{K}^2 T}{(4\pi)^4 k}
 \ln \biggl( 
     \frac{e^{{\kp} / {T}}+\sigma_0 e^{-\kp/T}-\sigma_1 - \sigma_4}
     {e^{{\km} / {T}}+\sigma_0 e^{-\km/T}-\sigma_1 - \sigma_4}
 \biggr)
 \biggl(
  \frac{1}{\epsilon} + 2 \ln\frac{\bmu^2}{\mathcal{K}^2} + 4    
 \biggr)
 \nn & & \quad + \, 
 (\sigma_1\leftrightarrow\sigma_2,\sigma_4\leftrightarrow\sigma_3)
 \;, \\[2mm] 
  \rho^{\rmii{$T$}}_{\mathcal{I}^{ }_\rmi{g}} & = & 
  \frac{\pi \mathcal{K}^2}{(4\pi)^4 k}
  \int_{\km}^{\kp} \! {\rm d}p \,
  \frac{ n^{ }_{\sigma_4}(\ko -p) n^{ }_{\sigma_1}(p)}{
     n^{ }_{\sigma_0}(\ko)}
  \biggl\{ 
       \ln \frac{ (\kp - p)(p - \km)}{k^2}  + 2 
 \nn & & \; + \, 
 \int_0^\infty \! \frac{{\rm d}q}{k} \,
    \bigl( n^{ }_{\sigma_2} + n^{ }_{\sigma_3}\bigr)(q)
  \, 
 \ln\biggl|\frac{(q-\kp)(q+\km)}{(q+\kp)(q-\km)} \biggr|
 \biggr\} 
 \nn & & \quad + \, 
 (\sigma_1\leftrightarrow\sigma_2,\sigma_4\leftrightarrow\sigma_3)
 \;. \la{Ig_thermal_splitup}
\ea

\begin{figure}[t]


\centerline{%
 \epsfysize=7.0cm\epsfbox{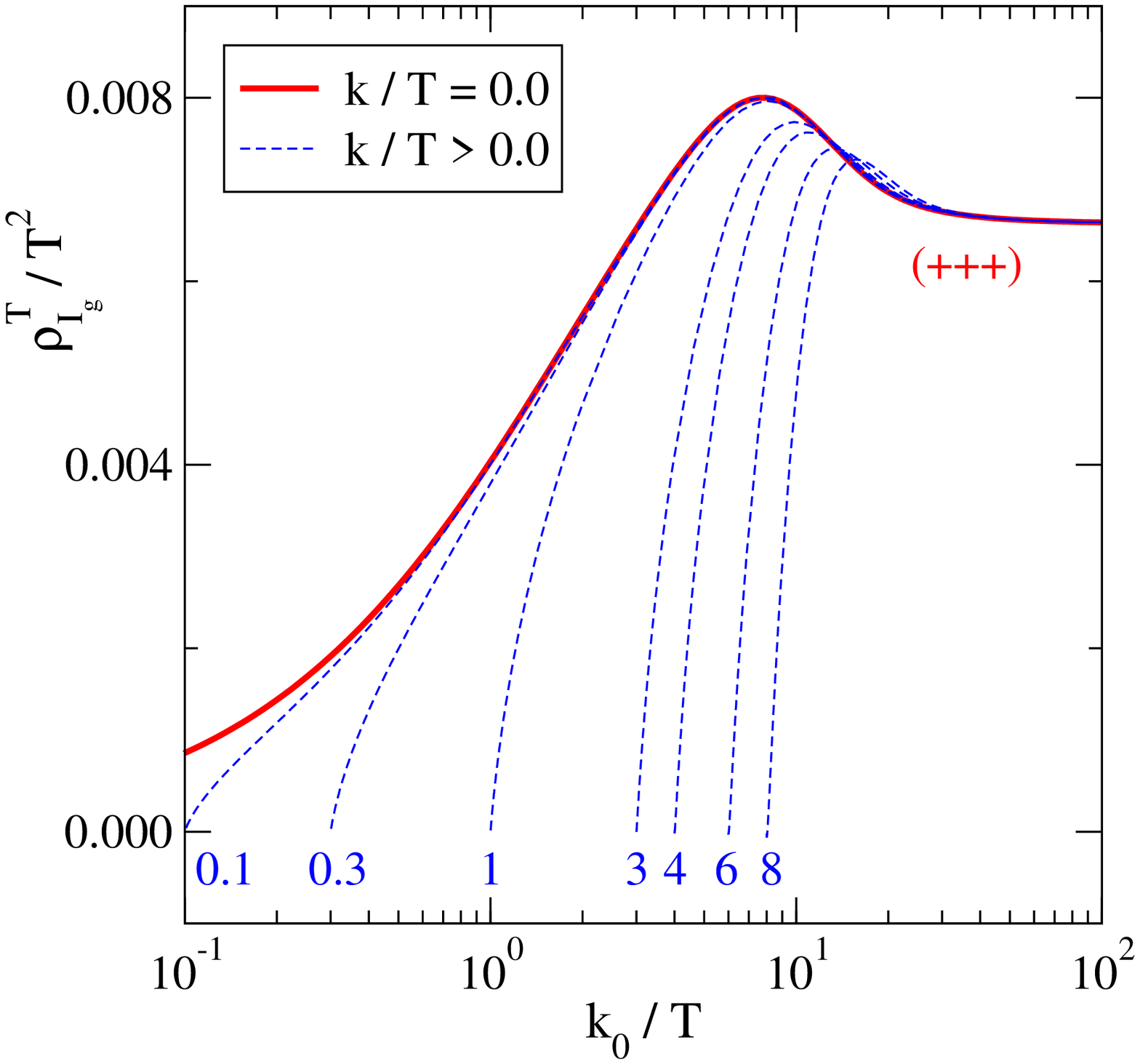}%
~~~\epsfysize=7.6cm\epsfbox{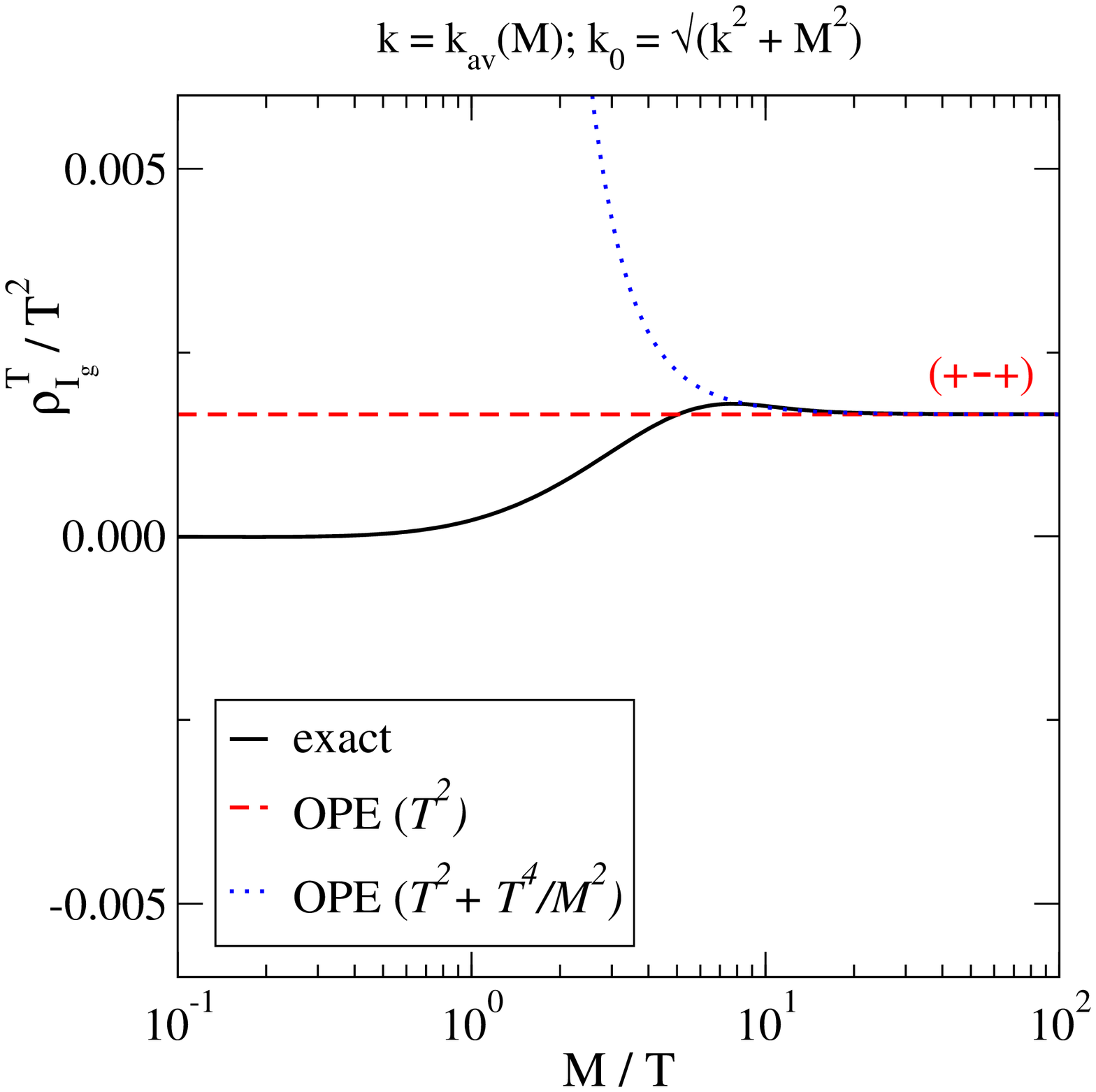}
}

\caption[a]{\small
Left: The thermal part of 
$\rho^{ }_{\mathcal{I}^{ }_\rmii{g}}$
(\eq\nr{Ig_thermal_splitup}) 
with the purely bosonic statistics $(\sigma_1\sigma_4\sigma_5) = (+++)$, 
for $\ko \ge k + 0.001 T$, 
compared with the zero-momentum limit determined in ref.~\cite{bulk_wdep}.
Right: The thermal part of 
$\rho^{ }_{\mathcal{I}^{ }_\rmii{g}}$ with the momentum of 
\eq\nr{kav} and statistics of 
\eq\nr{Ig_mp} as a function of $M/T$, 
compared with the OPE-asymptotics from \eq\nr{asy_Ig}. 
}

\la{fig:Ig}
\end{figure}

For $\kp, \km \gg \pi T$, the ultraviolet asymptotics reads~\cite{nonrel}
\ba
 \rho^{ }_{\mathcal{I}_\rmi{g}} &  \stackrel{\kp,\km \gg \pi T}{\approx}  & 
  -\frac{\mufac\mathcal{K}^2}{2(4\pi)^3} \biggl( 
   \frac{1}{\epsilon}+ 2\ln\frac{\bmu^2}{\mathcal{K}^2} + 4 \biggr)
 \, + \, 
  \int_\vec{p} \sum_{i = 1}^{4} \biggl\{
 \frac{n^{ }_{\sigma_i} }{16\pi p}
 +
 \frac{p\, n^{ }_{\sigma_i} }{4\pi}
 \frac{\ko^2 + \fr{k^2}3}{\mathcal{K}^4}
 \biggr\}
 \;. \hspace*{5mm} \la{asy_Ig}
\ea
The specific statistics for the current case are
\ba
 \mathcal{I}^{ }_\rmi{g} & \Leftrightarrow &  
 (\sigma_1\sigma_4\sigma_5 | \sigma_2 \sigma_3 \sigma_0) = (+-+|+--)
 \;. \la{Ig_mp}
\ea
The thermal part of $\rho^{ }_{\mathcal{I}_\rmi{g}}$ 
is plotted numerically in \fig\ref{fig:Ig}, and compared
with the OPE-asymptotics from \eq\nr{asy_Ig} as well as 
with the $k\to 0$ limit from ref.~\cite{bulk_wdep}.

%
\subsection{$\rho^{ }_{\mathcal{I}^{ }_\rmii{h}}$}
\la{app:Ih}

The spectral functions corresponding to 
$
 \mathcal{I}^{ }_\rmi{h}
$, 
$
 \widetilde{\mathcal{I}}^{ }_\rmi{h}
$, 
and 
$
  \widehat{\mathcal{I}}^{ }_\rmi{h}
$
can be handled simultaneously with the labelling of \eq\nr{labelling},
where now the line with the index $\sigma_3$ is absent.
The expression after carrying out the Matsubara sums reads
\ba
  \mathcal{I}^{ }_\rmi{h} & = &  \lim_{\lambda\to 0}
  \int_{\vec{p},\vec{q}} \frac{K^2} 
  { 8 \epsilon^{ }_q \epsilon^{ }_{pk} E^{ }_{qp} }
  \biggl\{ 
  \nn &&
  \frac{1}{-i k_n  + \epsilon^{ }_{pk} + \epsilon^{ }_q + E^{ }_{qp}}
  \; \frac{
   \bigl[1  + n^{ }_{\sigma_4}(\epsilon^{ }_{pk})
   + n^{ }_{\sigma_2}(\epsilon^{ }_q)
    \bigr] 
   \bigl[1 + n^{ }_{\sigma_5}(E^{ }_{qp}) \bigr] 
   + n^{ }_{\sigma_4}(\epsilon^{ }_{pk}) n^{ }_{\sigma_2}(\epsilon^{ }_q) }
  {\epsilon_p^2 - (\epsilon^{ }_q + E^{ }_{qp})^2}
  \nn & + &
  \frac{1}{-i k_n  - \epsilon^{ }_{pk} + \epsilon^{ }_q + E^{ }_{qp}}
  \; \frac{
  n^{ }_{\sigma_4}(\epsilon^{ }_{pk})
  \bigl[1 + n^{ }_{\sigma_2}(\epsilon^{ }_q)  +
     n^{ }_{\sigma_5}(E^{ }_{qp}) \bigr] 
  -  n^{ }_{\sigma_2}(\epsilon^{ }_q)  n^{ }_{\sigma_5}(E^{ }_{qp}) }
  {\epsilon_p^2 - (\epsilon^{ }_q + E^{ }_{qp})^2}
  \nn & + &
  \frac{1}{-i k_n  + \epsilon^{ }_{pk} - \epsilon^{ }_q + E^{ }_{qp}}
  \; \frac{
   n^{ }_{\sigma_2}(\epsilon^{ }_q) 
  \bigl[1 + n^{ }_{\sigma_4}(\epsilon^{ }_{pk}) 
  + n^{ }_{\sigma_5}(E^{ }_{qp})
   \bigr] 
  - n^{ }_{\sigma_4}(\epsilon^{ }_{pk}) n^{ }_{\sigma_5}(E^{ }_{qp}) }
  {\epsilon_p^2 - (\epsilon^{ }_q - E^{ }_{qp})^2}
  \nn & + &
  \frac{1}{-i k_n  + \epsilon^{ }_{pk} + \epsilon^{ }_q - E^{ }_{qp}}
  \; \frac{
   n^{ }_{\sigma_5}(E^{ }_{qp})
  \bigl[1 + n^{ }_{\sigma_2}(\epsilon^{ }_q) 
 + n^{ }_{\sigma_4}(\epsilon^{ }_{pk}) \bigr] 
  -  n^{ }_{\sigma_2}(\epsilon^{ }_q)  n^{ }_{\sigma_4}(\epsilon^{ }_{pk})}
  {\epsilon_p^2 - (\epsilon^{ }_q - E^{ }_{qp})^2}
  \; \biggr\} 
 \nn & + &  \lim_{\lambda\to 0}
  \int_{\vec{p}} \frac{K^2} 
  { 4 \epsilon^{ }_p \epsilon^{ }_{pk} }
  \biggl[ \frac{ 
    1 
   + n^{ }_{\sigma_1}(\epsilon^{ }_{p})  
   + n^{ }_{\sigma_4}(\epsilon^{ }_{pk}) 
    }{-i k_n + \epsilon^{ }_p + \epsilon^{ }_{pk}}
  + 
    \frac{ 
    n^{ }_{\sigma_1}(\epsilon^{ }_{p})  - 
       n^{ }_{\sigma_4}(\epsilon^{ }_{pk})
    }{i k_n - \epsilon^{ }_p + \epsilon^{ }_{pk}}
  \biggr]
  \; 
  \nn & & \hspace*{1.3cm} \times 
   \int_{\vec{q}}   \biggl[ 
  \frac{1 + n^{ }_{\sigma_2}(\epsilon^{ }_q)  +
     n^{ }_{\sigma_5}(E^{ }_{qp})}{4 \epsilon^{ }_q E^{ }_{qp} }
  \biggl(
  \frac{1}{\epsilon^{ }_p + \epsilon^{ }_q + E^{ }_{qp}}
  -\, \frac{1}{\epsilon^{ }_p - \epsilon^{ }_q - E^{ }_{qp}}
 \biggl)
  \nn[2mm] & & \hspace*{2.3cm} 
  +\, \frac{
   n^{ }_{\sigma_2}(\epsilon^{ }_q)  -
     n^{ }_{\sigma_5}(E^{ }_{qp})}{4 \epsilon^{ }_q E^{ }_{qp} }  
  \biggl(
  \frac{1}{\epsilon^{ }_p - \epsilon^{ }_q + E^{ }_{qp}}
  -\, \frac{1}{\epsilon^{ }_p + \epsilon^{ }_q - E^{ }_{qp}}
  \biggr)
  \; \biggr]  
  \nn 
 & + & (ik_n \to -i k_n) 
 \;. \la{Ih_full}
\ea
Taking the cut like in \eq\nr{cut2}, 
the first four structures here are real corrections, the last two 
are virtual corrections. The corresponding scattering processes
can be depicted like in \fig{3} of ref.~\cite{master}, with internal 
propagators carrying the index $\sigma_3$ shrunk to points. 

For the real corrections, \eq(4.28) of ref.~\cite{master} now reads
\ba
 \Bigl\langle 
   \Phi^{ }_\rmi{r1}(\ko - p_0 | q | p_0 - q | \cdot)
 \Bigr\rangle
 & = & - 
 \Bigl\langle 
   \Phi^{ }_\rmi{r2}(p_0 - \ko  | q | p_0 - q | \cdot)
 \Bigr\rangle
 \nn 
 & = &  - 
 \Bigl\langle 
   \Phi^{ }_\rmi{r3}(\ko - p_0 | -q | p_0 - q | \cdot)
 \Bigr\rangle
 \nn 
 & = &  - 
 \Bigl\langle 
   \Phi^{ }_\rmi{r4}(\ko - p_0 | q | q - p_0 | \cdot)
 \Bigr\rangle
 \nn[2mm]
 & = & 
 \frac{  
 \, n^{ }_{\sigma_4} (\ko - p_0) 
 \, n^{ }_{\sigma_2} (q) 
 \, n^{ }_{\sigma_5} (p_0 - q) } { n^{ }_{\sigma_0}(\ko)  } 
 \, \mathbbm{P}\, \biggl\{ 
 \, \frac{\mathcal{K}^2}{2 (p_0^2 - p^2 )}
 \biggr\}
 \;. \hspace*{1cm} 
\ea
The integration over $p$, with ranges as specified in ref.~\cite{master}, is 
readily carried out, leading to simple logarithms. Renaming
subsequently $p_0\to p$, and factoring out
\be
 \frac{\pi\mathcal{K}^2 
 \, n^{ }_{\sigma_4} (\ko - p) 
 \, n^{ }_{\sigma_2} (q) 
 \, n^{ }_{\sigma_5} (p - q)
 }{(4\pi)^4 k \,  n^{ }_{\sigma_0}(\ko) }
 \;, \la{Ih_factor}
\ee 
the $\lambda\to 0$ limits (in the sense specified
in ref.~\cite{master}) read: 
\ba
 (\mbox{a}) =  ({\mbox{l}}): & &  \frac{1}{p} \ln\biggl| 
 \frac{ 4 p (p-q)^2 }{ \lambda^2 q} \biggr|
  \;, \la{Ih_range_a} \\
 (\mbox{b}): & & \frac{1}{p} \ln\biggl| 
 \frac{ \kp (p-q)}{ q(p-\kp) } \biggr|
  \;, \\
 (\underline{\mbox{b}}): & & \frac{1}{p} \ln\biggl| 
 \frac{4 p (p-q)(p-\km) }{\lambda^2 \km } \biggr|
  \;, \\
 (\mbox{c}) = -  ({\mbox{h}}) = -  (\underline{\mbox{h}})
 = -  ({\mbox{j}}): & & \frac{1}{p} \ln\biggl| 
 \frac{ \kp (p-\km) }{\km(p-\kp) } \biggr|
  \;, \\
 (\tilde{\mbox{c}}): & & \frac{1}{p} \ln\biggl| 
 \frac{ 4 \km p (p-q)}{ \lambda^2( p -\km)} \biggr|
  \;, \\
 (\mbox{d}): & & \frac{1}{p} \ln\biggl| 
 \frac{q \kp }{ (p-q)(p-\kp) } \biggr|
  \;, \\
 (\mbox{e}) =  ({\mbox{f}}): & & \frac{1}{p} \ln\biggl| 
 \frac{ \kp \km }{ (p-\kp)(p-\km)} \biggr|
  \;, \\
 (\underline{\mbox{e}}) =  (\underline{\mbox{f}}): & & \frac{1}{p} \ln\biggl| 
 \frac{ 4 p q }{ \lambda^2 } \biggr|
  \;, \la{Ih_range_f} \\
 ({\mbox{g}}) =  (\underline{\mbox{g}}): & & \frac{1}{p} \ln\biggl| 
 \frac{q \km }{(p-q)(p-\km) } \biggr|
  \;, \\
 ({\mbox{i}}) =  (\underline{\mbox{k}}): & & \frac{1}{p} \ln\biggl| 
 \frac{\km(p-q) }{ q (p-\km)} \biggr|
  \;, \\
 (\underline{\mbox{i}}) =  ({\mbox{k}}): & & \frac{1}{p} \ln\biggl| 
 \frac{ 4 p (p-q)(p-\kp) }{ \lambda^2 \kp } \biggr|
  \;.  \la{Ih_range_uk}
\ea 

As far as the virtual corrections go, they include a divergent part: 
\ba
  \rho^\rmi{ }_{\mathcal{I}^{ }_\rmii{h}} 
  & \ni & 
  \int_{\vec{p}} 
  \frac{\pi \mathcal{K}^2  
  \delta\bigl( \ko - \epsilon^{ }_p - \epsilon^{ }_{pk} \bigr) 
  n^{ }_{\sigma_4} (\ko - p) n^{ }_{\sigma_1}(p) } 
  { 4  \epsilon^{ }_p \epsilon^{ }_{pk}\,  n_{\sigma_0}^{ }(\ko) }
  \int_{\vec{q}} \mathbbm{P} \biggl\{ 
  \frac{1}{2 \epsilon^{ }_q E^{ }_{qp}}
  \; \frac{\epsilon^{ }_q + E^{ }_{qp}}
  {\epsilon_p^2 - (\epsilon^{ }_q + E^{ }_{qp})^2} \biggr\}
 \nn & = & 
  \rho^{ }_{\mathcal{J}^{ }_\rmii{b}} 
  \times \re 
  \int_Q \frac{1}{Q^2[(Q-P)^2+\lambda^2]}
  \biggr|_{p_n = - i \epsilon^{ }_p}
  \;. \la{Ih_fz_vac_1}
\ea
Here we made use of the fact that the vacuum integral is 
independent of $p$: 
\be
 \re \int_Q \frac{1}{Q^2[(Q-P)^2+\lambda^2]}
  \biggr|_{p_n = - i \epsilon^{ }_p}
 = 
 \frac{1 \mufac}{(4\pi)^2}
 \biggl( 
  \frac{1}{\epsilon} + \ln\frac{\bmu^2}{\lambda^2} + 1 
 \biggr)  
 \;. \la{Ih_fz_vac_2}
\ee
It is helpful, however, not to separate the vacuum integral from 
the outset, but rather to treat the structures 
$
 \fr12 +  n^{ }_{\sigma_2}(\epsilon^{ }_q)  
$
and 
$
 \fr12 +  n^{ }_{\sigma_5}(E^{ }_{qp})
$
identifiable on the last two rows of \eq\nr{Ih_full}
as single entities for as long as possible. 
They can then be combined with the 
real corrections, cancelling all $\lambda$-dependence, 
which appears at moderate values of $|p|,|q| \lsim \ko$.
Only the large-$q$ range requires a more careful 
treatment, and we return to this presently. 

In order to implement this strategy, 
we first carry out angular integrals 
and substitute variables, obtaining
[the virtual correction part 
is denoted by $\rho^{(\rmi{v})}_{\mathcal{I}^{ }_\rmii{h}}$, 
and $\simeq$ is a reminder of the divergences appearing
at large $q$ and of the omission of terms of $\rmO(\epsilon)$]
\ba
 \rho^{(\rmi{v})}_{\mathcal{I}^{ }_\rmii{h}} & \simeq & 
 -\frac{\pi\mathcal{K}^2 }{(4\pi)^4 k}
 \int_{\km}^{\kp} \! {\rm d}p \, 
  \frac{ n^{ }_{\sigma_4}(\ko -p) n^{ }_{\sigma_1}(p)}
       { n^{ }_{\sigma_0}(\ko) }
 \nn 
 & & \; \times \, \biggl\{
 \int_{-\infty}^{\infty} \! \frac{ {\rm d}q }{p} \, 
 \Bigl[ \fr12 + n^{ }_{\sigma_2}(q)\Bigr]
 \ln\biggl| \frac{\lambda^2 + 4 p q}{\lambda^2} \biggr|
 \nn & & \; + \, 
 \biggl[\int_{-\infty}^{p-\lambda} + \int_{p+\lambda}^{\infty} \biggr]
 \! \frac{ {\rm d} q }{p} 
 \, \Bigl| \fr12 + n^{ }_{\sigma_5}(q-p)\Bigr|
 \ln \biggl| \frac{(\sqrt{(p-q)^2 - \lambda^2} + p)^2 - q^2}
 {(\sqrt{(p-q)^2 - \lambda^2} - p)^2 - q^2} \biggr|
 \biggr\}
 \;. \la{Ih_fz} \hspace*{5mm}
\ea
The first ``weight function''  
$\fr12 + n^{ }_{\sigma_2}(q)$ has a potential singularity
at $q = 0$ (if $\sigma_2 = +1$), 
the latter at $q = p$ (if $\sigma_5 = +1$); however, noticing that 
the combination in \eq\nr{Ih_factor} can be re-expressed as
\be
 n^{ }_{\sigma_4}(\ko -p) n^{ }_{\sigma_2}(q) n^{ }_{\sigma_5}(p-q)
 = 
 n^{ }_{\sigma_4}(\ko -p) n^{ }_{\sigma_1}(p)
 \Bigl[
 \fr12 + n^{ }_{\sigma_2}(q) - \fr12 - n^{ }_{\sigma_5}(q-p) 
 \Bigr]
 \;,  \la{w1w3}
\ee
these terms cancel exactly against the corresponding real corrections
within the domains ($\underline{\mbox{e}}$) and ($\underline{\mbox{f}}$)
as well as (a) and (l), respectively, which are adjacent to the singular
lines. An approximate form of the cancellation can be seen be 
rewriting \eq\nr{Ih_fz} in the limit $\lambda\to 0$ away from the 
singular lines: 
\ba
 \rho^{(\rmi{v})}_{\mathcal{I}^{ }_\rmii{h}} & \approx & 
 -\frac{\pi\mathcal{K}^2 }{(4\pi)^4 k}
 \int_{\km}^{\kp} \! {\rm d}p \, 
 \frac{ n^{ }_{\sigma_4}(\ko -p) n^{ }_{\sigma_1}(p) }
 { n^{ }_{\sigma_0}(\ko) }
 \nn 
 & & \; \times \,
 \int_{-\infty}^{\infty} \! \frac{ {\rm d}q }{p} \, 
 \biggl\{
 \Bigl[ \fr12 + n^{ }_{\sigma_2}(q)\Bigr]
 \ln\biggl| \frac{4 p q}{\lambda^2} \biggr|
  +  
 \, \Bigl[ \fr12 + n^{ }_{\sigma_5}(q-p)\Bigr]
 \ln \biggl| \frac{\lambda^2 q}
 {4 p (p-q)^2} \biggr|
 \biggr\}
 \;. \la{Ih_fz_2} \hspace*{5mm}
\ea
Summing this together with \eqs\nr{Ih_range_a}--\nr{Ih_range_uk} 
all $\lambda$'s cancel, and the result is integrable in the 
small-$q$ domain.  

It remains to deal with the ultraviolet divergence from the 
large-$q$ domain. 
The idea is to insert 
\be 
  0 = \int_{| \vec{q} | > \Lambda } \frac{1}{4 q^3} - 
      \int_{| \vec{q} | > \Lambda } \frac{1}{4 q^3}
\ee
inside the integrand representing virtual corrections. 
The individual terms have the same 
divergence as \eq\nr{Ih_fz_vac_2}: 
\be
 \int_{| \vec{q} | > \Lambda } \frac{1}{4 q^3} 
 = \frac{1 \mufac}{(4\pi)^2}
 \biggl( \frac{1}{\epsilon} + \ln\frac{\bmu^2}{4\Lambda^2} + 2 \biggr)
 \;. \la{asy_UV}
\ee
Separating the $1/\epsilon$-part hereof, 
together with finite terms chosen according
to the vacuum result (cf.\ \eqs(B.17-19) of ref.~\cite{nonrel}), and 
recalling the contribution from the $\rmO(\epsilon)$-term 
in \eq\nr{rho_Jb_final}, we obtain
\ba
 \rho^{(\rmi{v})}_{\mathcal{I}^{ }_\rmii{h}} & = & 
 -\frac{\pi\mathcal{K}^2 }{(4\pi)^4 k}
 \int_{\km}^{\kp} \! {\rm d}p \, 
 \frac{ n^{ }_{\sigma_4}(\ko -p) n^{ }_{\sigma_1}(p) }{
 n^{ }_{\sigma_0}(\ko) }
 \biggl( \frac{1}{\epsilon} + 2 \ln\frac{\bmu^2}{\mathcal{K}^2} + 5 \biggr)
 \nn 
 & &  + \, 
 \frac{\pi\mathcal{K}^2 }{(4\pi)^4 k}
 \int_{\km}^{\kp} \! {\rm d}p \,
 \frac{ n^{ }_{\sigma_4}(\ko -p) n^{ }_{\sigma_1}(p) }{
 n^{ }_{\sigma_0}(\ko) }
 \nn 
 & & \; \times \, 
 \biggl\{
   \ln \frac{4 (\kp - p)(p - \km) \Lambda^2 }{k^2 \mathcal{K}^2} + 3
 + \biggl( \int^{-\infty}_{-\Lambda} + \int_{\Lambda}^\infty 
 \biggr)\! \frac{ {\rm d}q }{q} 
 - (4\pi)^2 \int_{\vec{q}} [...] 
 \biggr\} 
 \;, \hspace*{6mm} \la{Ih_fz_3}
\ea
where $[...]$ refers to the original integrand from 
\eq\nr{Ih_full}. In addition it must be 
realized that going over to the shifted variables of \eq\nr{Ih_fz}
has introduced an ``error'' (because we have carelessly handled 
logarithmically divergent integrals) which must now be 
compensated for. Indeed, \eq\nr{asy_UV} and an infrared (IR) part
from $|q| < \Lambda$ only 
add up to the correct \eq\nr{Ih_fz_vac_2} if the IR part yields
\be
 \frac{1}{(4\pi)^2} \biggl( \frac{4 \Lambda^2}{\lambda^2} - 1 \biggr)
 \;. \la{Ihvac_should}
\ee
Yet the corresponding contribution from
the vacuum parts of \eq\nr{Ih_fz_2} reads
\be
 \frac{1}{(4\pi)^2}
 \int_{-\Lambda}^{\Lambda} \! \frac{{\rm d}q}{2 p}
 \biggl\{ {\rm sign}(q) \ln \biggl|\frac{4 p q}{\lambda^2}\biggr|
 + {\rm sign}(q-p) \ln \biggl| 
 \frac{ \lambda^2 q }{4 p (p-q)^2} \biggr| \biggr\}
 = 
 \frac{1}{(4\pi)^2} \biggl( \frac{4 \Lambda^2}{\lambda^2} + 1 \biggr)
 \;. \la{Ihvac_is}
\ee
The difference of \eqs\nr{Ihvac_should} 
and \nr{Ihvac_is} needs to be cancelled from 
the integrand of \eq\nr{Ih_fz_3} if we use the shifted
variables. Thereby the final expression reads
\ba
 \rho^{(\rmi{v})}_{\mathcal{I}^{ }_\rmii{h}} \!\! & = & \!\! 
  \rho^{\rmi{vac}}_{\mathcal{I}^{ }_\rmii{h}} 
 \, + \, 
 \frac{\pi\mathcal{K}^2 }{(4\pi)^4 k}
 \int_{\km}^{\kp} \! {\rm d}p \, 
 \frac{ n^{ }_{\sigma_4}(\ko -p) n^{ }_{\sigma_1}(p) }{
 n^{ }_{\sigma_0}(\ko) }
 \nn 
 & & \; \times \, 
 \biggl\{
   \ln \frac{4 (\kp - p)(p - \km) \Lambda^2 }{k^2 \mathcal{K}^2} + 5
 + \int_{-\infty}^{\infty} \! \frac{ {\rm d}q }{p} \, 
 \biggl[ \frac{p\, \theta(|q|-\Lambda)}{|q|}
 \nn & & \qquad + \, 
 \Bigl( \fr12 + n^{ }_{\sigma_2}(q)\Bigr)
 \ln\biggl| \frac{\lambda^2}{4 p q} \biggr|
   -   
 \, \Bigl( \fr12 + n^{ }_{\sigma_5}(q-p)\Bigr)
 \ln \biggl| \frac{\lambda^2 q}
 {4 p (p-q)^2} \biggr|
 \, \biggr]  \biggr\} 
 \;, \hspace*{6mm} \la{Ih_fz_final}
\ea
where a vacuum part has been defined as 
\be
  \rho^{\rmi{vac}}_{\mathcal{I}^{ }_\rmii{h}} 
 \equiv
 -\frac{\pi\mathcal{K}^2 T}{(4\pi)^4 k}
    \ln \biggl( 
     \frac{e^{{\kp} / {T}}+\sigma_0 e^{-\kp/T}-\sigma_1 - \sigma_4}
     {e^{{\km} / {T}}+\sigma_0 e^{-\km/T}-\sigma_1 - \sigma_4}
   \biggr)
 \biggl( \frac{1}{\epsilon} + 2 \ln\frac{\bmu^2}{\mathcal{K}^2} + 5 \biggr)
 \;. \la{Ih_vac}
\ee
There is no dependence on $\Lambda$ in \eq\nr{Ih_fz_final}, 
and the $1/q$-tails at $|q| > \Lambda$ 
cancel as well so that, when combined with 
the real corrections, the expression is integrable. 

For $\kp, \km \gg \pi T$, the ultraviolet asymptotics of 
$\rho^{ }_{\mathcal{I}_\rmi{h}}$ reads~\cite{nonrel}
\ba
 \rho^{ }_{\mathcal{I}_\rmi{h}} &  \stackrel{\kp,\km \gg \pi T}{\approx}  & 
  -\frac{\mufac\mathcal{K}^2}{4(4\pi)^3} \biggl( 
   \frac{1}{\epsilon}+ 2\ln\frac{\bmu^2}{\mathcal{K}^2} + 5 \biggr)
 \nn & & 
 \; + \, 
  \int_\vec{p} \biggl\{
 \frac{n^{ }_{\sigma_4} - ( n^{ }_{\sigma_2} + n^{ }_{\sigma_5})}{16\pi p}
 +
 \frac{p[3 n^{ }_{\sigma_4} - ( n^{ }_{\sigma_2} + n^{ }_{\sigma_5})]}{12\pi}
 \frac{\ko^2 + \fr{k^2}3}{\mathcal{K}^4}
 \biggr\}
 \;. \hspace*{5mm} \la{asy_Ih}
\ea
The specific statistics needed in the present case are
\ba
 \mathcal{I}^{ }_\rmi{h} & \Leftrightarrow &  
 (\sigma_1\sigma_4\sigma_5 | \sigma_2\sigma_0) = (+-+|+-)
 \;, \la{Ih_mpp}\\
 \widetilde{\mathcal{I}}^{ }_\rmi{h} & \Leftrightarrow &  
 (\sigma_1\sigma_4\sigma_5 | \sigma_2\sigma_0) = (+--|--)
 \;, \la{Ih_mpm} \\ 
 \widehat{\mathcal{I}}^{ }_\rmi{h} & \Leftrightarrow & 
 (\sigma_1\sigma_4\sigma_5 | \sigma_2\sigma_0) = (-+-|+-)
 \;, \la{Ih_pmm}
\ea
where only the first three indices are independent. 
For numerical evaluation, we have reflected the final 
integral to the domain defined in \fig{6} of ref.~\cite{master}. 
The corresponding integrand is not shown explicitly, since no 
substantial cancellations take place in the reflection. 
Results of numerical evaluations
(after subtracting the vacuum part, cf.\ 
\eq\nr{Ih_vac}) are shown in \fig\ref{fig:Ih},
and are seen to agree with the OPE-asymptotics from 
\eq\nr{asy_Ih} as well as with the $k\to 0$ limit
from ref.~\cite{bulk_wdep}.

\begin{figure}[t]


\centerline{%
 \epsfysize=7.0cm\epsfbox{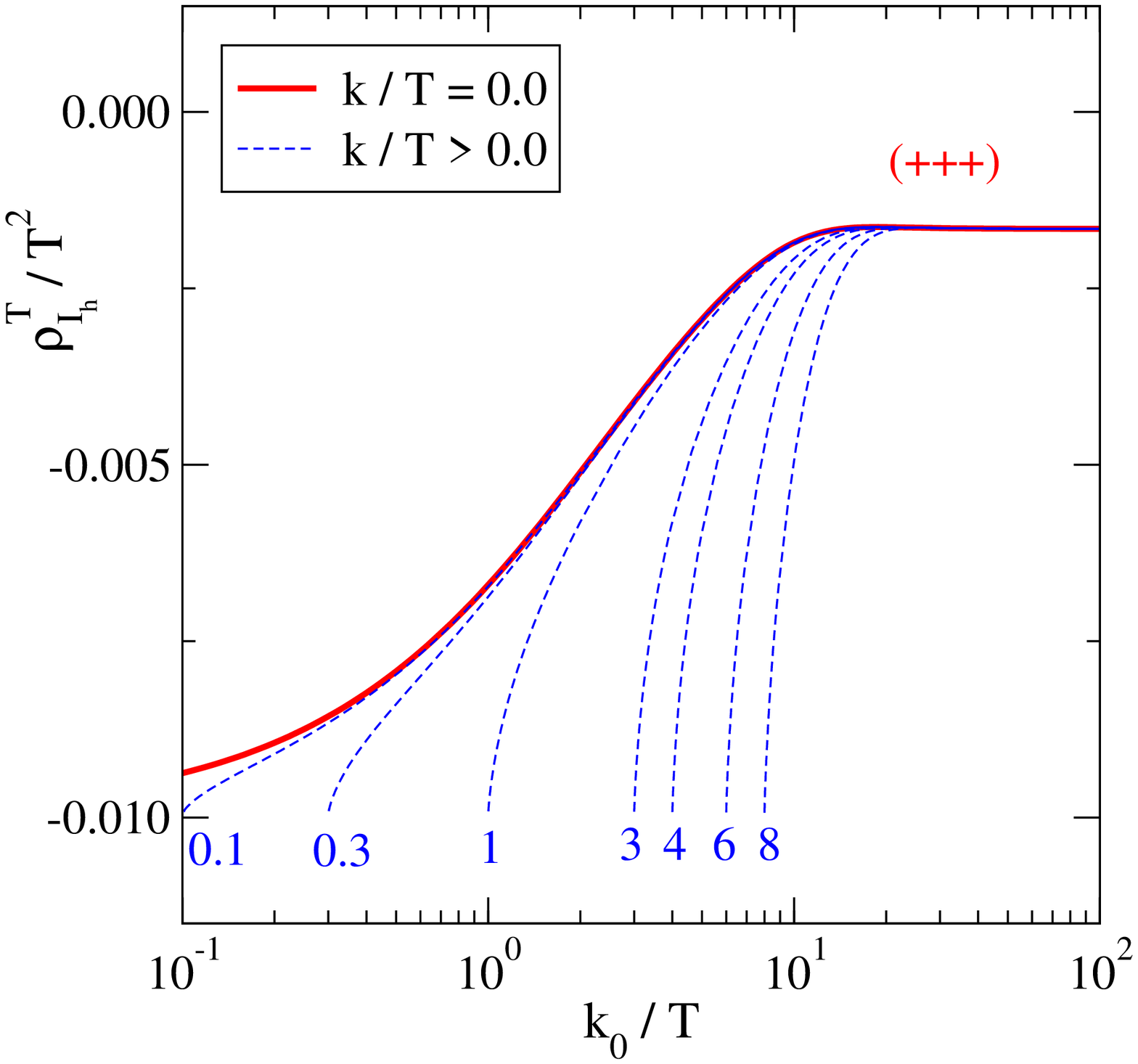}%
~~~\epsfysize=7.6cm\epsfbox{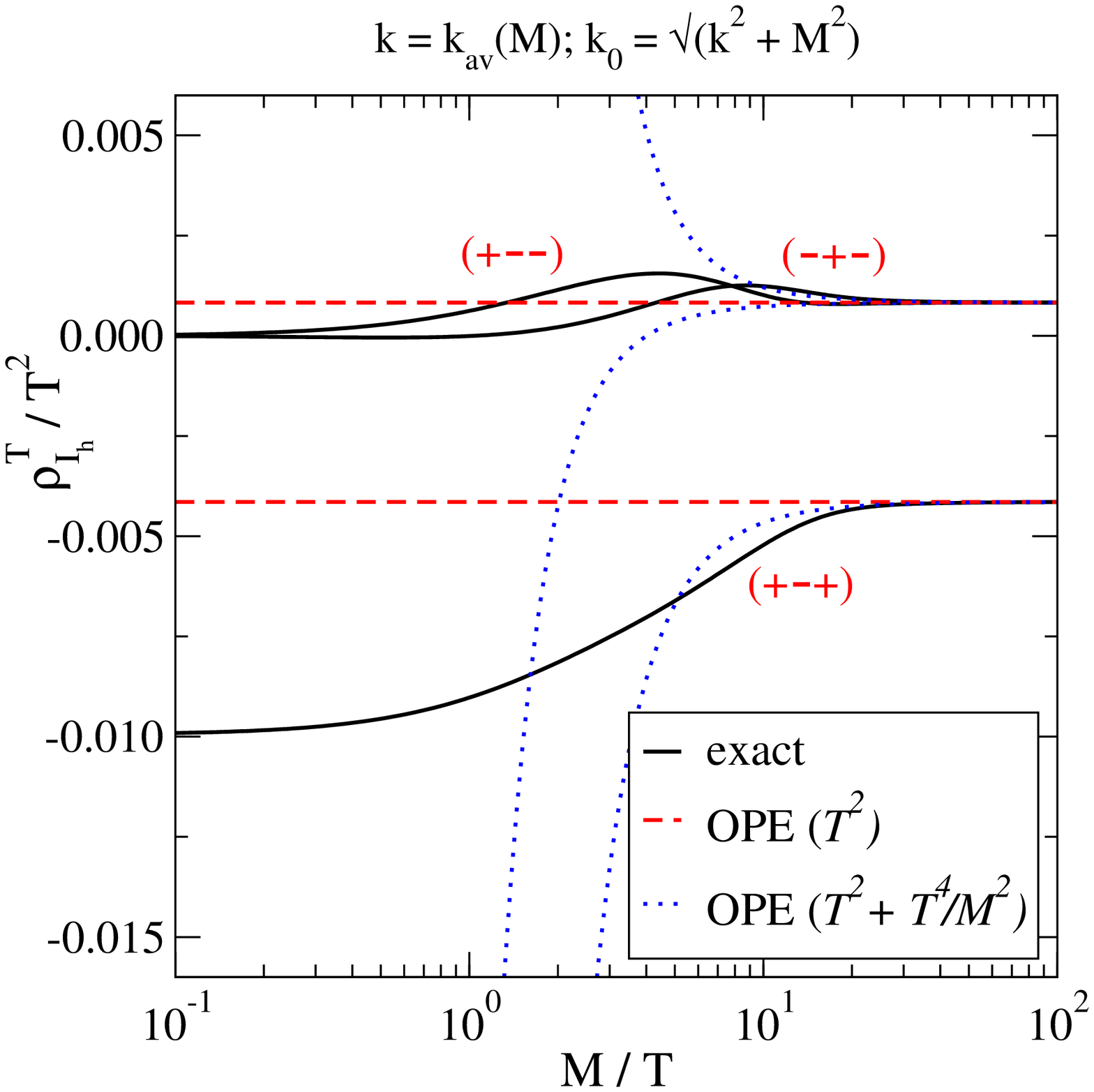}
}

\caption[a]{\small
Left: The thermal part of 
$\rho^{ }_{\mathcal{I}^{ }_\rmii{h}}$ 
with the purely bosonic statistics $(\sigma_1\sigma_4\sigma_5) = (+++)$, 
for $\ko \ge k + 0.001 T$, 
compared with the zero-momentum limit determined in ref.~\cite{bulk_wdep}.
Right: The thermal part of 
$\rho^{ }_{\mathcal{I}^{ }_\rmii{h}}$ with the momentum of 
\eq\nr{kav} and statistics of 
\eqs\nr{Ih_mpp}--\nr{Ih_pmm} as a function of $M/T$, 
compared with the OPE-asymptotics from \eq\nr{asy_Ih}. 
}

\la{fig:Ih}
\end{figure}

%
\subsection{$\rho^{ }_{{\mathcal{I}}^{ }_\rmii{h'}}$}

After carrying out 
the Matsubara sums for 
$
 {\mathcal{I}}^{ }_\rmii{h'}
$, we get 
\ba
  {\mathcal{I}}^{ }_\rmii{h'} & = &  \lim_{\lambda\to 0}
  \int_{\vec{p},\vec{q}} \frac{2} 
  { 8 \epsilon^{ }_q \epsilon^{ }_{pk} E^{ }_{qp} }
  \biggl\{ 
  \nn &&
  \frac{-i k_n \epsilon^{ }_q + \vec{k}\cdot\vec{q} }
       {-i k_n  + \epsilon^{ }_{pk} + \epsilon^{ }_q + E^{ }_{qp}}
  \; \frac{
   \bigl[1  + n^{ }_{\sigma_4}(\epsilon^{ }_{pk})
   + n^{ }_{\sigma_2}(\epsilon^{ }_q)
   \bigr] 
   \bigl[1 + n^{ }_{\sigma_5}(E^{ }_{qp}) \bigr] 
   + n^{ }_{\sigma_4}(\epsilon^{ }_{pk}) n^{ }_{\sigma_2}(\epsilon^{ }_q) }
  {\epsilon_p^2 - (\epsilon^{ }_q + E^{ }_{qp})^2}
  \nn & + &
  \frac{-i k_n \epsilon^{ }_q + \vec{k}\cdot\vec{q}}
    {-i k_n  - \epsilon^{ }_{pk} + \epsilon^{ }_q + E^{ }_{qp}}
  \; \frac{
  n^{ }_{\sigma_4}(\epsilon^{ }_{pk})
  \bigl[1 + n^{ }_{\sigma_2}(\epsilon^{ }_q)  +
     n^{ }_{\sigma_5}(E^{ }_{qp}) \bigr] 
  -  n^{ }_{\sigma_2}(\epsilon^{ }_q)  n^{ }_{\sigma_5}(E^{ }_{qp}) }
  { \epsilon_p^2 - (\epsilon^{ }_q + E^{ }_{qp})^2 }
  \nn & + &
  \frac{+i k_n \epsilon^{ }_q + \vec{k}\cdot\vec{q}} 
   {-i k_n  + \epsilon^{ }_{pk} - \epsilon^{ }_q + E^{ }_{qp}}
  \; \frac{
   n^{ }_{\sigma_2}(\epsilon^{ }_q) 
  \bigl[1 + n^{ }_{\sigma_4}(\epsilon^{ }_{pk})
   + n^{ }_{\sigma_5}(E^{ }_{qp})
   \bigr] 
  -  n^{ }_{\sigma_4}(\epsilon^{ }_{pk}) n^{ }_{\sigma_5}(E^{ }_{qp}) }
  {\epsilon_p^2 - (\epsilon^{ }_q - E^{ }_{qp})^2}
  \nn & + &
  \frac{-i k_n \epsilon^{ }_q + \vec{k}\cdot\vec{q}}
   {-i k_n  + \epsilon^{ }_{pk} + \epsilon^{ }_q - E^{ }_{qp}}
  \; \frac{
   n^{ }_{\sigma_5}(E^{ }_{qp})
  \bigl[1 + n^{ }_{\sigma_4}(\epsilon^{ }_{pk})
  + n^{ }_{\sigma_2}(\epsilon^{ }_q) 
  \bigr] 
  - n^{ }_{\sigma_4}(\epsilon^{ }_{pk}) n^{ }_{\sigma_2}(\epsilon^{ }_q)  }
  {\epsilon_p^2 - (\epsilon^{ }_q - E^{ }_{qp})^2}
  \; \biggr\} 
 \nn & + &  \lim_{\lambda\to 0}
  \int_{\vec{p}} \frac{2} 
  { 4 \epsilon^{ }_p \epsilon^{ }_{pk} }
  \biggl[ \frac{ 
    1  +  n^{ }_{\sigma_1}(\epsilon^{ }_{p})
     + n^{ }_{\sigma_4}(\epsilon^{ }_{pk})   
    }{-i k_n + \epsilon^{ }_p + \epsilon^{ }_{pk}}
  + 
    \frac{ 
     n^{ }_{\sigma_1}(\epsilon^{ }_{p})  - 
      n^{ }_{\sigma_4}(\epsilon^{ }_{pk})
    }{i k_n - \epsilon^{ }_p + \epsilon^{ }_{pk}}
  \biggr]
  \; 
  \nn & & \hspace*{1.3cm} \times \int_{\vec{q}} \biggl[ 
  \frac{1 + n^{ }_{\sigma_2}(\epsilon^{ }_q)  +
     n^{ }_{\sigma_5}(E^{ }_{qp})}{4 \epsilon^{ }_q E^{ }_{qp} }
  \biggl(
  \frac{ i k_n \epsilon^{ }_q + \vec{k}\cdot\vec{q}}
   {\epsilon^{ }_p + \epsilon^{ }_q + E^{ }_{qp}}
  -\, \frac{- i k_n \epsilon^{ }_q + \vec{k}\cdot\vec{q}}
   {\epsilon^{ }_p - \epsilon^{ }_q - E^{ }_{qp}}
 \biggl)
  \nn[2mm] & & \hspace*{2.3cm} 
  +\, \frac{
   n^{ }_{\sigma_2}(\epsilon^{ }_q)  -
     n^{ }_{\sigma_5}(E^{ }_{qp})}{4 \epsilon^{ }_q E^{ }_{qp} }  
  \biggl(
  \frac{ - i k_n \epsilon^{ }_q + \vec{k}\cdot\vec{q} }
    {\epsilon^{ }_p - \epsilon^{ }_q + E^{ }_{qp}}
  -\, \frac{ i k_n \epsilon^{ }_q + \vec{k}\cdot\vec{q} }
   {\epsilon^{ }_p + \epsilon^{ }_q - E^{ }_{qp}}
  \biggr)
  \; \biggr]  
  \nn 
 & + & (ik_n \to -i k_n) 
 \;. \la{Ihp_full}
\ea
Taking the cut like in \eq\nr{cut2}, 
the first four structures are real corrections, the last two 
are virtual corrections. The corresponding scattering processes
can be depicted like in \fig{3} of ref.~\cite{master}, with internal 
propagators carrying the index $\sigma_3$ shrunk to points. 

For the real corrections, \eq(4.28) of ref.~\cite{master} now reads
\ba
 && \hspace*{-1.5cm} 
 \Bigl\langle 
   \Phi^{ }_\rmi{r1}(\ko - p_0 | q | p_0 - q | \cdot)
 \Bigr\rangle
  =  - 
 \Bigl\langle 
   \Phi^{ }_\rmi{r2}(p_0 - \ko  | q | p_0 - q | \cdot)
 \Bigr\rangle
 \nn 
 & = &  - 
 \Bigl\langle 
   \Phi^{ }_\rmi{r3}(\ko - p_0 | -q | p_0 - q | \cdot)
 \Bigr\rangle
 \nn 
 & = &  - 
 \Bigl\langle 
   \Phi^{ }_\rmi{r4}(\ko - p_0 | q | q - p_0 | \cdot)
 \Bigr\rangle
 \nn[2mm]
 & = &    
 \frac{
    n^{ }_{\sigma_4} (\ko - p_0) 
 \, n^{ }_{\sigma_2} (q) 
 \, n^{ }_{\sigma_5} (p_0 - q) }
 {  n^{ }_{\sigma_0}(\ko) }
 \, \times \, \mathbbm{P}\, 
 \, \biggl\langle
   \frac{\ko q - \vec{k}\cdot\vec{q}}{p_0^2 - p^2 }
 \biggr\rangle^{ }_{(\ko - p_0 | q | p_0 - q | \cdot)} 
 \;, \hspace*{1cm} 
\ea
where the arguments ${(...|...|...|\cdot)}$  on the last line 
refer to $\epsilon^{ }_{pk}$, $\epsilon^{ }_q$, and $E^{ }_{qp}$, 
respectively. 
The azimuthal average, as defined in ref.~\cite{master}, yields
\be
 \bigl\langle \vec{k}\cdot\vec{q} \bigr\rangle^{ }_
  {(\epsilon^{ }_{pk} | q | E^{ }_{qp})} = 
  \frac{\bigl( p^2 + k^2 - \epsilon_{pk}^2 \bigr)
  \bigl( p^2 + q^2 + \lambda^2 - E_{qp}^2 \bigr)}{4p^2}
  \;.  
\ee
The subsequent $p$-integral leads to logarithms and fractions.  Renaming
finally $p_0\to p$ and factoring out
\be
 \frac{\pi
 \, n^{ }_{\sigma_4} (\ko - p) 
 \, n^{ }_{\sigma_2} (q) 
 \, n^{ }_{\sigma_5} (p - q)
 }{(4\pi)^4 k \,  n^{ }_{\sigma_0}(\ko)  }
 \;, 
\ee 
the $\lambda\to 0$ limits (in the sense specified
in ref.~\cite{master}) read: 
\ba
 & & \hspace*{-5mm}  (\mbox{a}) =  ({\mbox{l}}): \;\;\quad 
 2(p-q)\biggl[ 1 + \frac{(p-2\km)(p-2\kp)}{p^2} \biggr]
 + \frac{q \mathcal{K}^2}{p^2} \ln\biggl| 
 \frac{ 4 p (p-q)^2 }{ \lambda^2 q} \biggr|
  \;, \la{Ihp_range_a} \\
 & & \hspace*{-5mm} (\mbox{b}): \;\;\quad \frac{4 \km (\kp - q)}{p} 
 + \frac{q \mathcal{K}^2}{p^2} \ln\biggl| 
 \frac{ \kp (p-q)}{ q(p-\kp) } \biggr|
  \;, \\
 & & \hspace*{-5mm} (\underline{\mbox{b}}): \;\;\quad
 2(p-\km)\biggl[ 1 + \frac{(p-2q)(p-2\kp)}{p^2} \biggr]
 +  \frac{q \mathcal{K}^2}{p^2} \ln\biggl| 
 \frac{4 p (p-q)(p-\km) }{\lambda^2 \km } \biggr|
  \;, \\
 & &  \hspace*{-5mm} 
 (\mbox{c})  = - ({\mbox{h}}) = -  (\underline{\mbox{h}})
 = - ({\mbox{j}}): \;\;\quad
 \frac{4 q (\kp - \km)}{p}
 +  \frac{q \mathcal{K}^2}{p^2} \ln\biggl| 
 \frac{ \kp (p-\km) }{\km(p-\kp) } \biggr|
  \;, \\
 & & \hspace*{-5mm} (\tilde{\mbox{c}}): \;\;\quad
 2\km \biggl[ 1 - \frac{(p-2q)(p-2\kp)}{p^2} \biggr] 
  + \frac{q \mathcal{K}^2}{p^2} \ln\biggl| 
 \frac{ 4 \km p (p-q)}{ \lambda^2( p -\km)} \biggr|
  \;, \\
  & & \hspace*{-5mm} (\mbox{d}): \;\;\quad
 \frac{4(p - \km)(\kp - p + q)}{p}
  + \frac{q \mathcal{K}^2}{p^2}
 \ln\biggl| 
 \frac{q \kp }{ (p-q)(p-\kp) } \biggr|
  \;, \\
 & & \hspace*{-5mm} (\mbox{e}) = ({\mbox{f}}): \;\;\quad
 \frac{4(p-q)(\kp+\km-p)}{p}
  + \frac{q \mathcal{K}^2}{p^2} 
 \ln\biggl| 
 \frac{ \kp \km }{ (p-\kp)(p-\km)} \biggr|
  \;, \\
 & & \hspace*{-5mm} (\underline{\mbox{e}}) = 
 (\underline{\mbox{f}}): \;\;\quad
 2 q \biggl[ 1 - \frac{(p-2\km)(p-2\kp)}{p^2} \biggr]
  + \frac{q \mathcal{K}^2}{p^2} 
 \ln\biggl| 
 \frac{ 4 p q }{ \lambda^2 } \biggr|
  \;,  \la{range_e} \\
 & & \hspace*{-5mm} ({\mbox{g}}) =  (\underline{\mbox{g}}): \;\;\quad
 \frac{4 (p-\kp)(\km - p + q)}{p}
  + \frac{q \mathcal{K}^2}{p^2} 
  \ln\biggl| 
 \frac{q \km }{(p-q)(p-\km) } \biggr|
  \;, \\
 & & \hspace*{-5mm} ({\mbox{i}}) =  (\underline{\mbox{k}}): \;\;\quad
 \frac{4\kp(\km - q)}{p}
  + \frac{q \mathcal{K}^2}{p^2} 
 \ln\biggl| 
 \frac{\km(p-q) }{ q (p-\km)} \biggr|
  \;, \\
 & & \hspace*{-5mm} (\underline{\mbox{i}}) =  ({\mbox{k}}): \;\;\quad
  2 (p-\kp) \biggl[ 1 + \frac{(p-2q)(p-2\km)}{p^2} \biggr]
  + \frac{q \mathcal{K}^2}{p^2} 
 \ln\biggl| 
 \frac{ 4 p (p-q)(p-\kp) }{ \lambda^2 \kp } \biggr|
  \;. \hspace*{10mm}  \la{Ihp_range_uk}
\ea 

As far as the virtual corrections go, they include a divergent part: 
\ba
  \rho^\rmi{ }_{\mathcal{I}^{ }_\rmii{h'}} 
  & \ni & 
  \int_{\vec{p}} 
  \frac{2 \pi  
  \delta\bigl( \ko - \epsilon^{ }_p - \epsilon^{ }_{pk} \bigr)
   n^{ }_{\sigma_4} (\ko - p) n^{ }_{\sigma_1}(p) 
  } 
  { 4  \epsilon^{ }_p \epsilon^{ }_{pk}\, n_{\sigma_0}^{ }(\ko) }
 \nn & & \; \times \, 
  \int_{\vec{q}} \mathbbm{P} \biggl\{ 
  \frac{1}{4 \epsilon^{ }_q E^{ }_{qp}}
  \; \biggl( 
   \frac{\ko \epsilon^{ }_q + \vec{k}\cdot\vec{q}}
        {\epsilon^{ }_p + \epsilon^{ }_q + E^{ }_{qp} } 
  + 
   \frac{- \ko \epsilon^{ }_q + \vec{k}\cdot\vec{q}}
        {- \epsilon^{ }_p + \epsilon^{ }_q + E^{ }_{qp} } 
 \biggr) \biggr\} 
 \nn & = & - 
  \frac{ 2 \rho^{ }_{\mathcal{J}^{ }_\rmii{b}}}{\mathcal{K}^2} 
  \times \re
  \int_Q \frac{K\cdot Q}{Q^2[(Q-P)^2+\lambda^2]}
  \biggr|_{k_n = - i \ko,\; p_n = - i \epsilon^{ }_p,\; 
           \epsilon^{ }_{pk} = \ko - \epsilon^{ }_p  }
  \;. \la{Ihp_fz_vac_1}
\ea
Here we made use of the fact that the vacuum integral is 
independent of $p$: 
\be
  \re \int_Q \frac{K\cdot Q}{Q^2[(Q-P)^2+\lambda^2]}
  \biggr|_{k_n = - i \ko,\; p_n = - i \epsilon^{ }_p,\; 
           \epsilon^{ }_{pk} = \ko - \epsilon^{ }_p } \!\!\!
 = 
 - \frac{\mathcal{K}^2 \mufac}{4 (4\pi)^2}
 \biggl( 
  \frac{1}{\epsilon} + \ln\frac{\bmu^2}{\lambda^2} + \fr12  
 \biggr)  
 \;. \la{Ihp_fz_vac_2}
\ee
Like with $\rho^{ }_{\mathcal{I}^{ }_\rmii{h}}$
it is helpful, however, not to separate the vacuum integral from 
the outset, but rather to treat the structures 
$
 \fr12 +  n^{ }_{\sigma_2}(\epsilon^{ }_q)  
$
and 
$
 \fr12 +  n^{ }_{\sigma_5}(E^{ }_{qp})
$
identifiable on the last two rows of \eq\nr{Ihp_full} 
as single entities for as long as possible. 

In order to implement this, we first carry out angular integrals 
and substitute integration variables, obtaining
[the virtual correction part 
is denoted by $\rho^{(\rmi{v})}_{\mathcal{I}^{ }_\rmii{h'}}$, 
and $\simeq$ is a reminder of the divergences appearing
at large $q$ and of the omission of terms of $\rmO(\epsilon)$]
\ba
 \rho^{(\rmi{v})}_{\mathcal{I}^{ }_\rmii{h'}} & \simeq & 
 \frac{\pi }{(4\pi)^4 k}
 \int_{\km}^{\kp} \! {\rm d}p \,
 \frac{ n^{ }_{\sigma_4}(\ko -p) n^{ }_{\sigma_1}(p) }{
 n^{ }_{\sigma_0}(\ko) }
 \nn 
 & & \; \times \, \biggl\{
 \int_{-\infty}^{\infty} \! {\rm d}q \, 
 \Bigl[ \fr12 + n^{ }_{\sigma_2}(q)\Bigr]
 \biggl[
 \frac{2 q (\mathcal{K}^2 - 2 p \ko)}{p^2} 
 + \frac{2 q p \mathcal{K}^2 + \lambda^2 (\mathcal{K}^2 - 2 p \ko)}{2 p^3}
 \ln\biggl| \frac{\lambda^2}{\lambda^2 + 4 p q} \biggr|
 \biggr]
 \nn & & \; - \, 
 \biggl[\int_{-\infty}^{p-\lambda} + \int_{p+\lambda}^{\infty} \biggr]
 {\rm d} q  
 \, \Bigl| \fr12 + n^{ }_{\sigma_5}(q-p)\Bigr|
 \biggl[
 \frac{2 (\mathcal{K}^2 - 2 p \ko) \sqrt{(q-p)^2-\lambda^2} }{p^2} 
\nn & & \qquad\qquad + \,    
\frac{2 q p \mathcal{K}^2 + \lambda^2 (\mathcal{K}^2 - 2 p \ko)}{2 p^3}
 \ln \biggl| \frac{(\sqrt{(p-q)^2 - \lambda^2} + p)^2 - q^2}
 {(\sqrt{(p-q)^2 - \lambda^2} - p)^2 - q^2} \biggr|
 \biggr]
 \biggr\}
 \;. \la{Ihp_fz} \hspace*{5mm}
\ea
The first ``weight function''  
$\fr12 + n^{ }_{\sigma_2}(q)$ has a potential singularity 
at $q = 0$ (if $\sigma_2 = +1$), 
the latter at $q = p$ (if $\sigma_5 = +1$); 
however, making use of \eq\nr{w1w3},  
it can be seen that these terms cancel 
(apart from a harmless $\sim \sqrt{(q-p)^2-\lambda^2}$ in the latter case)
against the corresponding real corrections
within the domains ($\underline{\mbox{e}}$) and ($\underline{\mbox{f}}$)
as well as (a) and (l), respectively, which are adjacent to the singular
lines. An approximate form of the cancellation can be seen be 
rewriting \eq\nr{Ihp_fz} in the limit $\lambda\to 0$ away from the 
singular lines: 
\ba
 \rho^{(\rmi{v})}_{\mathcal{I}^{ }_\rmii{h'}} & \approx & 
 \frac{\pi }{(4\pi)^4 k}
 \int_{\km}^{\kp} \! {\rm d}p \, 
 \frac{ n^{ }_{\sigma_4}(\ko -p) n^{ }_{\sigma_1}(p) }{
  n^{ }_{\sigma_0}(\ko) }
 \nn 
 & & \; \times \,
 \int_{-\infty}^{\infty} \! {\rm d}q \, 
 \biggl\{
 \Bigl[ \fr12 + n^{ }_{\sigma_2}(q)\Bigr]
 \biggl[
 \frac{2 q (\mathcal{K}^2 - 2 p \ko)}{p^2} 
 + \frac{q  \mathcal{K}^2}{p^2}
 \ln\biggl| \frac{\lambda^2}{4 p q} \biggr|
 \biggr]
 \nn & & \; - \, 
 \, \Bigl[ \fr12 + n^{ }_{\sigma_5}(q-p) \Bigr]
 \biggl[
 \frac{2 (q-p) (\mathcal{K}^2 - 2 p \ko)}{p^2} 
 + 
 \frac{q  \mathcal{K}^2}{p^2}
 \ln \biggl| \frac{\lambda^2 q}
 {4 p (p-q)^2 } \biggr|
 \biggr]
 \biggr\}
 \;. \la{Ihp_fz_2} \hspace*{5mm}
\ea
Summing together with \eqs\nr{Ihp_range_a}--\nr{Ihp_range_uk}, 
all $\lambda$'s cancel, and the remainder is integrable 
in the IR domain $|p|,|q| \lsim \ko$. 

It remains to deal with the ultraviolet divergence. 
We add 
\be 
  0 = -\frac{\mathcal{K}^2}{4} 
      \int_{| \vec{q} | > \Lambda } \frac{1}{4 q^3}
      + \frac{\mathcal{K}^2}{4}
      \int_{| \vec{q} | > \Lambda } \frac{1}{4 q^3}
\ee
in the integrand of the virtual corrections. As seen from 
\eq\nr{asy_UV} the individual terms have the same 
divergence as \eq\nr{Ihp_fz_vac_2}.
Separating the $1/\epsilon$-part hereof, 
together with finite terms chosen according
to the vacuum result (cf.\ \eq(B.20) of ref.~\cite{nonrel}), and 
recalling the contribution from the $\rmO(\epsilon)$-term 
in \eq\nr{rho_Jb_final}, we obtain
\ba
 \rho^{(\rmi{v})}_{\mathcal{I}^{ }_\rmii{h'}} & = & 
 -\frac{\pi\mathcal{K}^2 }{2 (4\pi)^4 k}
 \int_{\km}^{\kp} \! {\rm d}p \,
 \frac{ n^{ }_{\sigma_4}(\ko -p) n^{ }_{\sigma_1}(p) }{
   n^{ }_{\sigma_0}(\ko) }
 \biggl( \frac{1}{\epsilon} + 2 \ln\frac{\bmu^2}{\mathcal{K}^2} + \fr92 \biggr)
 \nn 
 & &  + \, 
 \frac{\pi\mathcal{K}^2 }{2 (4\pi)^4 k}
 \int_{\km}^{\kp} \! {\rm d}p \,
 \frac{ n^{ }_{\sigma_4}(\ko -p) n^{ }_{\sigma_1}(p) }{
  n^{ }_{\sigma_0}(\ko)}
 \nn 
 & & \; \times \, 
 \biggl\{
   \ln \frac{4 (\kp - p)(p - \km) \Lambda^2 }{k^2 \mathcal{K}^2} + \fr52
 + \biggl( \int^{-\infty}_{-\Lambda} + \int_{\Lambda}^\infty 
 \biggr)\! \frac{ {\rm d}q }{q} 
 + \frac{4  (4\pi)^2}{\mathcal{K}^2} \int_{\vec{q}} [...] 
 \biggr\} 
 \;, \nn \la{Ihp_fz_n}
\ea
where $[...]$ refers to original integrand
in \eq\nr{Ihp_full}. In addition it must be 
realized that going over to the shifted variables of \eq\nr{Ihp_fz}
has introduced an ``error'' (because we have carelessly handled 
logarithmically divergent integrals) which must now be 
compensated for. In order for the ultraviolet contribution 
$ -\frac{\mathcal{K}^2}{4} 
      \int_{| \vec{q} | > \Lambda } \frac{1}{4 q^3} $
(cf.\ \eq\nr{asy_UV}) 
and the infrared contribution from $|q| < \Lambda$ 
to add up to the correct result
in \eq\nr{Ihp_fz_vac_2}, the vacuum terms of the 
latter should yield  
\be
 - \frac{\mathcal{K}^2}{4(4\pi)^2} 
     \biggl( \frac{4 \Lambda^2}{\lambda^2} - \fr32 \biggr)
  \;. \la{correct}
\ee
Explicit integration shows, however, that they yield
\ba
 && \hspace*{-2cm} \frac{1}{2 (4\pi)^2}
 \int_{-\Lambda}^{\Lambda} \! \! {\rm d}q \, 
 \biggl\{
 \frac{{\rm sign}(q)}{2} \,
 \biggl[
 \frac{2 q (\mathcal{K}^2 - 2 p \ko)}{p^2} 
 + \frac{q  \mathcal{K}^2}{p^2}
 \ln\biggl| \frac{\lambda^2}{4 p q} \biggr|
 \biggr]
 \nn & & \; - \, 
 \frac{ {\rm sign}(q-p)}{2}
 \biggl[
 \frac{2 (q-p) (\mathcal{K}^2 - 2 p \ko)}{p^2} 
 + 
\frac{q  \mathcal{K}^2}{p^2}
 \ln \biggl| \frac{\lambda^2 q}
 {4 p (p-q)^2 } \biggr|
 \biggr]
 \biggr\}
  \nn & = &    
 \frac{1}{(4\pi)^2}
 \biggl[
  \ko p  - \frac{ \mathcal{K}^2 }{4}
   \biggl( \ln \frac{4 \Lambda^2}{\lambda^2} +  \fr12 \biggr)
 \biggr]
 \;. \hspace*{10mm} \la{wrong}
\ea
The difference of \eqs\nr{wrong} and \nr{correct} needs to be 
cancelled from the integrand of \eq\nr{Ihp_fz_n} if we 
employ the shifted variables. This finally yields 
\ba
 \rho^{(\rmi{v})}_{\mathcal{I}^{ }_\rmii{h'}} & \!\! = \!\! & 
 \rho^{\rmi{vac}}_{\mathcal{I}^{ }_\rmii{h'}} 
 \; + \; 
 \frac{\pi\mathcal{K}^2 }{2(4\pi)^4 k}
 \int_{\km}^{\kp} \! {\rm d}p \, 
 \frac{ n^{ }_{\sigma_4}(\ko -p) n^{ }_{\sigma_1}(p) }{
  n^{ }_{\sigma_0}(\ko) }
 \nn 
 & & \; \times \, 
 \biggl\{
   \ln \frac{4 (\kp - p)(p - \km) \Lambda^2 }{k^2 \mathcal{K}^2} + \fr92
 - \frac{4 \ko p}{\mathcal{K}^2}
 +  \int_{-\infty}^{\infty} \! {\rm d}q \, \biggl[ 
    \frac{\theta(|q|-\Lambda)}{|q|}
 \nn & & \qquad + \, 
 \Bigl( \fr12 + n^{ }_{\sigma_2}(q)\Bigr)
 \biggl(
 \frac{4 q (\mathcal{K}^2 - 2 p \ko)}{p^2 \mathcal{K}^2} 
 + \frac{2 q }{p^2}
 \ln\biggl| \frac{\lambda^2}{4 p q} \biggr|
 \biggr)
 \nn & & \qquad - \, 
 \Bigl( \fr12 + n^{ }_{\sigma_5}(q-p) \Bigr)
 \biggl(
 \frac{4 (q-p) (\mathcal{K}^2 - 2 p \ko)}{p^2 \mathcal{K}^2} 
 + 
 \frac{ 2 q }{p^2}
 \ln \biggl| \frac{\lambda^2 q}
 {4 p (p-q)^2 } \biggr|
 \biggr) \, \biggr] \, 
 \biggr\}
 \;, \hspace*{10mm} \la{Ihp_fz_final}
\ea
where the vacuum part has been defined as 
\be
 \rho^{\rmi{vac}}_{\mathcal{I}^{ }_\rmii{h'}} 
 \; \equiv \; 
 -\frac{\pi\mathcal{K}^2T}{2 (4\pi)^4 k}
    \ln \biggl( 
     \frac{e^{{\kp} / {T}}+\sigma_0 e^{-\kp/T}-\sigma_1 - \sigma_4}
     {e^{{\km} / {T}}+\sigma_0 e^{-\km/T}-\sigma_1 - \sigma_4}
   \biggr)
 \biggl( \frac{1}{\epsilon} + 2 \ln\frac{\bmu^2}{\mathcal{K}^2} + \fr92 \biggr)
 \;. \la{Ihp_vac}
\ee
There is no dependence on $\Lambda$ in \eq\nr{Ihp_fz_final}, 
and the $1/q$-tails
at $|q| > \Lambda$ cancel as well, 
so that the expression is integrable once combined with 
the real corrections. 
(In fact the integrand also has a constant 
part at large $|q|$, but this cancels due to its antisymmetry 
in $q \to -q$.)

For $\kp, \km \gg \pi T$, the ultraviolet asymptotics of 
the spectral function reads~\cite{nonrel}
\ba
 \rho^{ }_{{\mathcal{I}}_\rmi{h'}}  
 &  \stackrel{\kp,\km \gg \pi T}{\approx}  & 
  -\frac{\mufac\mathcal{K}^2}{8(4\pi)^3} \biggl( 
   \frac{1}{\epsilon}+ 2\ln\frac{\bmu^2}{\mathcal{K}^2} + \fr92 \biggr)
 \nn & & 
 \; + \, 
  \int_\vec{p} \biggl\{
 \frac{n^{ }_{\sigma_4}}{16\pi p}
 +
 \frac{p[3 (n^{ }_{\sigma_4} -  n^{ }_{\sigma_2}) + n^{ }_{\sigma_5}]}{24\pi}
 \, \frac{\ko^2 + \fr{k^2}3}{\mathcal{K}^4}
 \biggr\}
 \;. \la{asy_Ihp}
\ea
The specific statistics needed in this paper are
\ba
 \widehat{\mathcal{I}}^{ }_\rmi{h'} & \Leftrightarrow & 
 (\sigma_1\sigma_4\sigma_5 | \sigma_2 \sigma_0) = (-+-|+-)
 \;. \la{Ihp_pmm}
\ea
For numerical evaluation, we have reflected the final
integral to the domain defined in \fig{6} of ref.~\cite{master}. 
The corresponding integrand is not shown explicitly, since no 
substantial cancellations take place in the reflection. 
Results of numerical evaluations are shown in \fig\ref{fig:Ihp}, 
and it can be seen that the OPE-asymptotics from \eq\nr{asy_Ihp}
as well as the $k\to 0$ limit from ref.~\cite{bulk_wdep}
are reproduced. For $\sigma_2 = \sigma_5$ we have also checked
that the identity
$ 
 \rho^{ }_{{\mathcal{I}}_\rmi{h'}}  
 = 
 \fr12 
 ( 
   \rho^{ }_{{\mathcal{I}}_\rmi{f}}  + 
   \rho^{ }_{{\mathcal{I}}_\rmi{h}}  
 ) 
$, obtained by substitutions of sum-integration variables,   
is satisfied. 

\begin{figure}[t]


\centerline{%
 \epsfysize=7.0cm\epsfbox{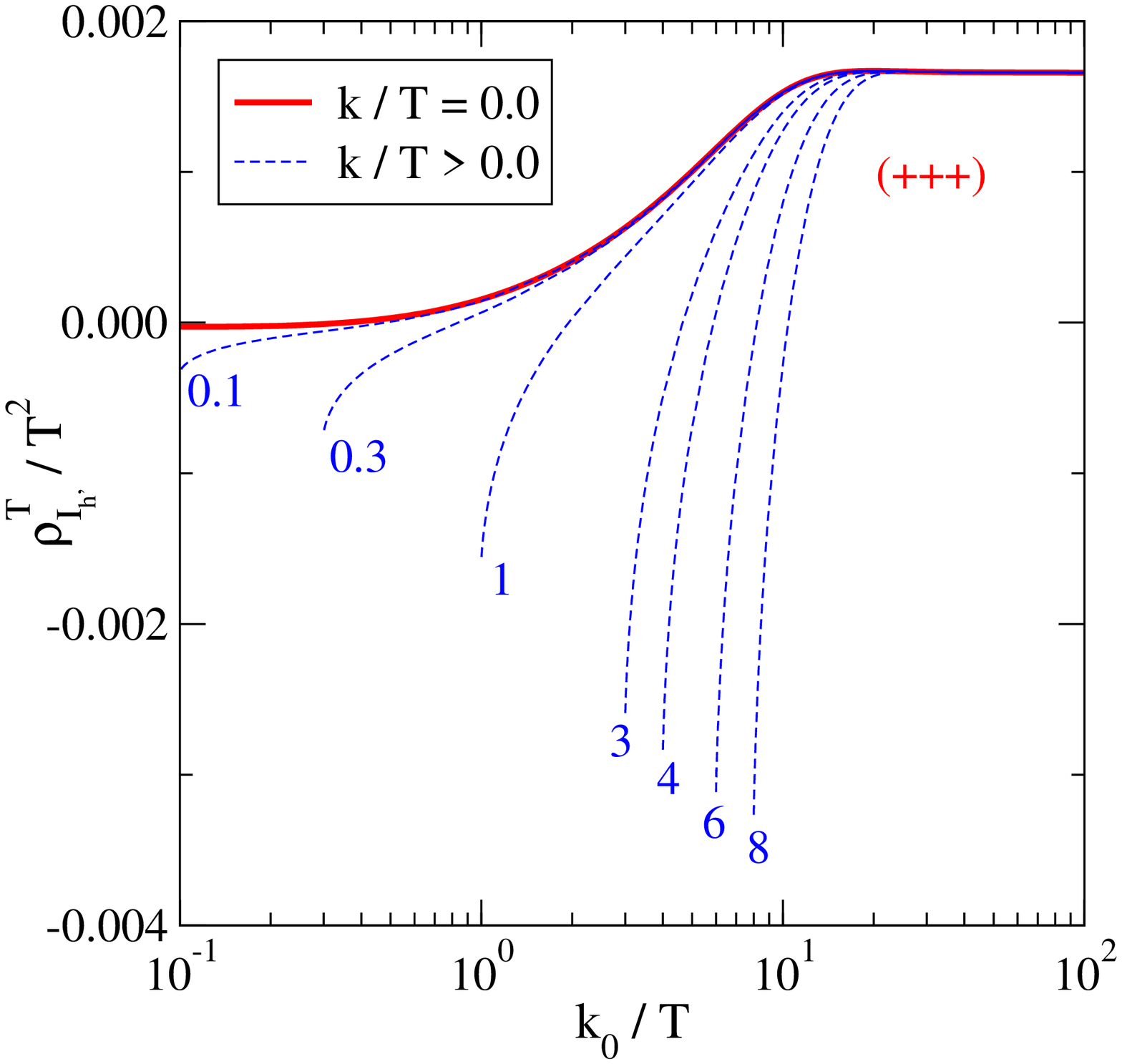}%
~~~\epsfysize=7.6cm\epsfbox{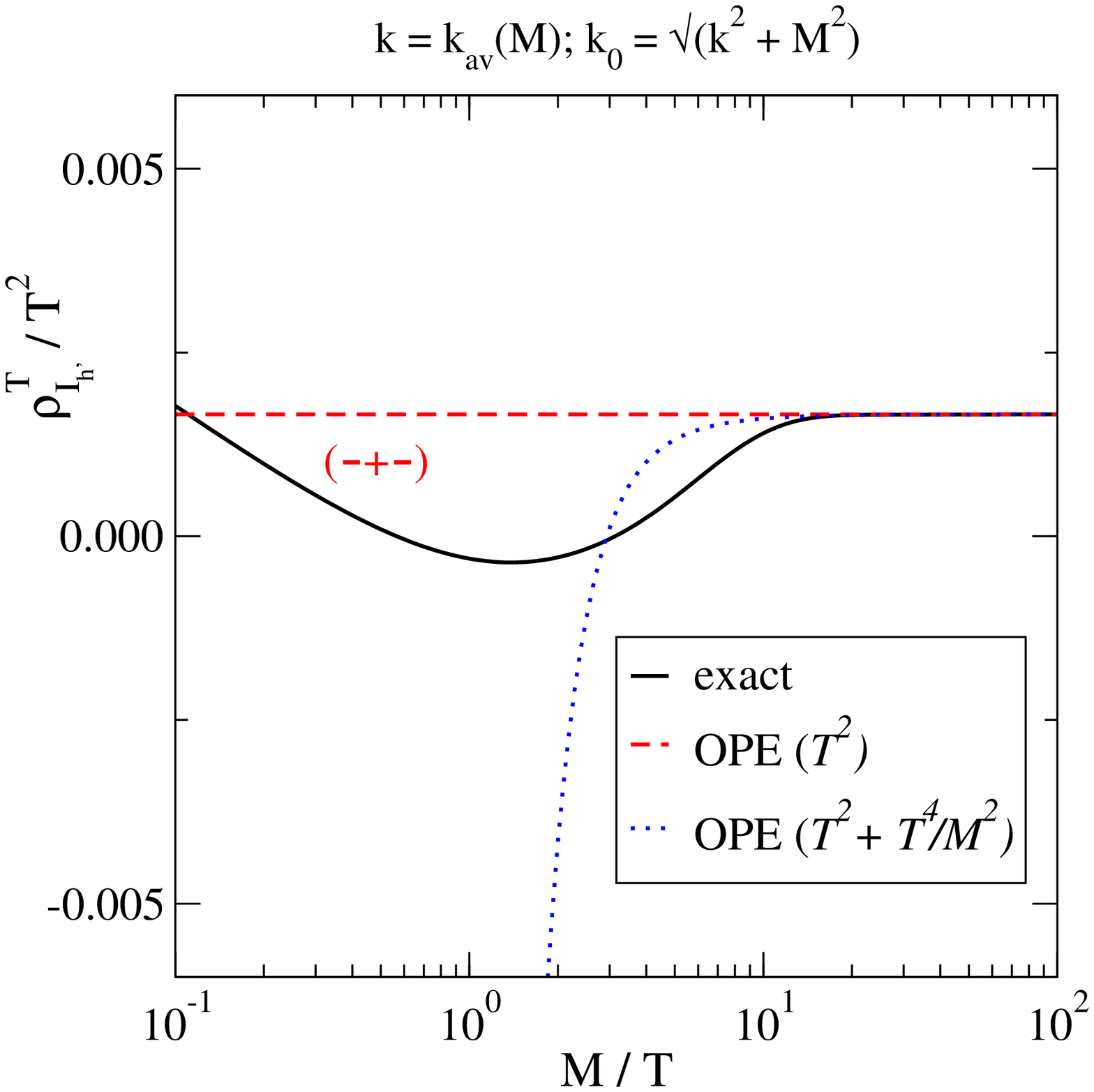}
}

\caption[a]{\small
Left: The thermal part of 
$\rho^{ }_{\mathcal{I}^{ }_\rmii{h'}}$ 
with the purely bosonic statistics $(\sigma_1\sigma_4\sigma_5) = (+++)$, 
for $\ko \ge k + 0.001 T$, 
compared with the zero-momentum limit determined in ref.~\cite{bulk_wdep}.
Right: The thermal part of 
$\rho^{ }_{\mathcal{I}^{ }_\rmii{h'}}$ with the momentum of 
\eq\nr{kav} and statistics of 
\eq\nr{Ihp_pmm} as a function of $M/T$, 
compared with the OPE-asymptotics from \eq\nr{asy_Ihp}. 
}

\la{fig:Ihp}
\end{figure}

%
\section{Choice of parameters}
\la{app:C}

For illustration we choose $M = 10^7$~GeV
for the numerics like in refs.~\cite{bb1,bb2}.
The physical Higgs mass is set to $m_H = 126$~GeV. 
In order
to convert pole masses and the muon decay constant to 
$\msbar$ scheme parameters at a scale $\bmu = \bmu_0 \equiv m_Z$
we employ 1-loop relations 
specified in ref.~\cite{generic}; 
subsequently, 1-loop renormalization group equations determine
the running of the couplings to a scale
\be
 \bmu_\rmi{ref} \equiv {\rm max}(M,\pi T)
 \;. \la{bmu_ref}
\ee
Within this approximation the U(1), SU(2) and SU(3) 
gauge couplings $g_1^2,g_2^2,g_3^2$ have explicit solutions
(we have set $\Nc = 3$ and considered 3 families), 
\ba
 g_1^2(\bmu)  =  \frac{48\pi^2}{41 
 \ln(\Lambda_1 / \bmu)}
 \;, \quad 
 g_2^2(\bmu)  =  \frac{48\pi^2}{19 
 \ln(\bmu/\Lambda_2)}
 \;, \quad
 g_3^2(\bmu)  =  \frac{24\pi^2}{21 
  \ln(\bmu/\Lambda_3)}
 \;, 
\ea
where $\Lambda_1,\Lambda_2,\Lambda_3$ are solved from the boundary 
values at $\bmu = \bmu_0$. The top Yukawa and the Higgs
self-coupling at $\bmu > \bmu_0$ are solved numerically from 
\ba
 \bmu \frac{{\rm d} h_t^2}{{\rm d}\bmu} & = & 
 \frac{h_t^2}{8\pi^2}
 \biggl[
  \fr92 
  h_t^2 
   - \frac{17}{12} g_1^2 - \fr94 g_2^2 - 8 
 g_3^2     
 \biggr]
 \;, \\ 
 \bmu \frac{{\rm d} \lambda}{{\rm d}\bmu} & = & 
 \frac{1}{8\pi^2}
 \biggl[ 
    \frac{3}{16} \Bigl( g_1^4 + 2 g_1^2 g_2^2 + 3 g_2^4 \Bigr)  
 - \frac{3}{2}\lambda \Bigl( g_1^2 + 3  g_2^2 \Bigr)
 + 12 \lambda^2
 + 6 
  \lambda h_t^2 
 - 3 h_t^4
 \biggr]
 \;. \hspace*{10mm}
\ea
Given many confusions in leptogenesis literature concerning
the physical 
value of $\lambda$ (numbers too large by a factor of 2 or 4
can be found even though everyone uses the same \eq\nr{mphi}), let us 
recall that at tree level 
$\lambda \approx g_2^2 m_H^2 / (8 m_W^2) \approx 0.13$.

%

\end{document}